\documentclass[aps,11pt,showkeys,preprint,superscriptaddress]{revtex4}
\usepackage[caption = false]{subfig}
\usepackage[final]{graphicx}
\usepackage{graphicx}
\usepackage{float}
\usepackage{graphicx}
\usepackage{bm}

\usepackage{eqnarray,amsmath}
\usepackage{tabularx}
\usepackage{amsmath}
\usepackage{txfonts}
\usepackage{epstopdf}
\usepackage{amsmath}
\usepackage{cleveref}
\usepackage{slashed}
\usepackage{graphicx,pstricks}
\usepackage{amsmath,amssymb}
\usepackage{amsfonts}
\newcommand{\hypgeo}[2]{%
  {\vphantom{F}}_{#1}\kern-\scriptspace F_{#2}%
}
\usepackage[english]{babel}
\usepackage{afterpage}
\usepackage{graphicx}
\usepackage{dcolumn}
\usepackage{bm}
\usepackage{color}
\usepackage{slashed}
\usepackage{latexsym}

\newcommand{\ii}{\textrm {i}}
\newcommand{\ee}{\textrm {e}}

\newcommand{\de}[1]{\partial_{#1}}

\newcommand{\br}[1]{\left( #1 \right)}

\newcommand{\bet}[1]{\left \lvert #1 \right \rvert}

\newcommand{\Hyper}[1]{ {}_2 \mathcal F_1 \br{#1}}

\usepackage{bm}
\usepackage{color}
\usepackage{slashed}
\usepackage{latexsym}
\usepackage{eqnarray,amsmath}
\usepackage{tabularx}
\usepackage{amsmath}
\usepackage{txfonts}
\usepackage{epstopdf}
\usepackage{amsmath}
\usepackage{slashed}
\usepackage{graphicx,pstricks}
\usepackage{amsmath,amssymb}
\usepackage{amsfonts}

\newcommand{\dd}{\mathrm{d}}  
\newcommand{\op}[1]{\hat{#1}} 
\newcommand{\bra}[1]{\langle{#1}\vert} 
\newcommand{\ket}[1]{\vert{#1}\rangle} 

\newcommand{\bea}{\begin{eqnarray}}
\newcommand{\eea}{\end{eqnarray}}
\newcommand{\eeq}{\end{equation}}
\newcommand{\bes}{\begin{subequations}}
\newcommand{\ees}{\end{subequations}}

\DeclareMathAlphabet\mathbfcal{OMS}{cmsy}{b}{n}

\usepackage{natbib}
\begin{document}
\title{Dynamical Scaling in Pair Production for Scalar QED}
\author{Deepak Sah\footnote{Corresponding author.\\E-mail address: deepakk@rrcat.gov.in \& dsah129@gmail.com (Deepak Sah).}}
\author{Manoranjan P. Singh}
\affiliation{Homi Bhabha National Institute, Training School Complex, Anushakti Nagar, Mumbai 400094, India}
\affiliation{Theory and Simulations Lab, Raja Ramanna Centre for Advanced
Technology,  Indore-452013, India}
\begin{abstract}
We report on the dynamical scaling of momentum spectra for particle-antiparticle pairs at finite times within the framework of scalar Quantum Electrodynamics (QED). The analysis focuses on the momentum spectra in two different choices of adiabatic mode functions, which are related by a Wronskian normalization condition. Oscillations in the momentum spectra are attributed to quantum interference effects in the adiabatic number basis. A novel dynamical scaling behavior emerges when examining the oscillatory momentum spectra of pairs created by a Sauter pulsed field at intermediate times. While the oscillatory spectra are observed at distinct times in the two different choices, they overlap when time is rescaled by the point marking the initiation of the first occurrence of the Residual Particle-Antiparticle Plasma (RPAP) stage (or end of the transient stage) for the central momentum case. This scaling identifies the approximate time at which real particle-antiparticle pair formation becomes possible, shifting the focus from asymptotic times to finite-time dynamics. Additionally, in the multi-photon regime, we find that the momentum spectra exhibit a multi-modal profile structure at finite times, consistent across both choices and also follow the dynamical scaling in this case as well.
\end{abstract}
\maketitle
\tableofcontents
\section{Introduction}
The spontaneous creation of particle-antiparticle pairs in the presence of strong electromagnetic fields is known as the Schwinger effect. This phenomenon was initially proposed by F. Sauter \cite{Sauter:1931zz,Sauter:1932gsa} and first studied through the effective action of a charged particle in a strong electromagnetic background by Heisenberg and Euler \cite{Heisenberg:1936nmg}, as well as Weisskopf \cite{Weisskopf:1936hya}. Schwinger later provided a comprehensive explanation within the framework of Quantum Electrodynamics for slowly varying fields \cite{Schwinger:1951nm}. 
The phenomenon of particle creation is generally characteristic of situations where an external agent strongly influences the vacuum state of a quantum field. In quantum field theory (QFT) in curved spacetimes, the curvature of spacetime serves as this agent, as seen in the Hawking effect, which describes radiation from black holes \cite{Hawking:1974rv}. However, direct experimental verification of particle creation due to gravitational fields remains elusive due to the minute nature of these effects. Nonetheless, quantum acoustic Hawking radiation from analogue black holes in atomic Bose-Einstein condensates has been observed \cite{Leonhardt:2016qdi}.
Similarly, empirical verification of the Schwinger effect requires generating electromagnetic fields exceeding the Schwinger limit,
$E_{c} \approx 10^{18} \text{V/m} $ \cite{Yakimenko:2018kih}, posing significant technical challenges. However, experimental proposals, such as those involving ultraintense lasers \cite{Marklund:2022gki,Vincenti:2018nzj}, hold promise for making the Schwinger effect one of the first non-perturbative phenomena to be tested. For a recent review, see \cite{Fedotov:2022ely}. Additionally, the Schwinger effect has been observed in an analogue mesoscopic experiment in graphene \cite{Berdyugin:2021njg,Schmitt:2022pkd}.
\newline
Particle creation occurs when external fields break symmetries in the quantum field theory. In flat spacetime, free fields exhibit Poincaré symmetry, and the canonical quantum theory remains invariant under this symmetry group. As a result, the solutions to the equations of motion expand uniquely in the plane wave basis, which defines the annihilation and creation operators. These operators, in turn, determine the Fock vacuum, known as the Minkowski vacuum. When a time-dependent external field interacts with matter fields, the classical Hamiltonian loses its invariance under time translations, leading to a loss of symmetry. This loss introduces flexibility in defining the annihilation and creation operators, making the vacuum state for Fock quantization non-unique. Different vacuum choices lead to different interpretations of physical quantities, such as the time evolution of particle-antiparticle pairs and the energy density. Understanding these dynamics is essential, as it reveals how quantum systems evolve under external influences and how physical observables depend on the chosen reference basis. The choice of vacuum depends on the system and the specific properties one wishes to impose on the quantum theory. Among the most commonly used vacua are \textit{adiabatic vacua}, introduced by Parker \cite{Parker:1969au} and formalized by Lüders and Roberts\cite{Lueders:1990np}. These vacua utilize the WKB approximation to extend the plane wave solutions from flat spacetime to scenarios with slowly varying external fields. Other approaches include diagonalizing the Hamiltonian \cite{Fulling:1979ac,Fahn:2018ahm}, minimizing the renormalized stress-energy tensor instantaneously \cite{Agullo:2014ica}, minimizing the time evolution of particle numbers in the Schwinger effect \cite{Dabrowski:2016tsx}, and among many others.
This paper focuses on exploring the dynamics of particle creation to provide a time-dependent picture of pair production. By diagonalizing the Hamiltonian using time-dependent Bogoliubov transformations. Additionally, kinetic phase-space methods derive a single-particle distribution function \( f(p, t) \), offering a dynamic view of particle creation in momentum space \cite{Schmidt:1998vi,Schmidt:1998zh}.
Understanding the time evolution of the distribution function $f(\bm{p}, t)$ is critical for connecting the behavior of quasiparticles to observable quantities \cite{Aleksandrov:2020mez,Smolyansky:2019dqd}. This includes examining how  momentum spectra evolve over time, as well as the dynamics of real particle-antiparticle pairs that survive after the external field ceases. This investigation provides insights into post-interaction phenomena such as the generation of pair annihilation photons \cite{Smolyansky:2010as,Blaschke:2010vs,Blaschke:2011af} and birefringence effects \cite{DiPiazza:2006pr}. Approximate solutions to the mode equations are introduced using the WKB approximation, which establishes a particle picture by decomposing the field operator into positive and negative frequency components \cite{Anderson:2005hi,Habib:1999cs,Kluger:1998bm,Kluger:1992md}. This decomposition produces time-dependent creation and annihilation operators, which define the instantaneous vacuum and particle number evolution. While the final particle distribution at late times is independent of the chosen basis, significant variations occur at intermediate times. This emphasizes the importance of studying the dynamics to understand how different reference bases influence physical interpretations. Rather than focusing on finding an ideal basis for pair creation, this paper emphasizes the broader physics that emerges from studying multiple bases. By analyzing the dynamics of momentum distribution functions in different reference frames, we can gain a deeper understanding of the physical observables and phenomena associated with these distributions. Recently, the study of quasiparticle excitations in graphene under the influence of an electric field has gained significant attention. These excitations exhibit behaviors similar to the electron-hole pair creation observed in QED (like -Schwinger effect). Recent works have used distribution functions to investigate these phenomena in graphene \cite{Blaschke:2022ppg, Panferov:2019stq, Smolyansky:2019dqd, Jiang:2024ilt}.
In this paper, we investigate  pair creation in a spatially homogeneous, time-dependent Sauter pulsed electric field, a widely studied configuration due to its analytic traceability \cite{Gavrilov:1996pz, Gelis:2015kya, Popov:1971iga}. Through the Bogoliubov transformation, we find that the single-particle distribution function \( f(\bm{p}, t) \) depends on the choice of mode function, particularly in the adiabatic basis. In this basis, the adiabatic mode function is expressed as a WKB-approximate solution, satisfying a normalized Wronskian condition, and is governed by two time-dependent functions: \( \Omega(\bm{p}, t) \) and \( V(\bm{p}, t) \).To analyze the particle distribution function across different bases, we consider two adiabatic bases, leveraging the freedom in defining \( \Omega(\bm{p}, t) \) and \( V(\bm{p}, t) \). We develop an analytical theory that is valid for finite times \( t > \tau \). In particular, we find an analytical expression for the one-particle distribution function in the power series of the small parameter \( (1 - y) \). The interesting dynamical features of the momentum distribution function at finite times are attributed to the function that appears in this expansion. Comparing the temporal evolution for both choices reveals three distinct stages: the quasi-particle particle-antiparticle plasma (QPAP) stage, the transient stage, and the residual particle-antiparticle plasma (RPAP) stage. These stages are consistent with well-known vacuum pair production in intense fields for fermions (Spinor QED) discussed in Ref.~\cite{Blaschke:2014fca}.The momentum spectra's time evolution shows interesting oscillatory behavior at non-asymptotic times. This oscillation arises due to quantum interference effects. For the second choice of adiabatic basis, the momentum spectra exhibit different behavior at early times but converge to similar oscillatory behavior near \( t = 2\tau \), particularly around the Gaussian peak at \( p_\parallel = 0 \), albeit at slightly different times. To better understand the time scales, we study the temporal evolution of the distribution function. The pair production process follows three distinct stages: the QPAP stage, the transient stage, and the RPAP stage.We then explore the oscillations in the momentum spectra hold physical significance by relating them to a specific time scale that provides an equivalent picture across both bases. For this, we define the time \( t_{\text{out}} \) as the first occurrence of the RPAP stage (or the end of the transient stage), where \( f(\bm{p} = 0, t = t_{\text{out}}) = f_{\text{const.}} \). Using this dynamical time scaling, we examine the momentum spectra at \( t = \frac{3}{4}t_{\text{out}}, \frac{5}{4}t_{\text{out}}, \frac{7}{4}t_{\text{out}} \) to visualize the behavior within the transient and RPAP stages. Our findings reveal that just before reaching the RPAP stage, the two choices show similar qualitative features but with differences in the oscillation amplitudes. In the RPAP stage, the results overlap, though fine details differ due to the dependence of \( t_{\text{out}} \) on momentum values. Furthermore, we propose that \( t = 2t_{\text{out}} \) can be identified as the time where both approaches converge exactly. This provides a finite time scale for interpreting real particle production instead of relying solely on asymptotic time scales.
Finally, we confirm that this dynamical scaling works well for multiphoton pair production in the regime \( \gamma > 1 \).
\newline
The structure of this article is outlined as follows: Section II provides a comprehensive theoretical framework, largely derived from \cite{Kluger:1998bm,Cooper:1989kf,}. In Section III, we introduce expressions for the particle momentum distribution function, leveraging the exact analytical solution for the mode function within a Sauter-pulsed electric field scenario. Section IV delves into the discussion of our findings, while Section V presents the concluding remarks of the article.
\newline 
Throughout the paper, we use natural units and set $ \hslash = c = m =|e|= 1 $, the electric charge $e < 0$, and express all variables in terms of the electron mass unit.

\section{Theory}
In this section, we will consider a comprehensive overview of the theoretical formulation of our problem by examining the canonical quantization of the charged scalar field in a uniform electric field within the semiclassical limit \cite{Kluger:1991ib}. Here, we fully quantize the matter field while treating the electromagnetic field classically, following the original references\cite{Cooper:1989kf,Kluger:1991ib, Kluger:1998bm}.
\newline
Consider an electric field along the $z-$direction.
It is related to the gauge potential through $E_{z}(t)= - \frac{\partial A_{z}(t)}{\partial t}.$
For a scalar particle of mass m and charge $e,$ the Klein-Gordon equation on the four dimensional Minkowski spacetime with the metric $(+,-,-,-)$ is given by 
\begin{equation}
    [ (\partial_{\mu} - \ii e A_{\mu})(\partial^{\mu} - \ii e  A^{\mu} ) + m^2] \hat{\Phi}(\bm{x},t) =0
    \label{eq1}
\end{equation}
    
where, $A_{\mu} = (0,0,0,A_{z} (t))$ and $\hat{\Phi}(\bm{x},t)$ is the scalar field.
and we Fourier expand the charged  scalar field operator as 
\begin{equation}
\hat{\Phi}(\bm{x},t)=\int\frac{\dd^3{\bm p}}{(2\pi)^3}\Bigl(\Phi_{\bm p}^{(+)}(x)\hat{b}_{\bm p}+\Phi_{\bm p}^{(-)}(x)\hat{d}_{-{\bm p}}^\dagger\Bigr)  ,
\label{bf1}
\end{equation}
where $\Phi_{\bm{p}}^{(\pm)}(x)$ are the one-particle solutions of the Klein-Gordon equation, and $\hat{b}_{\bm{p}}, \hat{b}_{\bm{p}}^\dagger, \hat{d}_{\bm{p}}, \hat{d}_{\bm{p}}^\dagger$ are the usual time-independent creation and annihilation operators. These operators define the vacuum in-state through the conditions $\hat{b}_{\bm{p}} \ket{0_{\text{in}}} = 0$ and $\hat{d}_{\bm{p}} \ket{0_{\text{in}}} = 0$. These time-independent operators obey the commutation relations:

\begin{equation}
[\op{b}_{\bm p},\op{b}_{{\bm p}'}^\dagger]=[\op{d}_{\bm p},\op{d}_{{\bm p}'}^\dagger]=\delta^{(3)}({\bm p}-{\bm p}'),
\label{bf2}
\end{equation}
with the remaining commutators equal to zero. 
\newline
The start point is the Klein-Gordon equation for a uniform and time-dependent electric field
 \begin{equation}
     [ ( \ii \partial - e A) - m^2 ] \Phi(x) = 0
     \label{kgeq1}
 \end{equation}
Due to the spatial homogeneity, we can write the $\Phi(x)$ in the convenient form 
\begin{equation}
\Phi(x) = \ee^{\ii {\bm p}\cdot {\bm x}}  \Phi_{\bm p}(t)
     \label{kgeq2}
\end{equation}
where, $\Phi_{\bm p}(t)$ satisfies the ordinary differential equation
\begin{equation}
\ddot{ \Phi}_{\bm p}(t) +  \omega_{\bm p}^2(t)  \Phi_{\bm p}(t) =0,
\label{kg3}
\end{equation}
with the time-dependent frequency $\omega_{\bm p}(t):$ 
\begin{equation}
\omega(\bm p,t)=\sqrt{ m^2 + p_\perp^2 +(p_{\|}-eA(t))^2}.
\label{kg4}
\end{equation}
Here, $p_\perp = |{\bm p}_\perp|$ is the modulus of vector ${\bm p}_\perp$ perpendicular to the field vector, and $p_\parallel =p_z$ is the momentum component parallel to the field.
\newline
To understand the solutions of Eq.~\eqref{kg3}, we observe that in the early time limit, as $t \to -\infty$, this equation simplifies to:
\begin{equation}
\ddot{ \Phi}_{\bm p}(t) +  \omega_{\bm p}^2 \Phi_{\bm p}(t) =0,
\label{kg6}
\end{equation}
where 
\begin{equation}
\omega_{\bm p}=\sqrt{m^2+{\bm p}^2}.
\label{kg7}
\end{equation}
Therefore,  there exist two linearly independent solutions of Eq.~\eqref{kg6} which we will label by the parameter $\sigma$,
\begin{equation}
\Phi_{\bm p}^{(\sigma)}(t)\underset{t\rightarrow -\infty}{\sim}\ee^{-\ii\sigma\omega_{\bm p}t}.
\label{kg8}
\end{equation}
The one corresponding to a positive energy (with $\sigma=+$) will be interpreted as a particle whereas the other one (with $\sigma=-$) as
an antiparticle. In this manner, we have identified two sets of solutions to the Klein-Gordon equation, as presented in Eq.
\eqref{kgeq2},
\begin{equation}
\Phi_{\bm p} ^{(\sigma)} (x) = \ee^{\ii {\bm p}\cdot {\bm x}}  \Phi_{\bm p}^{(\sigma)}(t)
     \label{kgeq2n1}
\end{equation}
where  $\Phi_{\bm p}^{(\sigma)}(t) $ solves Eq.\eqref{kg3} and asymptotically behaves according to \eqref{kg8}.
Note that the mode function $\Phi_{\bm{p}}^{(\sigma)}(t)$ is normalized using the Wronskian condition,

\begin{align}
     [\Phi_{\bm p}^{(\sigma)}(t)]^* \dot{\Phi}_{\bm p}^{(\sigma)} (t)  -     [\dot{\Phi}_{\bm p}^{(\sigma)}(t)]^* \Phi_{\bm p}^{(\sigma)} (t) &= -\ii
     \label{eq4}
\end{align}
Up to this point, creation and annihilation operators, as well as mode functions, have been considered time-independent. This approach assumes a clear separation into positive and negative energy solutions, facilitating a straightforward interpretation of particles and antiparticles. However, in a time-dependent background field, solving the equation of motion for the mode functions, reveals that such a separation is no longer apparent. This ambiguity makes it challenging to interpret the solutions in terms of particles and antiparticles. Furthermore, the non-diagonal nature of the Hamiltonian in prior calculations complicates the interpretation further, as it indicates that the particle number is not conserved. To address these issues and achieve a meaningful interpretation, the system is reformulated to incorporate time-dependent quantities. This involves transitioning from time-independent particle numbers and operators to time-dependent creation and annihilation operators, along with time-dependent mode functions. This transformation allows for a more accurate representation of the system, capturing dynamic processes such as particle creation and annihilation over time. This transformation can be achieved through the time-dependent Bogoliubov transformation \cite{Greiner:1992bv,Reinhardt:1994aha}. 
To achieve a meaningful definition, we can define a number basis that depends on the adiabatic mode function. This approach takes advantage of the fact that adiabatic mode functions are solutions to the equations of motion under slowly varying background fields. The resulting particle number basis is intrinsically linked to the adiabatic approximation, representing the instantaneous state of the system at any given time. This dependence ensures that the particle interpretation remains consistent with the chosen adiabatic reference frame, effectively capturing the dynamics of particle creation and annihilation under varying external conditions.

\subsection{Mode functions and Bogolubov transformation}
The Bogoliubov transformation defines a set of time-dependent creation and annihilation operators, $\op{b}_{\bm p}(t)$and $\op{d}_{\bm p}(t)$, which are related to the original time-independent operators, $\op{b}_{\bm p}$and $\op{d}_{\bm p}$, through a linear transformation:
\begin{equation}
\begin{pmatrix}
\op{b}_{\bm p}(t) \\
\op{d}^\dagger_{-{\bm p}} (t)
\end{pmatrix}
=
\begin{pmatrix}
\alpha_{\bm p}(t) & \beta_{\bm p}^*(t) \\
\beta_{\bm{p}}(t) & \alpha_{\bm{p}}^*(t)
\end{pmatrix}
\begin{pmatrix}
\op{b}_{\bm p}\\
\op{d}^\dagger_{-{\bm p}} 
\end{pmatrix}.
\end{equation}
with the unitarity   condition 
    \begin{equation}
|\alpha_{\bm p}(t)|^2-|\beta_{\bm p}(t)|^2=1.
\label{bf11}
\end{equation}
be satisfied at all times.
Thus the new operators $\op{b}_{\bm p}(t)$and $\op{d}_{\bm p}(t)$  describe quasiparticles at the time t with the instantaneous vacuum $ \ket{0_{\text{t}}}  $.
Applying the Bogoliubov transformation to Eq.\eqref{bf1} yields a new representation for the field operators.
\begin{equation}
\op{\Phi}(\bm{x},t)=\int\frac{\dd^3{\bm p}}{(2\pi)^3}\Bigl(\phi_{\bm p}^{(+)}(x)\hat{b}_{\bm p}(t)+ \phi_{\bm p}^{(-)}(x)\hat{d}_{-{\bm p}}^\dagger(t)\Bigr),
\label{bf18}
\end{equation}
 The correspondence  between the new  $\phi_{\bm p}^{(\pm)}(x)$ and the former  $\Phi_{\bm p}^{(\pm)} (x)$ functions  is defined by a canonical transformation 
 \begin{align}
\phi_{\bm p}^{(+)}(x)&=\alpha_{\bm p}^*(t)\Phi_{\bm p}^{(+)}(x)-\beta_{\bm p}^*(t)\Phi_{\bm p}^{(-)}(x),\label{bf19}\\
\phi_{\bm p}^{(-)}(x)&=\alpha_{\bm p}(t)\Phi_{\bm p}^{(-)}(x)-\beta_{\bm p}(t)\Phi_{\bm p}^{(+)}(x),\label{bf20}
\end{align}
Therefore it is justified to assume that 
 \begin{align}
      \phi_{\bm p}^{(\sigma)}(x) &= \ee^{\ii {\bm p}\cdot {\bm x}}  \tilde{\phi}_{\bm p}^{(\sigma)}(t)
      \label{bf221}
 \end{align}
Where \(\tilde{\phi}_{\bm{p}}^{(\sigma)}(t)\) are yet unknown functions. These time-dependent mode functions, \(\tilde{\phi}_{\bm{p}}^{(\sigma)}(t)\), are not exact solutions to the equation of motion for \(\Phi_{\bm{p}}^{(+)}(t)\). However, they can still be employed as mode functions for constructing the time-dependent (but adiabatic) number basis.
 The  substituation  of Eq.\eqref{bf19}  \eqref{bf20}into Eq.\eqref{bf18}  leads  to the relations
  \begin{align}
\Phi_{\bm p}^{(+)}(t)=\alpha_{\bm p}(t)&\tilde{\phi}_{\bm p}^{(+)}(t)+\beta_{\bm p}^*(t)\tilde{\phi}_{\bm p}^{(-)}(t),\label{bf22}\\
\Phi_{\bm p}^{(-)}(t)=\alpha_{\bm p}^*(t)&\tilde{\phi}_{\bm p}^{(-)}(t)+\beta_{\bm p}(t)\tilde{\phi}_{\bm p}^{(+)}(t).
\label{bf23}
\end{align}
We now able to construct the distribution function of particle with momentum $\bm{p}$ \cite{Banerjee:2018fbw},
 \begin{align}
   f(\bm{p},t) &=
\bra{0_{in}}\op{b}_{\bm p}^\dagger(t)\op{b}_{\bm p}(t)\ket{0_{in}}
     \label{disfun0}
\end{align}
and anti-particle
\begin{align}
   \bar{f}(\bm{p},t) &=
   \bra{0_{in}}\op{d}_{\bm -p}^\dagger(t)\op{d}_{\bm -p}(t)\ket{0_{in}}
      \label{disfun0}
\end{align}
Charge conservation implies that \( f(\bm{p}, t) = \bar{f}(\bm{p}, t) \), ensuring that the summation over momentum provides the normalization for the total time-dependent adiabatic particle number. 

 \begin{align}
      \sum_{p} f(\bm{p},t) &=    \sum_{p} \bar{f}(\bm{p},t) = N(t).
 \end{align}
In other words, it represents the number of quasiparticles in the system at a given time \( t \).
\newline
By knowing the exact solution $\Phi_{\bm p}^{(\pm)}(t)$ and reference  adiabatic mode functions $\tilde{\phi}_{\bm p}^{(\pm)}(t)$ , which allows for the calculation $\beta_{\bm p}(t).$
Therefore, we can obtain an expression  for $f(\bm{p},t)$ that rely on the choice of reference basis functions  $\tilde{\phi}_{\bm p}^{(\pm)}(t)$.
\subsubsection{The Adiabatic mode function and choices}
\label{adibatic vacuum}
In a slowly changing time-dependent background field, the adiabatic vacuum often provides a more meaningful notion of particle number\cite{Kluger:1998bm,Winitzki:2005rw,Dabrowski:2014ica, Dabrowski:2016tsx}.
The procedure relies on the WKB approximation to solve mode equation,
\begin{align}
     \ddot{\tilde{\phi}}_{\bm p}(t) + \omega^2_{\bm{p}}(t) \tilde{\phi}_{\bm p}(t) &=0.
     \label{wkb0}
\end{align}

For the ansatz
\begin{equation}
    \tilde{\phi}_{\bm p}(t)=\frac{1}{\sqrt{2 \Omega_{\bm{p}} (t)}} \ee^{-{\ii}\int_{t_0}^t \Omega_{\bm{p}} (t') dt'}
     \label{wkb1}
\end{equation}
where $\Omega_{\bm{p}} (t)$ is an unspecified function. 
By substituting the WKB-ansatz  Eq.\eqref{wkb1} into the differential Eq.\eqref{wkb0}, one obtains an equation for $\Omega_{\bm{p}} (t)$  as
\begin{align}
     0 &=  \omega_{\bm{p}}^2(t) -  \Omega_{\bm{p}}^2(t) + \left[\frac{3}{4}\left(\frac{\dot{\Omega}_{\bm{p}}(t)}{\Omega_{\bm{p}}(t)}\right)^2 - \frac{\ddot{\Omega}_{\bm{p}}(t)}{2\Omega_{\bm{p}}(t)}\right]
\end{align}
Assuming the adiabatic behavior of \(\omega_{\bm{p}}(t)\), characterized by the condition 
\[
\left|\frac{\dot{\omega}_{\bm{p}}(t)}{\omega_{\bm{p}}^2(t)}\right| \ll 1,
\]
the equation can be solved iteratively as:

\begin{equation}
   ( \Omega_{\bm{p}}^{(j+1)})^2=\omega_{\bm{p}}^2-\left[\frac{\ddot{\Omega}_{\bm{p}}}{2 \Omega_{\bm{p}}}-\frac{3}{4}\left(\frac{\dot{\Omega}_{\bm{p}}}{\Omega_{\bm{p}}}\right)^2\right]\Biggr|_{\Omega_{\bm{p}}=\Omega_{\bm{p}}^{(j)}},
   \label{j-th}
\end{equation}
where $j=0,1,2,\cdots$ with $\Omega_{\bm{p}}^{(0)}=\omega_{\bm{p}}$ as the lowest solution. At the $(j+1)$-th order, one has to take the terms containing at most $2(j+1)$ time derivatives. For example, the first order is given by
\begin{equation}
    ( \Omega_{\bm{p}}^{(1)})^2=\omega_{\bm{p}}^2-\left[\frac{\ddot{\omega}_{\bm{p}}}{2\omega_{\bm{p}}}-\frac{3}{4}\left(\frac{\dot{\omega}_{\bm{p}}}{\omega_{\bm{p}}}\right)^2\right].
\end{equation}
It can be seen as a generalization of the WKB approximation, where the WKB approximation corresponds to the zeroth order of the adiabatic expansion.
For any adiabatic order, we can construct the solution of the mode equation \eqref{bf23} using auxiliary functions $\alpha_{\bm p}(t)$ and  $\beta_{\bm p}(t)$ . Considering the $j$-th order, we formally write the mode function as follows:
\begin{equation}
\Phi_{\bm p}^{(+)}(t)=\alpha_{\bm p}(t) \tilde{\phi}_{\bm p}^{+(j)}(t)+\beta_{\bm p}^*(t)\tilde{\phi}_{\bm p}^{-(j)}(t)
\label{mdj}
\end{equation}
where
\begin{equation}
    \tilde{\phi}_{\bm p}^{\pm (j)} (t)\equiv  \frac{1}{\sqrt{2 \Omega_{\bm{p}}^{(j)}} (t)} \exp\left[\mp{\rm i}\int_{t_0}^t\Omega_{\bm{p}}^{(j)} (t')dt'\right]
\end{equation}
are the $j$-th order basis functions. 
\newline
We require the first time-derivative of $\Phi_{\bm p}^{(+)}(t)$ to be
\begin{equation}
    \dot{\Phi}_{\bm{p}}^{(+)}(t)= Q_{\bm{p}} (t) \alpha_{\bm p}(t)\tilde{\phi}_{\bm p}^{+(j)}(t)+ Q^*_{\bm{p}} (t) \beta_{\bm p}^*(t)\tilde{\phi}_{\bm p}^{-(j)}(t),
    \label{1stder}
\end{equation}
Here, $ Q_{\bm{p}} (t) = - \ii \Omega_{\bm{p}} (t) + V_{\bm{p}} (t),$
and $ V_{\bm{p}} (t)$ is a real  time-dependent function . 
We can verify that the normalization condition $\dot{\Phi}_{\bm p}^{(-)}(t) \Phi_{\bm p}^{(+)}(t) -\Phi_{\bm p}^{(-)}(t) \dot{\Phi}_{\bm p}^{(+)}(t)= -\ii$ holds true if $\alpha_{\bm{p}}(t)|^2 - |\beta_{\bm{p}}(t)|^2 = 1$ not dependent on real function $ V_{\bm{p}} (t).$  
The flexibility in selecting $ \Omega_{\bm{p}} (t)$ and $V_{\bm{p}} (t)$ reflects the arbitrary nature of defining positive and negative energy states at non-asymptotic times.
The suggested ``natural choice" for this degree of freedom is proposed in \cite{Habib:1999cs,Dabrowski:2014ica} as
\begin{align}
    V_{\bm{p}}^{(j)} (t)= -\frac{\Omega_{\bm{p}}^{(j)}(t)} {2 \Omega_{\bm{p}}^{(j)}(t)}.
    \label{V_p}
\end{align}
Therefore, the time-dependent Bogoliubov coefficients may be found explicitly:
\bes
\bea
\alpha_{\bm p}(t)&=& i \tilde \phi_{\bm p}^{-(j)}(t)
\left[\dot \Phi_{\bm p}^{(+)}(t) - \left( i \Omega_{\bm p}^{(j)}(t)+ V_{\bm p}^{(j)} (t) \right) \Phi_{\bm p}^{(+)}(t)
\right]\\
\beta_{\bm p} (t)&=& -i \tilde \phi_{\bm p}^{+(j)}(t)
\left[ \dot \Phi_{\bm p}^{(+)} (t)+
\left( i\Omega_{\bm p}^{(j)}(t) - V_{\bm{p}}^{(j)} (t)\right) \Phi_{\bm p}^{(+)}(t) \right]  \label{genbet}
\eea
\label{genalpbet}
\ees
and in particular
\begin{equation}
    |\beta_{\bm p}(t)|^2 = \frac{1}{2 \Omega_{\bm p}(t)}  \left\vert\dot \Phi_{\bm p}^{(+)}(t) +
\left( i \Omega_{\bm p}^{(j)} (t)- V_{\bm p}^{(j)} (t)\right) \Phi_{\bm p}^{(+)}(t)\right\vert^2
\label{betsq}
\end{equation}

is determined in terms of the adiabatic frequency functions $(\Omega_{\bm{p}} (t), V_{\bm{p}} (t))$
and the exact mode function solution $\Phi_{\bm p}^{(+)}$ of the oscillator Eq.\eqref{kg3}, which is specified by initial point $(\Phi_{\bm p}^{(+)}, \dot{\Phi}_{\bm p}^{(+)})$
at $t= t_0$.
\newline
The choice of $(\Omega_{\bm{p}}^{(j)}(t),V_{\bm{p}}^{(j)}(t))$ is not unique, but it is constrained by the need to match the adiabatic behavior of the asymptotic expansion \eqref{j-th} to a sufficiently high order. 
While the detailed time dependence of $f(\bm{p},t)$ is influenced by the specific choice of $(\Omega_{\bm{p}}^{(j)}(t), V_{\bm{p}}^{(j)}(t))$, the main characteristics in asymptotic time limit remain largely unaffected by these choices.
\newline
In this paper, we consider the two common choices for adiabatic frequency functions  $(\Omega_{\bm{p}}(t), V_{\bm{p}}(t))$  that extensively discussed in the literature in concern of the time-dependent particle number \cite{Barry:1989zz, Dumlu:2010ua}.
In this context, we study the single-particle distribution function defined concerning two different adiabatic bases based on these choices.
Two common choices are as follows.
\newline
$(1)$ $\Omega_{\bm{p}}(t) = \omega_{\bm{p}}(t)$ and $V_{\bm{p}}(t) =0.$ This approach is taken, for example, in \cite{Brezin:1970xf,Dumlu:2010ua,Anderson:2017hts}.
\newline
$(2)$ $\Omega_{\bm{p}}(t) = \omega_{\bm{p}}(t)$ and $V_{\bm{p}}(t) =-\frac{\dot{\omega}_{\bm{p}}(t)}{2 \omega_{\bm{p}}(t)}.$ This approach is taken, for example, in \cite{Barry:1989zz,Kluger:1998bm,Anderson:2013ila}.

\section{Particle distribution function for Sauter-pulse electric field}
\label{pair}
Let us consider a simple model of  the electric  field,
\begin{align}
     E_{z}(t)=E_{0} sech^{2}(t/\tau)
     \label{Et}
\end{align}
where,  $\tau $ is the  duration of pulse and $E_0$ is field strength. 
 The corresponding vector potential is
\begin{align}
A_{z}(t)=-E_{0}\tau \tanh \Bigl(\frac{t}{\tau} \Bigr)
\label{At}
\end{align}

Now, Eq. (\ref{kg3}) can be rewritten in the presence of electric field Eq.\eqref{Et} 
\begin{equation}
\ddot{\Phi}_{\bm p}(t) +\omega_{\bm p}^2(t) \Phi_{\bm p}(t)=0,
\label{4.3}
\end{equation}
where,
 \begin{align}
     \omega_{\bm p}^2(t)&= \sqrt{ \Biggl(p_{\parallel}- e E_{0}\tau \tanh \Bigl(\frac{t}{\tau} \Bigr) \Biggr)^{2}+p^{2}_{\perp}+ m^{2}}.
     \label{4.4}
 \end{align}
 This equation can be solved by converting it into a hypergeometric differential equation\cite{abramowitz}, by changing the time variable to $y = \frac{1}{2} \left(1 +\tanh \Bigl(\frac{t}{\tau} \Bigr)\right).$ 
 \newline
 The new variable $y$ transforms the equation as
 \begin{align}
    \br{\frac{4}{\tau^2} y \br{1-y} \de{y} y \br{1-y} \de{y} + \omega^2(\bm{p},y) } \Phi_{\bm p}(y)  = 0.
\label{4.5}
\end{align}
In this case, solutions can be written in terms of
hypergeometric functions \cite{abramowitz} .
The two linearly independent solutions of Eq. (\ref{4.5}):
\begin{align}
     \Phi_{\bm p}^{(+)}(y)  = C^{(+)} y^{-i\tau\omega_0/2} (y-1)^{i\tau\omega_1/2}  \Hyper{a,b,c;y}
     \label{4.6}
\end{align}
\begin{align}
      \Phi_{\bm p}^{(-)}(y)  = C^{(-)} y^{i\tau\omega_0/2} (y-1)^{i\tau\omega_1/2}  \Hyper{a-c+1,b-c+1,2-c;y}
      \label{4.7}
\end{align}
where , $C^{(\pm)}(p)$ are some normalization constants
and $\Hyper{a,b,c;y}$ is the hypergeometric function.
\bea
a&=&\frac{1}{2}+\frac{i}{2}(\tau\omega_{1}-\tau\omega_{0})-i\lambda,\nonumber\\
 b&=&\frac{1}{2}+\frac{i}{2}(\tau\omega_{1}-\tau\omega_{0})+i\lambda,\nonumber\\
 c&=&1-i\tau \omega_{0},\nonumber\\
 \lambda&=&\sqrt{(e E_{0}\tau^{2})^{2}-\frac{1}{4}}.
 \label{4.8}
 \eea
 in which $\omega_{0}$ and $\omega_{1}$ are the kinetic energies of the field modes at asymptotic initial and final times.
 \bea
 \omega_{0}&=&\sqrt{(p_\parallel+ e E_{0}\tau)^{2}+p^{2}_{\perp}+m^{2}},\nonumber\\
 \omega_{1}&=&\sqrt{(p_\parallel-e E_{0}\tau)^{2}+p^{2}_{\perp}+m^{2}}.
 \label{4.9}
 \eea
\vspace{2.5mm}
To obtain the single-particle distribution function using Eq.~\eqref{disfun0}, we convert all functions in $\beta(\bm{p},t)$, given by Eq.~\eqref{betsq}, to the new time variable $y $. This transformation yields:
\begin{align}
    | \beta(\bm{p},y) |^2 =\frac{1}{ 2 \Omega(\bm{p},y) } 
    \left| \Bigl(  \frac{2}{\tau} y (1-y) \partial_y  + \ii Q(\bm{p},y) \Bigr)\Phi^{(+)}(\bm{p},y) \right|^2
\end{align}
Using Eq.~\eqref{4.6}, the analytical expression for the single-particle distribution function in terms of the transformed time variable $y$ can be written as 
\begin{align}
      f(\bm{p},y)  &=  \frac{|C^{(+)}(\bm{p})|^2 }{ 2\Omega(\bm{p},y) }  \bigg| \frac{2}{\tau} y (1-y)  \frac{a b}{c} g_1 + \ii (Q(\bm{p},y) - (1-y) \omega_0 - y \omega_1 ) g_2 \bigg|^2 
      \label{eqfp}
\end{align}
where,
$ g_1 = \Hyper{1+a,1+b,1+c;y}$, $g_2 =\Hyper{a,b,c;y}.$ and 
$Q(\bm{p},y)  = \ii \Omega(\bm{p},y) - V(\bm{p},y).$ 
\newline
This expression for the time-dependent single-particle momentum distribution function provides insights into pair creation at various dynamical stages \cite{Blaschke:2011is}. 
Using this expression, we investigate the dynamical behavior of $f(\bm{p},t)$ using two different choices of adiabatic frequency functions, $\Omega(\bm{p},y)$ and $V(\bm{p},y)$, as discussed in Section \ref{adibatic vacuum}. This analysis is based on the different adiabatic frequency functions chosen. A detailed discussion of these findings is presented in Section \ref{result} to understand the role of the adiabatic choice during pair formation at finite times.

\subsection{Approximate analytical expression for $f(\bm{p},t)$ }
\label{Approximate analytical}
We use  approximations based on Gamma and Gauss-hypergeometric functions to analyze the behavior of the function $f(\bm{p}, t)$ in the late-time limit. These approximations help us derive simplified analytical expressions for $f(\bm{p}, t)$. We start by approximating the Gauss-hypergeometric function as $y \to 1$. Achieving smooth convergence to the limit $\Hyper{a,b,c;y \rightarrow 1}$ requires a solid understanding of this limit. The complexity of the variables $a$, $b$, and $c$ in this context demands careful handling. Thus, we transform the argument by substituting $y$ with $(1-y)$ using the following mathematical identity \cite{abramowitz}.
\begin{multline}
\Hyper{a,b,c;z} = \frac{\Gamma \br{c} \Gamma \br{c-a-b}}{\Gamma \br{c-a} \Gamma \br{c-b}} \Hyper{a,b,a+b-c+1;1-z} \\
  + \br{1-z}^{c-a-b} \frac{\Gamma \br{c} \Gamma \br{a+b-c}}{\Gamma \br{a} \Gamma \br{b}} \Hyper{c-a, c-b, c-a-b+1; 1-z} .\\
\bet{\textrm{arg}(1- z )} < \pi
\end{multline}
In general Gauss-Hypergeometric function,
\begin{align}
    \Hyper{a,b,c;z}  &=  1 +  \frac{ a b }{c } z +  \frac{ a(a+1)  b(b+1) }{c (c+1) } \frac{z^2}{2!} +  \frac{ a(a+1)(a+2)   b(b+1)(b+2) }{c (c+1)(c+2) } \frac{z^3}{3!}+...
    \label{1.69}
 \end{align}
 The series continues with additional terms involving higher powers of $z.$ Each term in the series involves the parameters $a, b,$ and $c$, as well as the variable $z$ raised to a specific power.
To approximate the particle distribution function at finite time  Eq.\eqref{eqfp}, we truncate the power series of the Gauss-hypergeometric functions $g_1$ and $g_2$  to a certain order. The truncation order depends on the required accuracy and the specific finite-time behavior under study.
.\newline
Let's start by approximating the different terms present in the particle distribution relation \eqref{eqfp}:
\begin{align}
    \frac{2}{\tau} y(1-y)  \frac{ab}{c} g_1 &= \frac{2}{\tau}y(a+b-c)  \Gamma_2 (1-y)^{(c-a-b)} +(1-y)
\Bigl(\frac{2 } {\tau}  y a b \Gamma_1 -  \frac{2 y }{\tau}(c-a)(c-b)  \Gamma_2 (1-y)^{(c-a-b)}
\Bigr) \nonumber \\
  & + (1-y)^2 \Biggl( \frac{2}{\tau}   y \Gamma_1 \frac{a(1+a) b(1+b)}{(2+a+b-c)}  + \frac{2}{\tau} y \Gamma_2 (1-y)^{(c-a-b)} \frac{(c-a)(c-b)(c-a+1)(c-b+1)}{(a+b-c-1)}\Biggr)
\end{align}
 where, $\Gamma_1  =\frac{ \Gamma ( c) \Gamma (c-a-b-1) }{\Gamma (c-a) \Gamma (c-b)}$ and $\Gamma_2 =\frac{ \Gamma ( c) \Gamma (a+b-c) }{\Gamma (a) \Gamma (b)} $
 \newline
 Similarly,
 \begin{align}
     \Bigl(Q(\bm{p},y) - (1-y) \omega_0- y \omega_1 \Bigr) f_2 =   (Q(\bm{p},y) - (1-y) \omega_0 - y \omega_1) \Biggl(\Gamma_1 (c-a-b-1) 
     \nonumber \\
\Bigl(1 + (1-y) \frac{ a b}{(1+a+b-c)} \Bigr) +  (1-y)^{(c-a-b)} \Gamma_2 \Bigl(1 + (1-y) \frac{(c-a)(c-b)}{(1+c-a-b)} \Bigr)
\end{align}
Also, it is possible to write down the time-dependent quasi-energy $\omega(\bm{p},y)$ as the following series expansion near 
$y \rightarrow 1:$ 
\begin{align}
     \omega(\bm{p},y) & \approx \omega_1 +  w_1 (1-y) + w_2 (1-y)^2 
\end{align}
up to second order.
Here, \begin{align}
     w_1 &=-\frac{2 E_0 e \tau (-p_\parallel + E_0 e \tau)}{\sqrt{1 + p_\parallel^2 - 2 E_0 p_\parallel e \tau + E_0^2 e^2 \tau^2}}\nonumber \\
w_2 &=\frac{2 E_0^2 e^2 \tau^2}{(1 + p_\parallel^2 - 2 E_0 p_\parallel e \tau + E_0^2 e^2 \tau^2)^{3/2}}
\end{align}
Similarly,
\begin{align}
      V(\bm{p},y) & \approx V_1 (1-y) + V_2 (1-y)^2  
\end{align}
Here, \begin{align}
     V_1 &=\frac{4 E_0 e (p_\parallel - E_0 e \tau)}{1 + (p_\parallel - E_0 e \tau)^2}\nonumber \\
V_2 &=-\frac{4 E_0 e \left(p_\parallel + p_\parallel^3 - E_0 \left(3 + p_\parallel^2\right) e \tau - E_0^2 p_\parallel e^2 \tau^2 + E_0^3 e^3 \tau^3\right)}{\left(1 + \left(p_\parallel - E_0 e \tau\right)^2\right)^2}
\end{align}
 therefore,
 \begin{align}
      \Bigl(\omega(\bm{p},y) - (1-y) \omega_0(\bm{p})- y \omega_1(\bm{p}) \Bigr) f_2 = (1-y) \Biggl((\omega_1 - \omega_0) - q_1 \Biggr) \Biggl(\Gamma_1 (c-a-b-1)  + e^{- \ii \tau\omega_1(\bm{p}) \ln{(1-y)})} \Gamma_2\Biggr)
      \nonumber \\
      + (1-y)^2 \Biggl( \Gamma_1 (c-a-b-1)  \Biggl(  q_2  +     \frac{ a b ( (\omega_1 - \omega_0) + q_1)}{ (1+a+b-c)} \Biggl)
      \nonumber \\
      + \Gamma_2  e^{-\ii \tau \omega_1(\bm{p}) \ln{(1-y)}} ( q_2 + \Bigl((\omega_1 - \omega_0) + q_1  \frac{(c-a)(c-b)}{(1+c-a-b)} \Bigr)\Biggr)
 \end{align}
 Using the above relation, we get
\begin{align}
   \bigg| \frac{2}{\tau} y (1-y)  \frac{a b}{c} g_1 + \ii (\omega - (1-y) \omega_0 -y \omega_1) f_2 \bigg|^2 
    \nonumber \\
      \simeq \Bigg|\frac{2}{\tau} y (a+b-c) \Gamma_2 e^{- \ii \tau \omega_1 \ln{(1-y)}}  + (1-y) \Biggl[\Gamma_1 \Biggl(  \frac{2}{\tau} y a b + \ii (c-a-b-1) ( \omega_1 - \omega_0 - \frac{ 2 E_0 \tau P_1}{\omega_1}) \Biggr)  
  \nonumber \\
  + \Gamma_2  e^{-\ii \tau \omega_1 \ln{(1-y)}}\Biggl(\frac{2}{\tau} y (a+b-c) \frac{(c-a)(c-b)}{(c-a-b)} + \ii ( \omega_1 - \omega_0 - \frac{ 2 E_0 \tau P_1}{\omega_1})   \Biggr) \Biggl] + (1-y)^2 \Biggl[ \Gamma_1 \Biggl(\frac{2}{\tau} y a b 
  \nonumber \\
   \frac{(1+a)(1+b)}{(2+a+b-c)} + \ii  \biggl( \frac{2 E_0^2 \tau^2 \epsilon_\perp^2}{\omega_1^3} + \frac{ a b }{(1+a+b-c)}( \omega_1 - \omega_0  - \frac{ 2 E_0 \tau P_1}{\omega_1})\biggr) (c-a-b-1)  \Biggr)
  \nonumber \\
   + \Gamma_2 e^{-\ii \tau \omega_1 \ln{(1-y)}} \Biggl( \frac{2}{\tau} y (a+b-c) \frac{ (c-a)(c-b)(c-a+1)(c-b+1)}{(c-a-b)(c-a-b+1)}+ \frac{2 E_0^2 \tau^2 \epsilon_\perp^2}{\omega_1^3}  
    \nonumber \\
    +  ( \omega_1 - \omega_0  - \frac{ 2 E_0 \tau P_1}{\omega_1})  \frac{(c-a) (c-b)}{(1+c-a-b)} \Biggr) \Biggr]\Bigg|^2
    \label{absapp}
 \end{align}

As result, an approximate expression for the particle distribution function \eqref{eqfp}, which can then be re-expressed using Eq.\eqref{absapp}  as follows:
  \begin{align}
     f(\bm{p},y)  \simeq |N^+(\bm{p})|^2 \Bigg|\frac{2}{\tau} y (a+b-c) \Gamma_2 e^{- \ii \tau \omega_1 \ln{(1-y)}}  + (1-y) \Biggl[\Gamma_1 \Biggl(  \frac{2}{\tau} y a b + \ii (c-a-b-1) ( \omega_1 - \omega_0 - \frac{ 2 E_0 \tau P_1}{\omega_1}) \Biggr)  
  \nonumber \\
  + \Gamma_2  e^{-\ii \tau \omega_1 \ln{(1-y)}}\Biggl(\frac{2}{\tau} y (a+b-c) \frac{(c-a)(c-b)}{(c-a-b)} + \ii ( \omega_1 - \omega_0 - \frac{ 2 E_0 \tau P_1}{\omega_1})   \Biggr) \Biggl] + (1-y)^2 \Biggl[ \Gamma_1 \Biggl(\frac{2}{\tau} y a b 
  \nonumber \\
   \frac{(1+a)(1+b)}{(2+a+b-c)} + \ii  \biggl( \frac{2 E_0^2 \tau^2 \epsilon_\perp^2}{\omega_1^3} + \frac{ a b }{(1+a+b-c)}( \omega_1 - \omega_0  - \frac{ 2 E_0 \tau P_1}{\omega_1})\biggr) (c-a-b-1)  \Biggr)
  \nonumber \\
   + \Gamma_2 e^{-\ii \tau \omega_1 \ln{(1-y)}} \Biggl( \frac{2}{\tau} y (a+b-c) \frac{ (c-a)(c-b)(c-a+1)(c-b+1)}{(c-a-b)(c-a-b+1)}+ \frac{2 E_0^2 \tau^2 \epsilon_\perp^2}{\omega_1^3}  
    \nonumber \\
    +  ( \omega_1 - \omega_0  - \frac{ 2 E_0 \tau P_1}{\omega_1})  \frac{(c-a) (c-b)}{(1+c-a-b)} \Biggr) \Biggr]\Bigg|^2 \Biggl(1 +  \frac{P (p_\parallel,y)}{\omega(\bm{p},y)}    \Biggr)
    \label{apppdf1}
\end{align}
To explore the behavior of the particle distribution function at finite times $( t > \tau )$, we aim to express $ f(\bm{p},y) $ in a series involving $(1-y)$. We can then consider truncating higher-order terms to simplify the analysis while still capturing essential features.
We provide detailed calculations for the approximate analytical expression of the Gamma function, which appears in Eq.~\eqref{apppdf1}, in Appendix \ref{appendix1}. For simplicity, we present here the approximate expression for the distribution function in terms of the small parameter $(1-y)$ up to the second order :
\begin{align}
        f(\bm{p},y) &\approx |N^+(\bm{p})|^2 \Bigl( \mathcal{C}_0(\bm{p},y) +  (1-y) \mathcal{C}_1(\bm{p},y)+ (1-y)^2 \mathcal{C}_2(\bm{p},y) \Bigr)
   \label{appdisfun}
\end{align}

\begin{align}
     \mathcal{C}_0(\bm{p},y) &=  4 y^2 \nu_0 \omega_1^2 |\Gamma_2|^2
\label{pdfC0}
\end{align}

\begin{align}
    \mathcal{C}_1(\bm{p},y) &=  4 |\Gamma_2|^2  \omega_1 y (  y \nu_1 \omega_1  - \nu_0 ( \omega_1 (1+y)))+  4   \nu_0  \omega_1 |\Gamma_1 \overline{\Gamma_2}|  \Biggl [  - (  \tau \omega_1  \frac{V_1}{2} - w_1))   \sin{\Upsilon}   \nonumber \\ &
    + \frac{y + 4 \lambda^2 y + \tau ( V_{1} + 2 \omega_1 \tau (\omega_1 - \omega_0  + w_1) - (\omega_0 - \omega_1 )^2 \tau y )}{2 \tau} \cos{\Upsilon} \Biggr]
    \label{pdfC1}
\end{align}
 \begin{align}
    \mathcal{C}_2 (\bm{p},y) &= \frac{-\cos{\Upsilon} |\Gamma_1 \overline{\Gamma_2}|}{2 \tau^2 (4 + \tau^2 \omega_1^2)}
    \Biggl[
        2 \nu_0 \omega_0^3 \omega_1 \tau^4 (4 + \tau^2 \omega_1^2)
        - 2 \nu_0 \omega_1^5 \tau^6 w_1 (1 + 2y)
        - \nu_0 \omega_1^6 \tau^6 y(1 + 2y) \nonumber \\
        & \quad + \nu_0 \omega_0^4 \tau^4 y \big(8y + \omega_1^2 \tau^2 (1 + 2y)\big)
        + 8 \nu_0 \big(\tau^2 (V_1^2 + 4w_1^2) + 2y \tau V_1 (1 + 4\lambda^2)
        + (y + 4\lambda^2 y)^2\big) \nonumber \\
        & \quad - 2 \omega_0 \tau^2 (4 + \omega_1^2 \tau^2)
        \big(2y \omega_1^2 \tau^2 \nu_0 w_1 + y \omega_1^3 \tau^2 \nu_0 
        + \nu_1 \omega_1 (9 + 4\lambda^2 - 12y + \tau V_1 y)\big) \nonumber \\
        & \quad + 2 \omega_1^3 \tau^4 
        \big(4 \nu_1 w_1 + 4 \nu_0 w_2 + \nu_0 w_1 (5 - 6y + \lambda^2 (4 + 8y))\big)
        + 8 \omega_1 \tau^2 
        \big(4V_1 w_1 + 4 \nu_0 w_2 \nonumber \\
        & \quad + \nu_0 w_1 (9 + 2y + \lambda^2 (4 + 8y))\big)
        - 2 \omega_1^4 \tau^4 
        \big(2 \nu_1 \tau V_1 + \nu_0 (-1 + y(9 + 2y) + \tau(V_1 + 2v_2 + 4y V_1))\big) \nonumber \\
        & \quad + \omega_1^2 \tau^2 
        \big(-16 \nu_1 \tau V_1 
        + \nu_0 \big(40 + 2\tau^2(V_1^2 + 4w_1^2) - 23y(1 + 2y) + 16\lambda^4 y(1 + 2y) \nonumber \\
        & \quad - 4\tau(2V_1 + 4v_2 + 7V_1 y) + 8\lambda^2(4 + (5 + 2\tau V_1 - 6y)y)\big)\big) \nonumber \\
        & \quad - 2 \omega_0^2 \tau^2 \nu_0 
        \big(4\omega_1 \tau^2 w_1 (1 + 2y) + \omega_1^3 \tau^4 w_1 (1 + 2y) 
        + 8(-2 + y(4 + \tau V_1 - y + 4\lambda^2 y)) \nonumber \\
        & \quad + \omega_1^2 \tau^2 
        \big(8 + y(-13 + 2\tau V_1 + 6y + \lambda^2 (4 + 8y))\big)\big)
    \Biggr] \nonumber \\
    & \quad + \frac{\sin{\Upsilon} |\Gamma_1 \overline{\Gamma_2}| \omega_1}{2 \tau (4 + \omega_1^2 \tau^2)}
    \Biggl[
        8 \nu_0 \tau^2 (V_1^2 + 4w_1^2) 
        + 2 \nu_0 \omega_0^4 \tau^4 y 
        + 16 \omega_1 \tau^2 
        \big(2 \nu_1 w_1 + 2 \nu_0 w_2 + \nu_0 w_1 (5 + 2y)\big) \nonumber \\
        & \quad + 4 \omega_1^3 \tau^4 
        \big(2 \nu_1 w_1 + 2 \nu_0 w_2 + \nu_0 w_1 (5 + 2y)\big) 
        + 2y(1 + 4\lambda^2)
        \big(8 \nu_1 + \nu_0 (9 + 4\lambda^2 + 16y)\big) \nonumber \\
        & \quad - 2 \omega_0 \tau^2 (4 + \tau^2 \omega_1^2) 
        \big(2 \nu_0 w_1 (5 - 6y) + \omega_1 \nu_0 (\tau V_1 y - 4y)\big) 
        + 4 \tau 
        \big(4 \nu_1 V_1 + \nu_0 (V_1 + 4v_2 + 10V_1 y \nonumber \\
        & \quad + \lambda^2 V_1 (4 + 8y))\big)
        - \omega_1^4 \tau^4 
        \big(-4 \nu_1 + \nu_0 (8y^2 - 2(6 + y) + \tau(V_1 + 2V_1 y))\big)
    \Biggr] \nonumber \\
    & \quad + \frac{\nu_0 |\Gamma_1|^2}{4 \tau^2}
    \Biggl(
        \tau^2 (1 + \tau^2 \omega_1^2)(V_1^2 + 4(\omega_1 - \omega_0 + w_1)^2)
        + 2 \tau y 
        \Bigl(
            2 \omega_0^3 \omega_1 \tau^3 - 2 \tau^3 \omega_1^4 + V_1 + 4\lambda^2 V_1 \nonumber \\
            & \quad + \tau \omega_1^2 (-2 + 8\lambda^2 + \tau V_1)
            + 2(-1 + 4\lambda^2) \omega_1 \tau w_1
            - 2 \omega_1^3 \tau^3 w_1
            - \omega_0^2 \tau 
            (4 + \tau (V_1 + 2 \omega_1 \tau (3\omega_1 + w_1))) \nonumber \\
            & \quad + 2 \omega_0 \tau 
            (2w_1 + \omega_1(3 - 4\lambda^2 + \omega_1 \tau^2 (3\omega_1 + 2w_1)))
        \Bigr)
        + \big(1 + (2\lambda + (\omega_0 - \omega_1)\tau)^2\big)
        \big(1 + (2\lambda + (-\omega_0 + \omega_1)\tau)^2\big)y^2
    \Biggr)
 \nonumber \\ 
  & + 
  |\Gamma_2|^2   \frac{1}{ 4 ( \tau^2 + \omega_1^2 \tau^4)}   \Biggl (  
  \tau^2 \nu_0  ( 1 + \omega_1^2 \tau^2 ) ( V_1^2 + 4  ( w_1 +\omega_1 -\omega_0)^2) - \tau  ( 2 \omega_0^3  \omega_1 \tau^3  \nu_0  - 2 \nu_0 V_1  
  ( 1 +  4 \lambda^2)  
  \nonumber \\ &
  + \omega_1^4 \tau^3  ( 16 \nu_1 + 3 \nu_0  ( -2  + \tau V_1)) +  \omega_1^2 \tau  ( 16 \nu_1  + \nu_0  ( -6 + \tau V_1 + \lambda^2 (8 -12 \tau V_1))) 
  \nonumber \\
  &+ \omega_0^2 \tau \nu_0  ( 8 +\tau  ( 2 V_1 + \tau \omega_1 ( 3 \omega_1 ( 2 + \tau V_1 ) -2 w_1))) -  2 \omega_0 \tau  ( \omega_1^3 \tau^2  ( \nu_0 + 8 \nu_1 -3 \nu_0 \tau V_1   )   + \omega_1 ( \nu_0 + 4 \lambda^2 \nu_0  + 8 \nu_1 - 3 \tau  \nu_0 V_1 ) 
  \nonumber \\&
  + 4 \nu_0 w_1 + 4 \omega_1^2 \tau^2 \nu_0 w_1 ) +  2 \omega_1^3 \tau^3 ( -3  \nu_0  w_1  + 8 \nu_1  w_1 +  4 \nu_0 w_2 ) + 2 \omega_1 \tau  (   (-3 + 4 \lambda^2) \nu_0 w_1 + 8 \nu_1 w_1  + 4 \nu_0 w_2 )) y 
  \nonumber \\&
  + ((( 1 + 4 \lambda^2 )^2  + 2 ( ( 2 - 4 \lambda^2)  \omega_0^2   - 2  \omega_0 \omega_1 ( 5 +  12 \lambda^2)  + 2 (-3 + 2  \lambda^2 + 8 \lambda^4)  \omega_1^2) 
  \tau^2 + ( \omega_1 + \omega_0)^2 ( \omega_0^2  + 10 \omega_0 \omega_1 
  \nonumber \\&- ( 11 + 16 \lambda^2) \omega_1^2  ) \tau^4 + 2 \omega_1^2 (\omega_0 + \omega_1)^4  \tau^6 ) \nu_0 + 16 \omega_1 \tau^2 ( 1 + \tau^2 \omega_1^2)
(-(\omega_0 + \omega_1) \nu_1  + \omega_1 \nu_2 )) y^2   \Biggr)
  \label{pdfC2}
\end{align}
Here, $ \Upsilon = \mathrm{\varrho} + \tau \omega_1 \ln{(1-y)}.$ and also note that $\mathcal{C}_0$, $\mathcal{C}_1$, and $\mathcal{C}_2$ remain unchanged regardless of the function of $(1-y)$.

\section{Result and Discussion}
\label{result}
In this section, we explore how the choice of adiabatic basis functions influences the analysis of pair creation dynamics in a pulsed electric field. The single-particle momentum distribution function, as described in Eq.~\eqref{eqfp}, depends on the choices of $\Omega(\bm{p},t)$ and $V(\bm{p},t)$. These choices, guided by certain conditions and the inherent arbitrariness in defining positive and negative energy states, allow for an infinite number of consistent configurations.

For our analysis, we focus on two specific adiabatic bases. The first choice corresponds to the standard leading-order WKB solutions of the mode equation, where $\Omega(\bm{p},t) = \omega(\bm{p},t)$ and $V(\bm{p},t) = 0$. The second choice sets $\Omega(\bm{p},t) = \omega(\bm{p},t)$ and $V(\bm{p},t) = -\frac{\dot{\omega}(\bm{p},t)}{2 \omega(\bm{p},t)}$.
\subsection{Momentum distributions of created particles}
\label{LMS_choice}
\begin{figure}[t]
\begin{center}
{
\includegraphics[width =  1.9358802in]{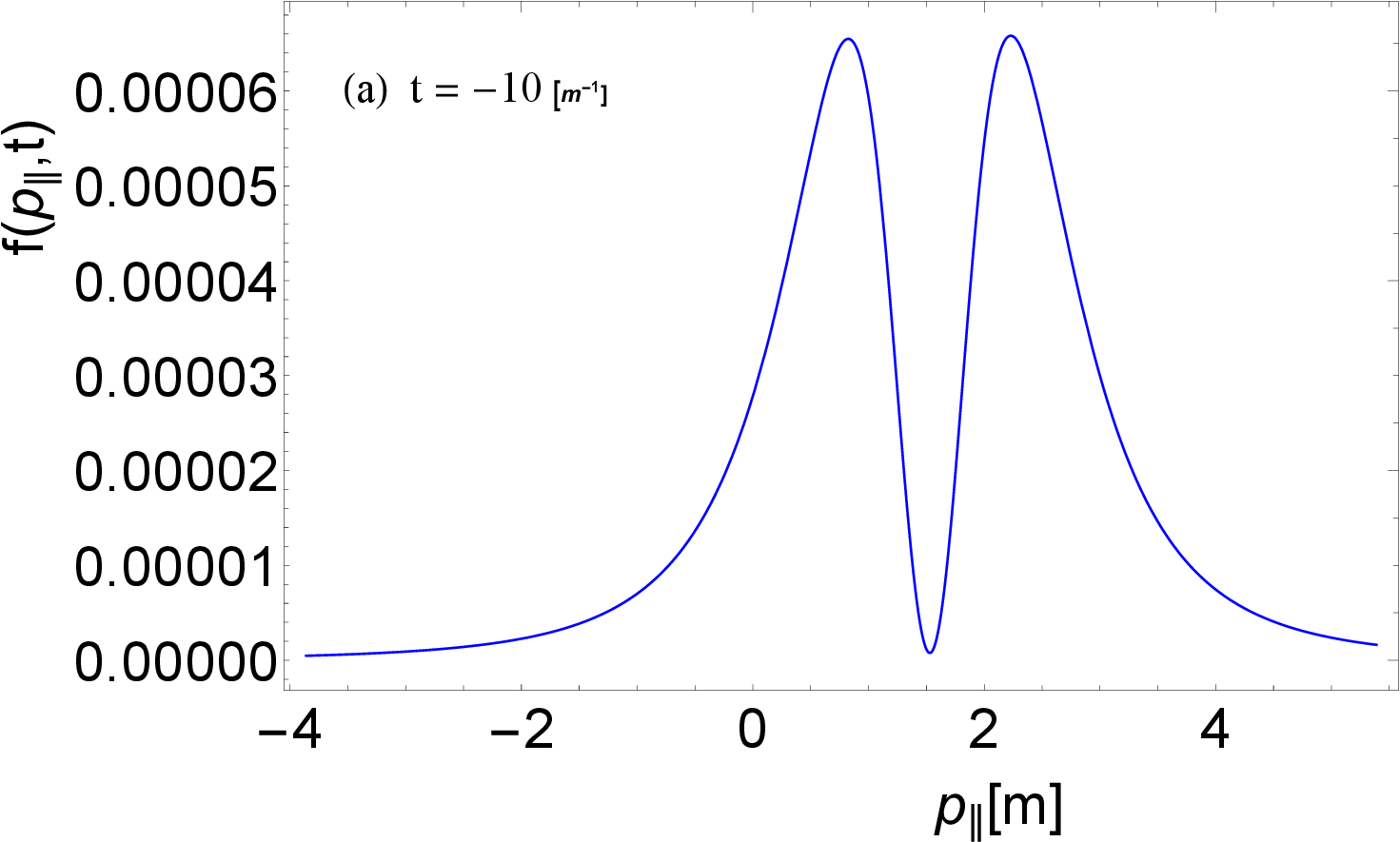}
\includegraphics[width =  1.9358802in]{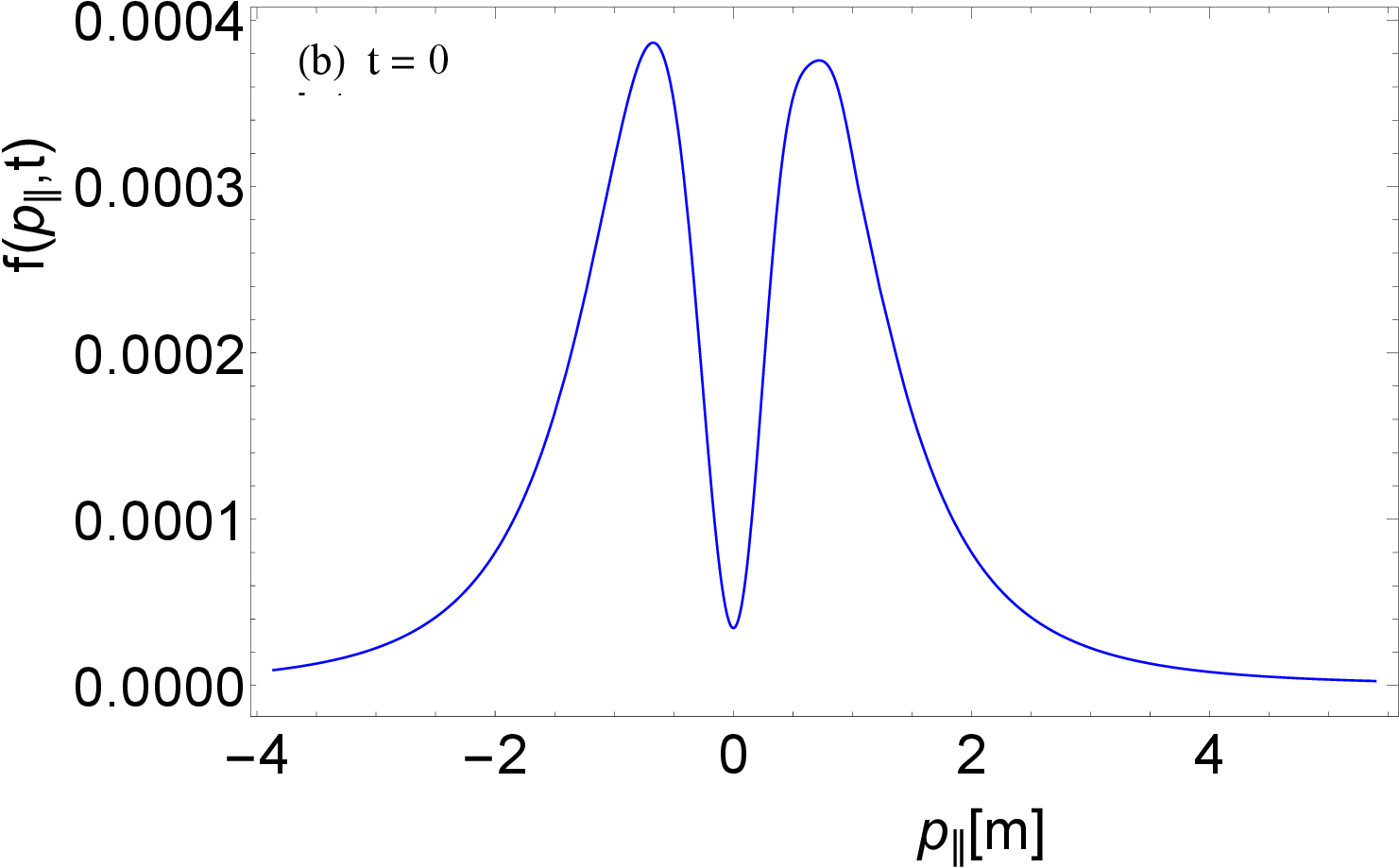}
\includegraphics[width =  1.9358802in]{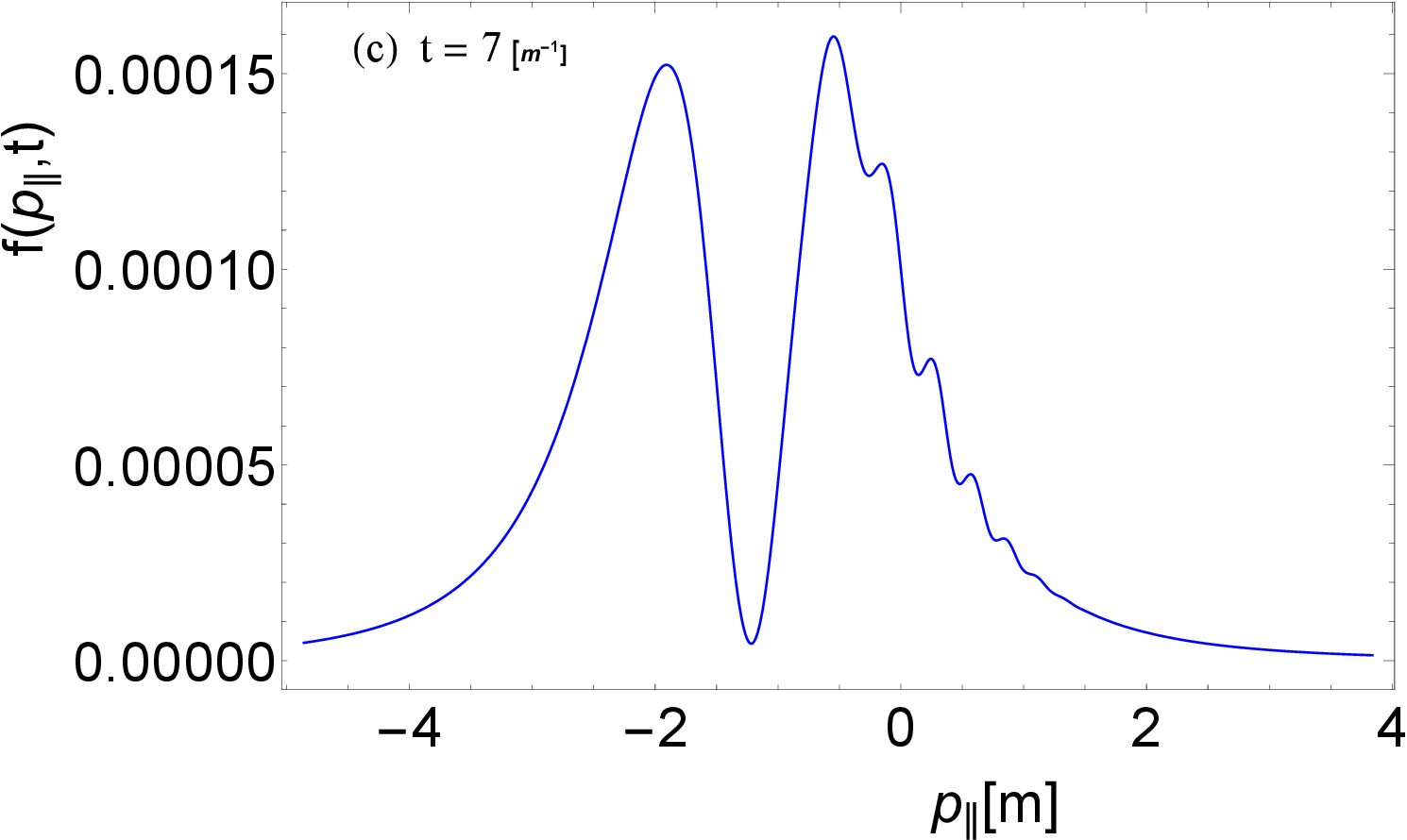}
\includegraphics[width =  1.9358802in]{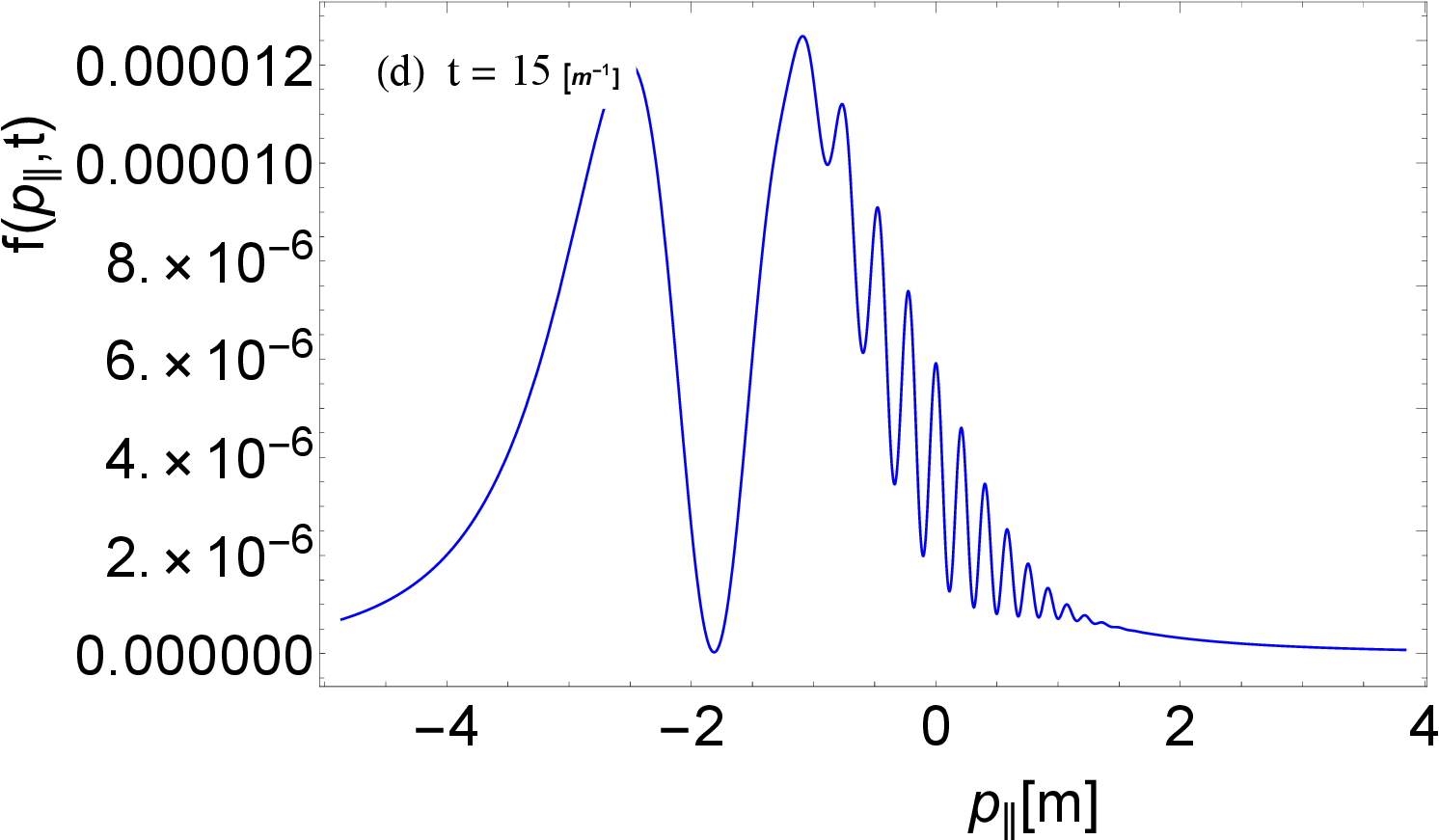}
\includegraphics[width =  1.9358802in]{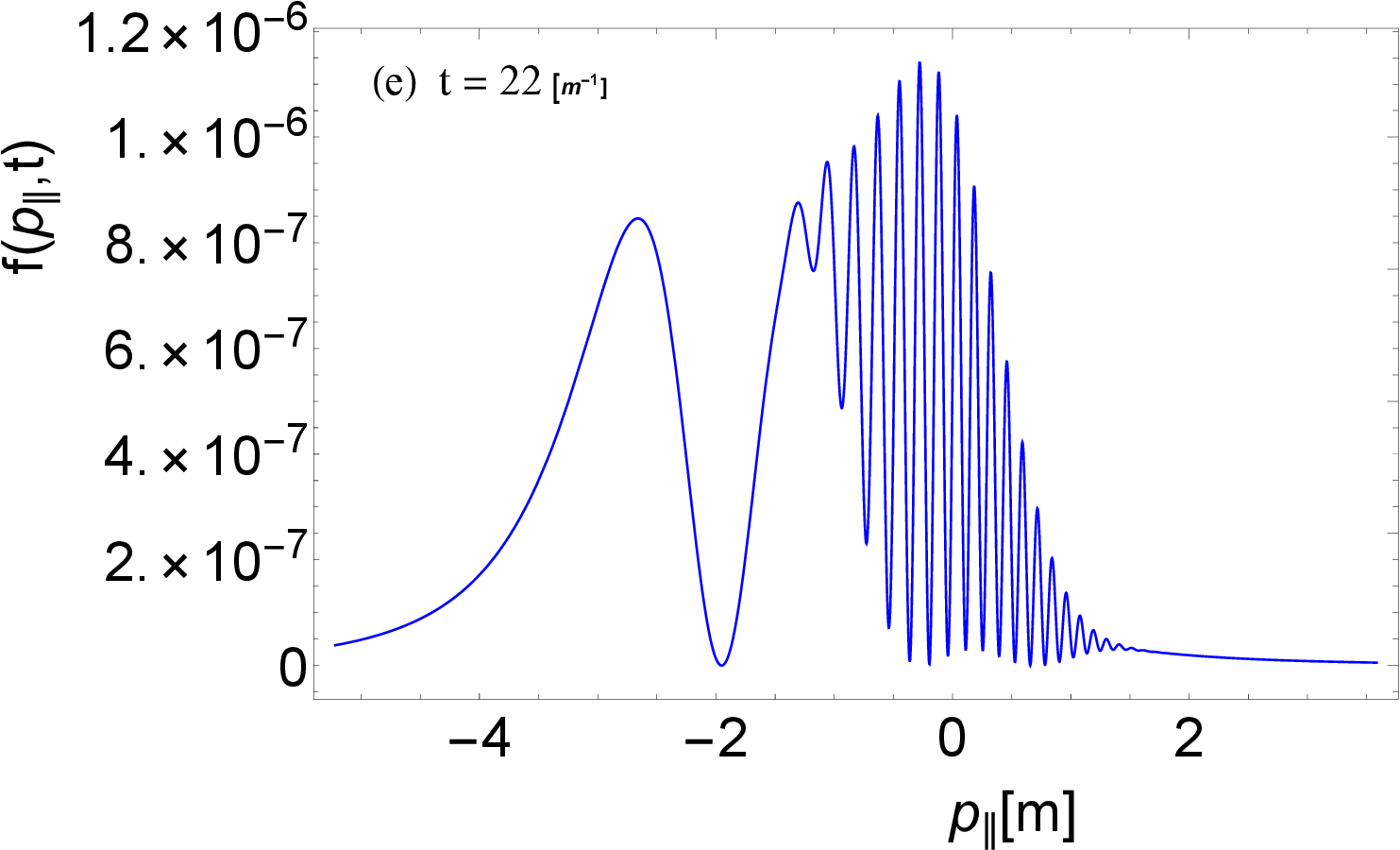}
\includegraphics[width =  1.9358802in]{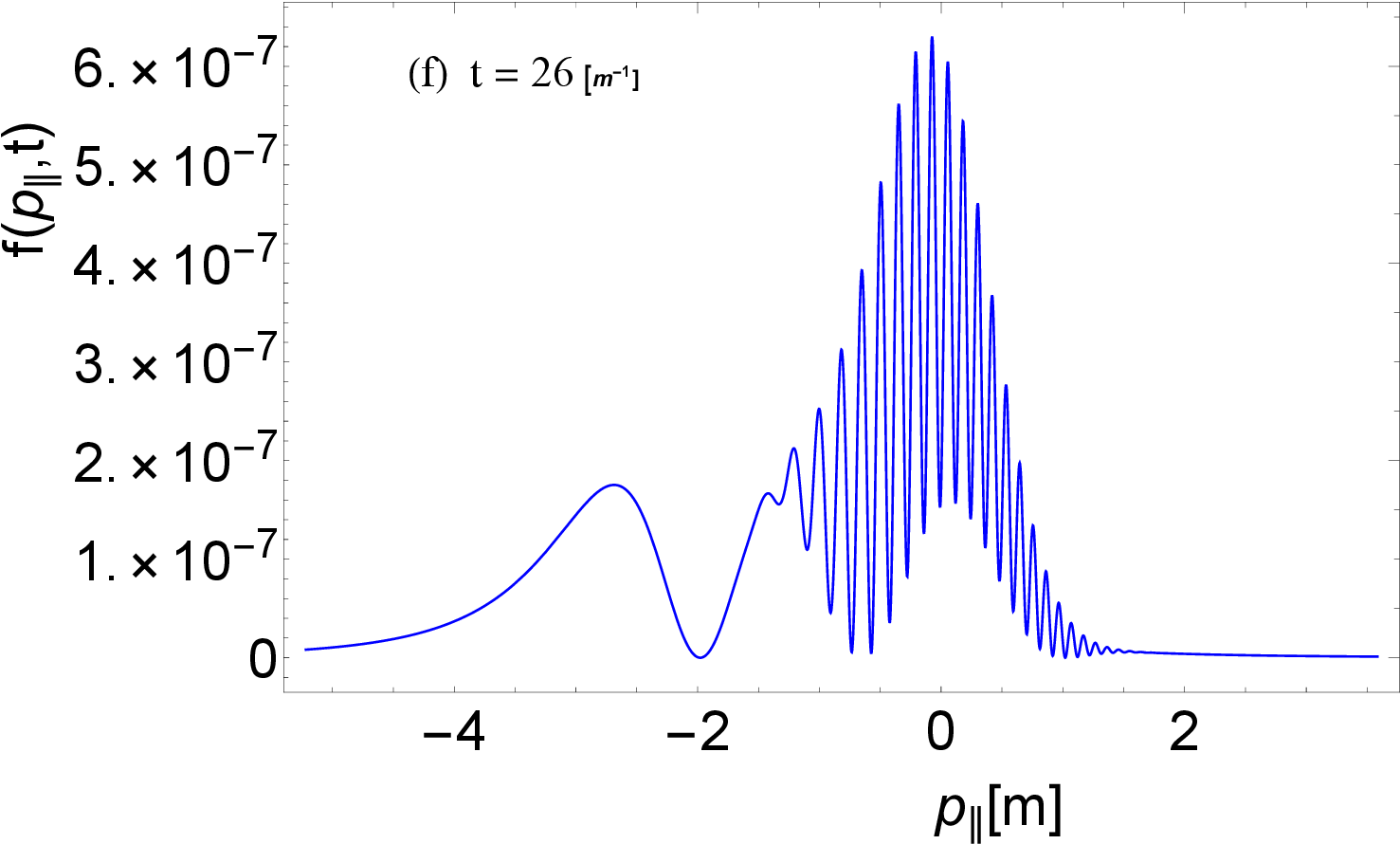}
\includegraphics[width =  1.9358802in]{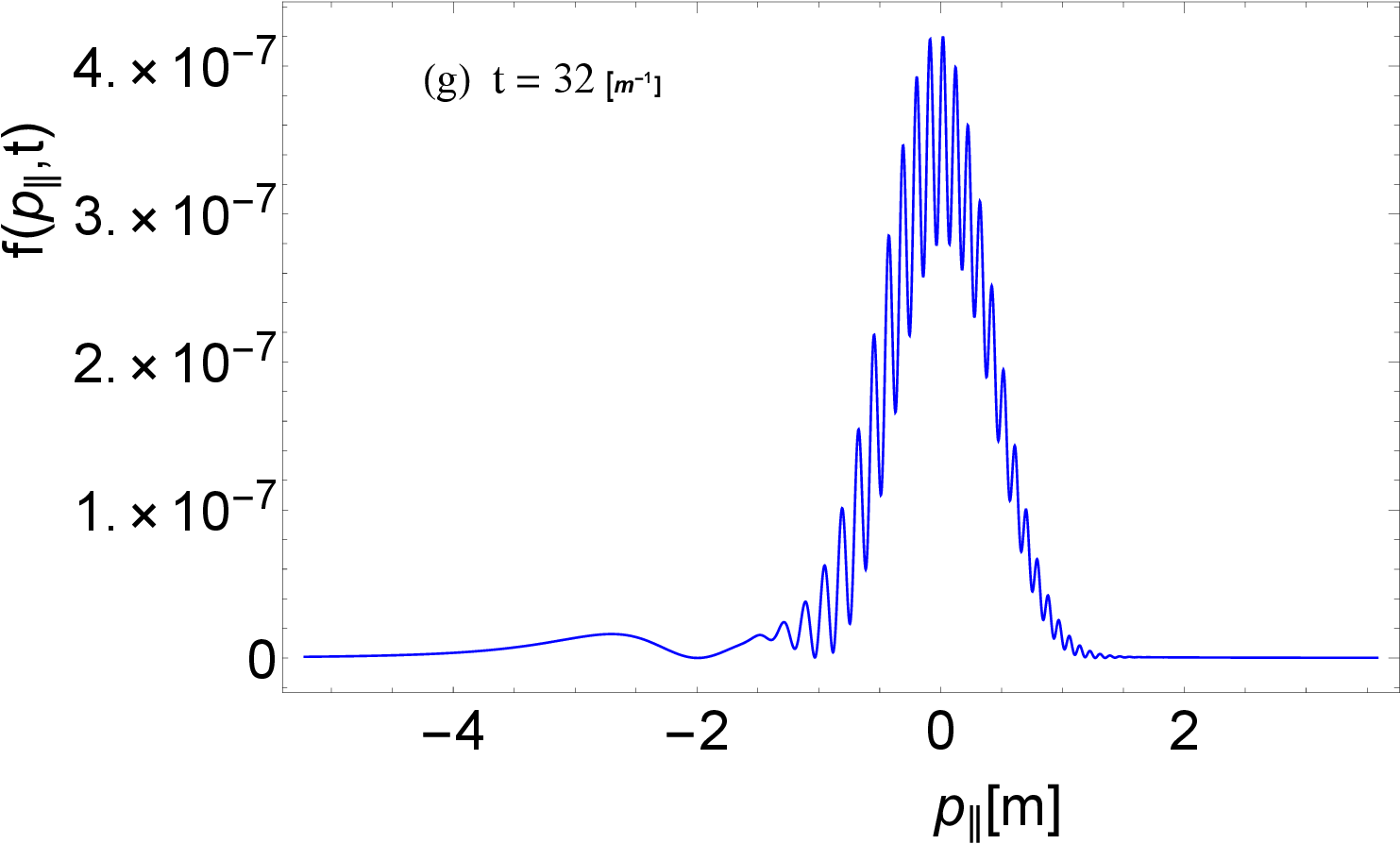}
\includegraphics[width =  1.9358802in]{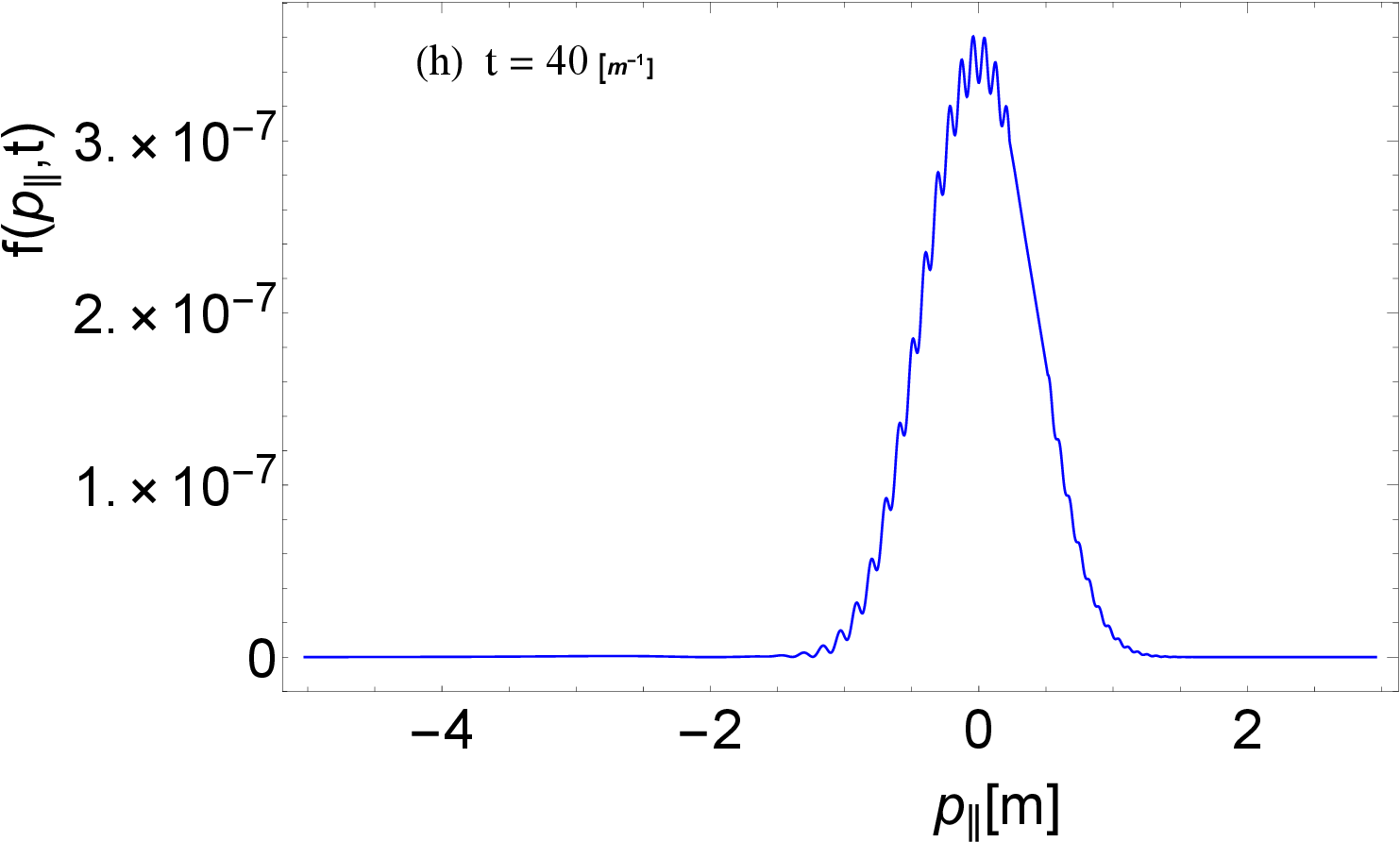}
\includegraphics[width =  1.9358802in]{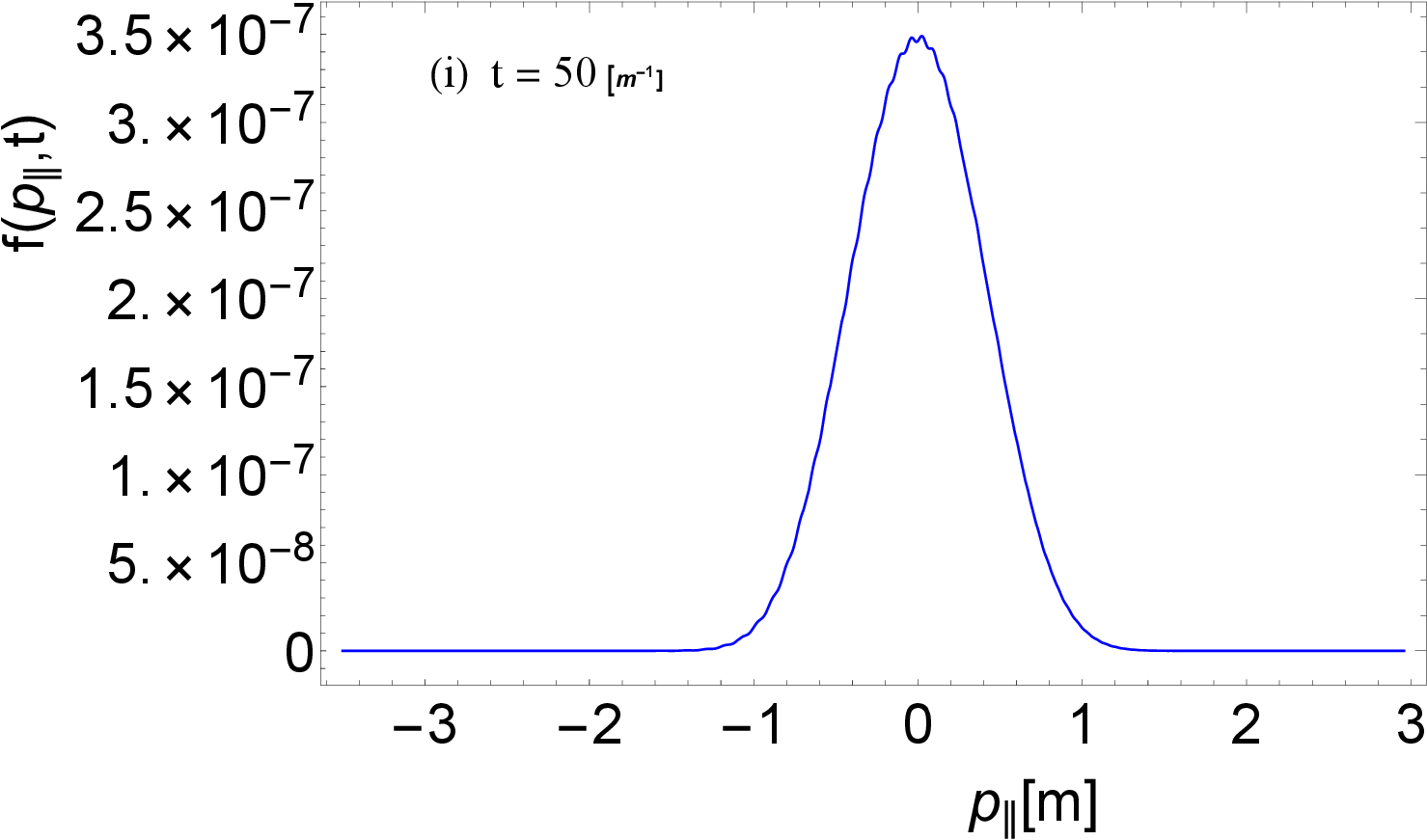}
}
\caption{
Time evolution of momentum distribution function for first choice $\Omega(p_\parallel, t) = \omega(p_\parallel, t)$ and $V(p_\parallel, t) = 0$. The transverse momentum is considered to be zero, and all the units are taken in the electron mass unit.The field parameters are  $E_0=0.2 E_c$ and $ \tau =10 [m^{-1}].$}
   	\label{lms_zero}
\end{center}
\end{figure}

In this section, we present the momentum spectrum of the created particles based on the distribution function given by Eq.\eqref{eqfp} for a time-dependent Sauter pulse \eqref{Et}. We analyze the time evolution of the momentum spectra and describe the observations in simple words. The pulse parameters chosen are $\tau =10 [m^{-1}]$ and $E_0 = 0.2 E_{cr}.$ The resulting Keldysh parameter , $\gamma < 1 $, which signifies that the Schwinger effect is dominant \cite{Keldysh:1965ojf}.
\newline
The time evolution of the momentum distribution for the first choice, in which the adiabatic frequency functions are \(\Omega(\bm{p}, t) = \omega(\bm{p}, t)\) and \(V(\bm{p}, t) = 0\), is shown in Figure~\ref{lms_zero}. 
The momentum distribution is almost symmetric in the direction of the applied electric field (or z-axis direction). 
Momentum spectra show two nearly separate bell-shaped profiles, with particles concentrated at the momentum values \( p_\parallel \approx 1 \, [m] \) and \( p_\parallel \approx 2 \, [m] \). Due to the changing electric field strength over time, the location of the peaks and the magnitude of the momentum distribution of the particles also change, as seen in Figs.~\ref{lms_zero}(a) to (b). At \( t = 7 \, [m^{-1}] \), the first deformations appear on the right flank of the bell-shaped profile peak at \( p_\parallel \approx 0 \); see Fig.~\ref{lms_zero}(c). This deformation becomes stronger as time progresses; see Fig.~\ref{lms_zero}(d). At \( t = 22 \, [m^{-1}] \), a dip is observed at \( p_\parallel \approx -2 \, [m] \), while the right-side profile, with a peak at \( p_\parallel \approx 0 \), becomes dominant; see Fig.~\ref{lms_zero}(e). These oscillations strengthen over time, eventually revealing a growing white area on the right side near \( p_\parallel \approx 0 \) (see Fig.~\ref{lms_zero}(f)). This area remains unaffected by the oscillations. Over time, the left-side peak gradually decreases in height.At \( t = 32 \, [m^{-1}] \), an outgoing distribution is evident, as shown in Fig.~\ref{lms_zero}(g), with a Gaussian-like profile and onset oscillations whose peak lies at \( p_\parallel = 0 \).
This oscillatory behavior observed in the momentum spectra of pairs created at finite times was reported for the first time for fermion particles (Spinor QED) in \cite{deepak2022,10.1007/978-981-97-0289-3_275} and other references \cite{Diez:2022ywi}. The oscillations originate from quantum interference effects, as explained in \cite{Sah:2023jlz}, which also discusses the mathematical origin in greater detail using an approximate expression for the distribution function in Subsection~\ref{approximate}. A further decrease in the oscillation amplitude follows the weakening of the interference effect over time. Around \( t = 40 \, [m^{-1}] \), the left-side profile peak diminishes. Only the dominant central peak at \( p_\parallel = 0 \) persists, with a faint onset of an oscillatory effect superimposed on a Gaussian-like profile. Eventually, as depicted in Figs.~\ref{lms_zero}(h) to (i), the oscillations are barely visible at \( t = 50 \, [m^{-1}] \). This corresponds to five times the pulse duration of the applied electric field. At this time, the electric field strength has decreased to approximately \( 10^{-4} \), and the spectra show a nearly smooth Gaussian-like structure. Numerous studies, including Dumlu et al.~\cite{Dumlu:2011rr}, have shown that the momentum spectra for a Sauter pulsed field lack an oscillatory structure as time approaches infinity, instead displaying a single-peaked Gaussian-like profile. This is consistent with the results shown in Fig.~\ref{lms_zero}(i).
It is worth mentioning that in the literature, momentum spectra at asymptotic times (\( t \to \infty \)) are physically interpreted as corresponding to real particle formation. Here, we highlight that the occurrence or absence of oscillatory effects in the momentum spectra can be related to asymptotic times. This allows for a physical interpretation and helps quantify time scales that are crucial for describing the different subprocesses involved in pair production in time-dependent electric fields.
\subsubsection{Approximate expression for momentum distribution function }
\label{approximate}
To analyze the momentum distribution function analytically, we start with the approximate expression of the distribution function $f(\bm{p}, t)$ (Eq.\eqref{appdisfun}) and set $p_\perp = 0$. This allows us to derive an approximate expression for the longitudinal momentum distribution function $f(p_\parallel, t)$. Utilizing this expression, we unveil that momentum spectrum structure mainly comprises three distinct functional behaviors.
\begin{figure}[t]
\begin{center}
{
\includegraphics[width =  2.7629358802in]{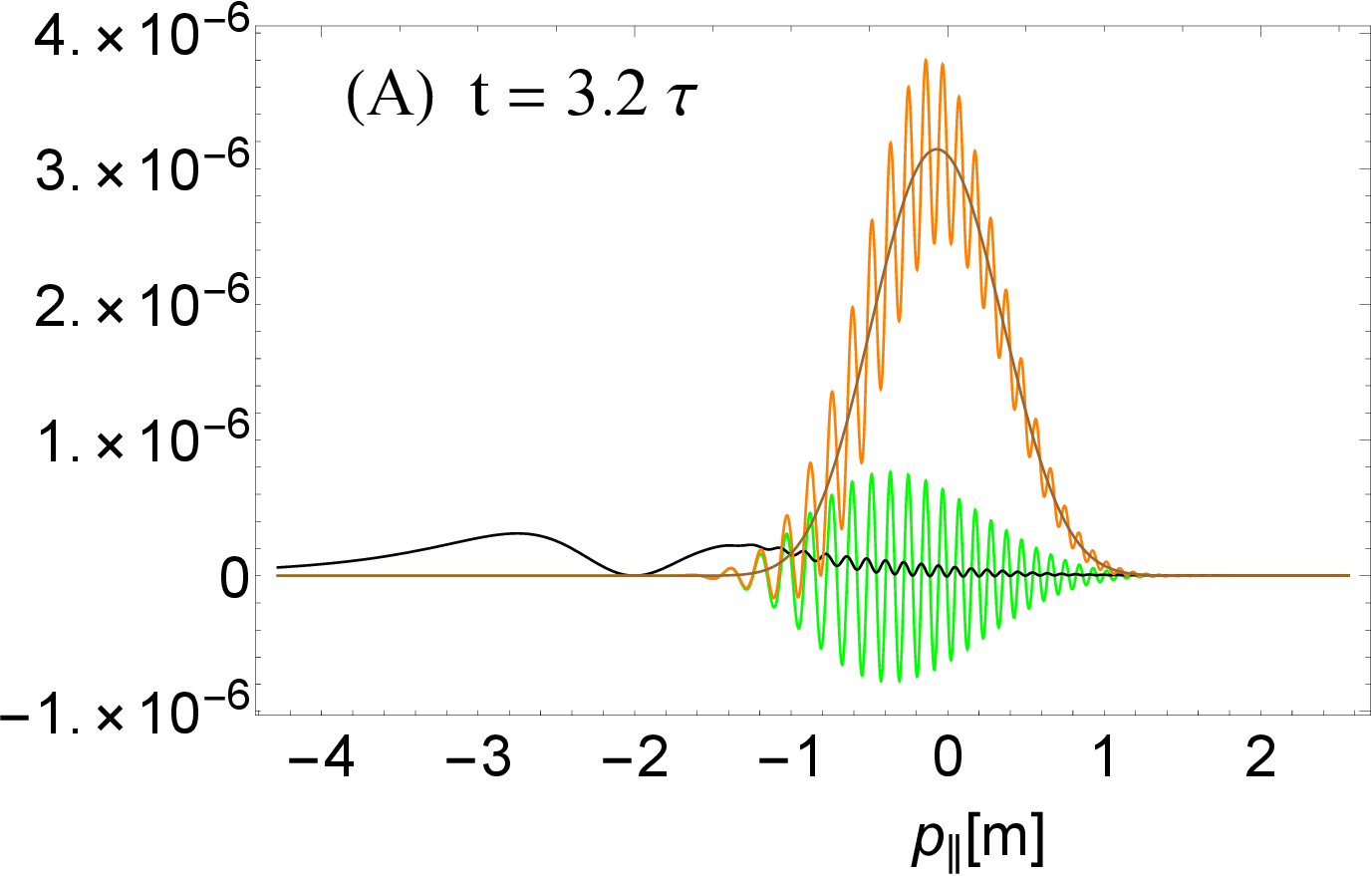}
\includegraphics[width =  2.7629358802in]{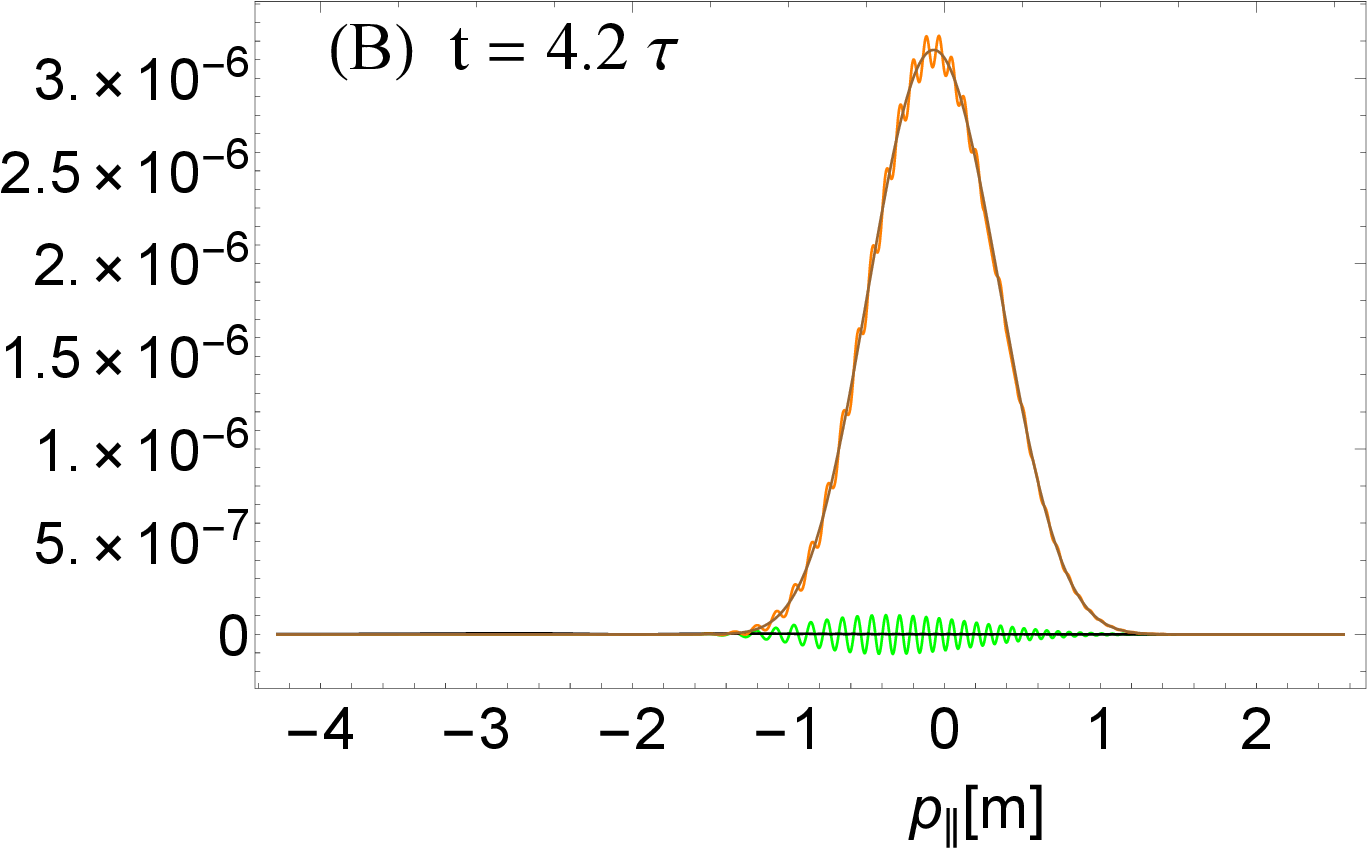}
}
\caption{ The second, first, and zeroth order terms are defined in Eq.\eqref{appdisfun} as functions of the longitudinal momentum for different times. The curves are as follows: Brown: $\mathrm{C}_0$, Green: $(1-y) \mathrm{C}_1$, Black: $(1-y) \mathrm{C}_2$, and Orange: $((1-y) \mathrm{C}_1 + \mathrm{C}_0)$.
The transverse momentum is considered to be zero, and all the units are taken in the electron mass unit.The field parameters are  $E_0=0.2 E_c$ and $ \tau =10 [m^{-1}].$}
   	\label{appr_terms}
\end{center}
\end{figure}
Upon carefully examining the approximate expression for the distribution function, we find that in the quasi-particle stage, the behavior of the distribution function is primarily governed by the $(1-y)^2 \mathcal{C}_2$ term. This term leads to a two-peak structure, where we observe oscillations on the right flank of the right-side peak. As time progresses, this peak structure diminishes.
The central peak at $p_\parallel = 0$ in the spectrum can be mathematically understood by the zeroth-order term $\mathcal{C}_0$ and the first-order term $(1-y) \mathcal{C}_1$, which is responsible for the onset of oscillations in that peak. The oscillation pattern of $\mathcal{C}_1$ undergoes a transformation over time, primarily due to the presence of $\ln(1-y)$ in the sinusoidal and cosine functions. As time progresses towards infinity, the $(1-y) \mathcal{C}_1$ term leads to suppression. Consequently, we observe only a central peak at $p_\parallel = 0$ due to the dominance of the $\mathcal{C}_0$ term. This observation is explicitly confirmed in Figure \ref{appr_terms}.
It's important to note that $\mathcal{C}_1$ represents an oscillatory finite function whose magnitude depends on $t$. The magnitude of this function plays a crucial role in determining the dynamics of $f(p_\parallel, t)$ in $p_\parallel$-space at finite times.

\subsubsection{Second choice}
\begin{figure}[t]
\begin{center}
{
\includegraphics[width =  1.9358802in]{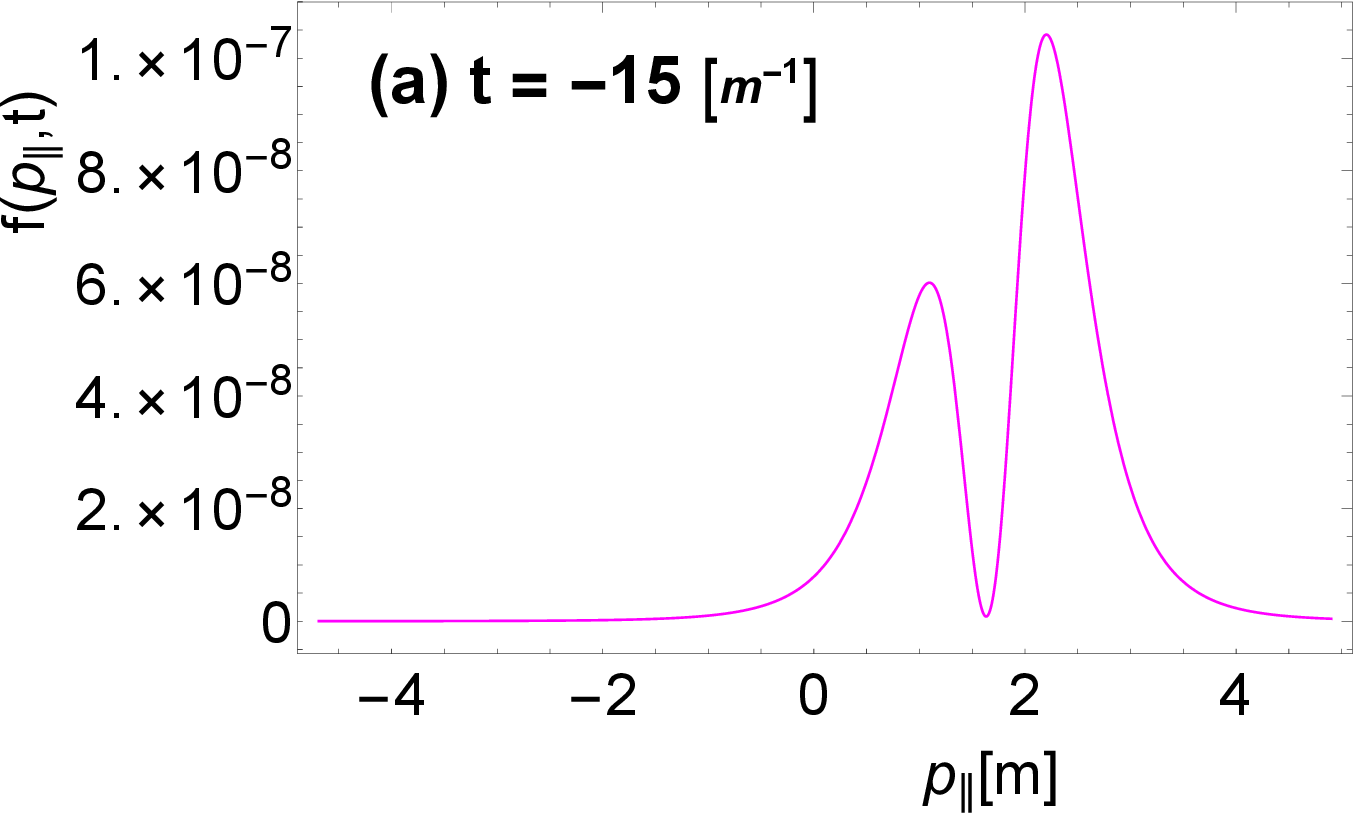}
\includegraphics[width =  1.9358802in]{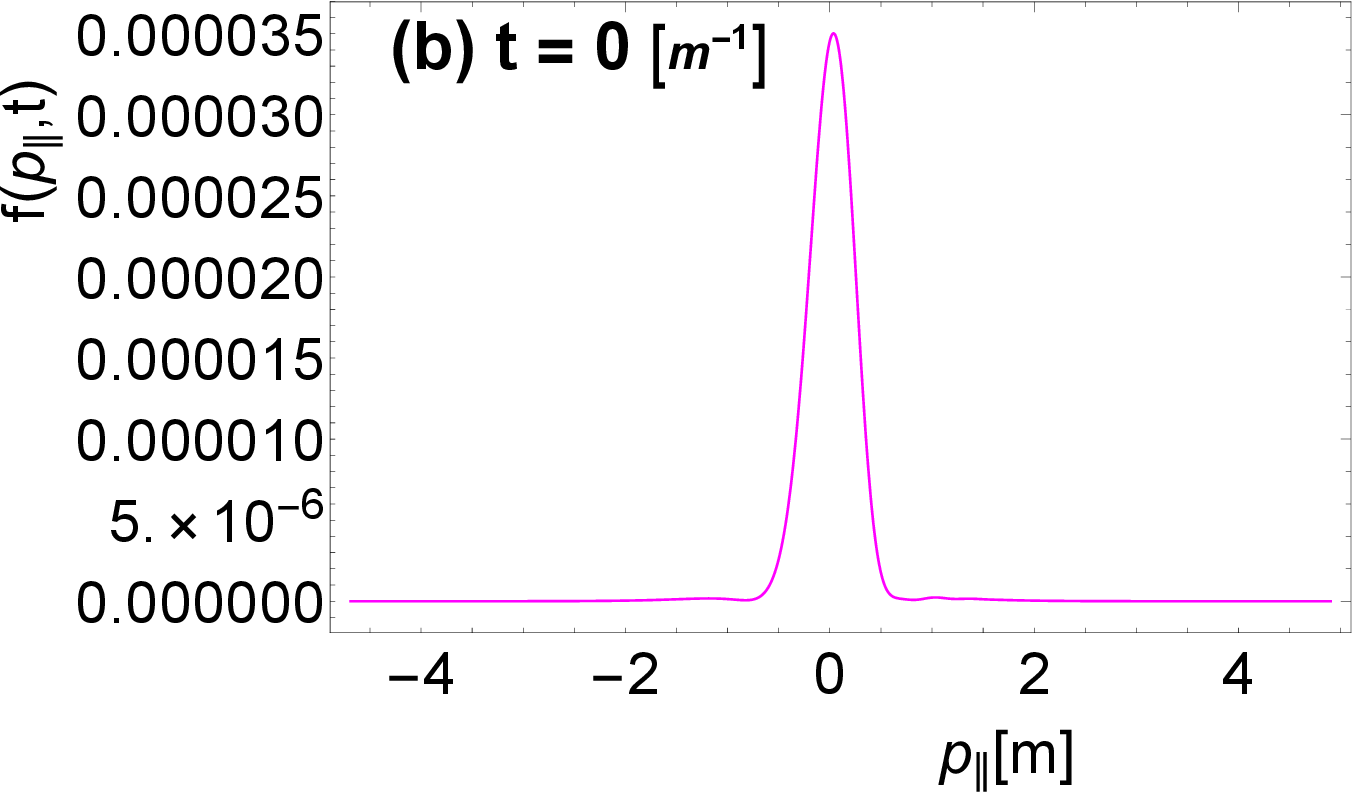}
\includegraphics[width =  1.9358802in]{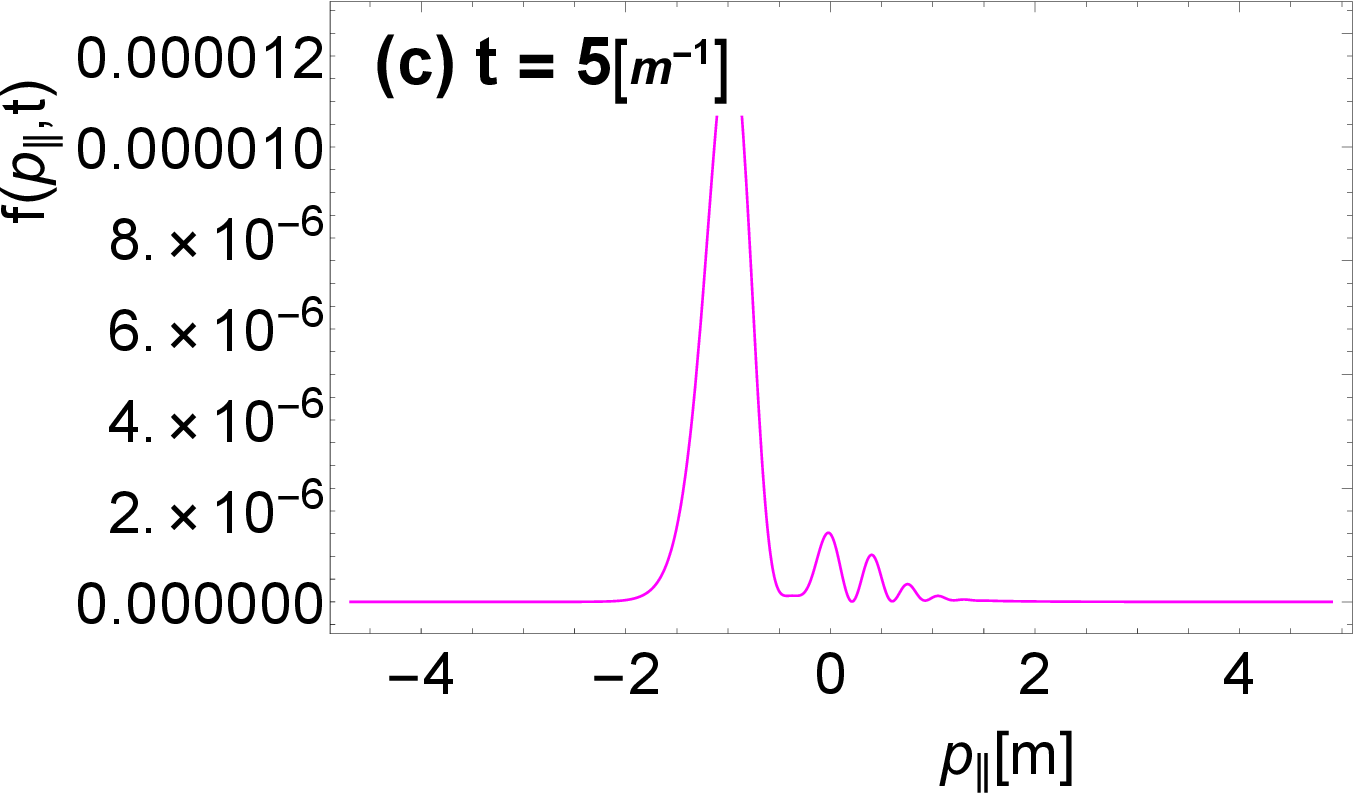}
\includegraphics[width =  1.9358802in]{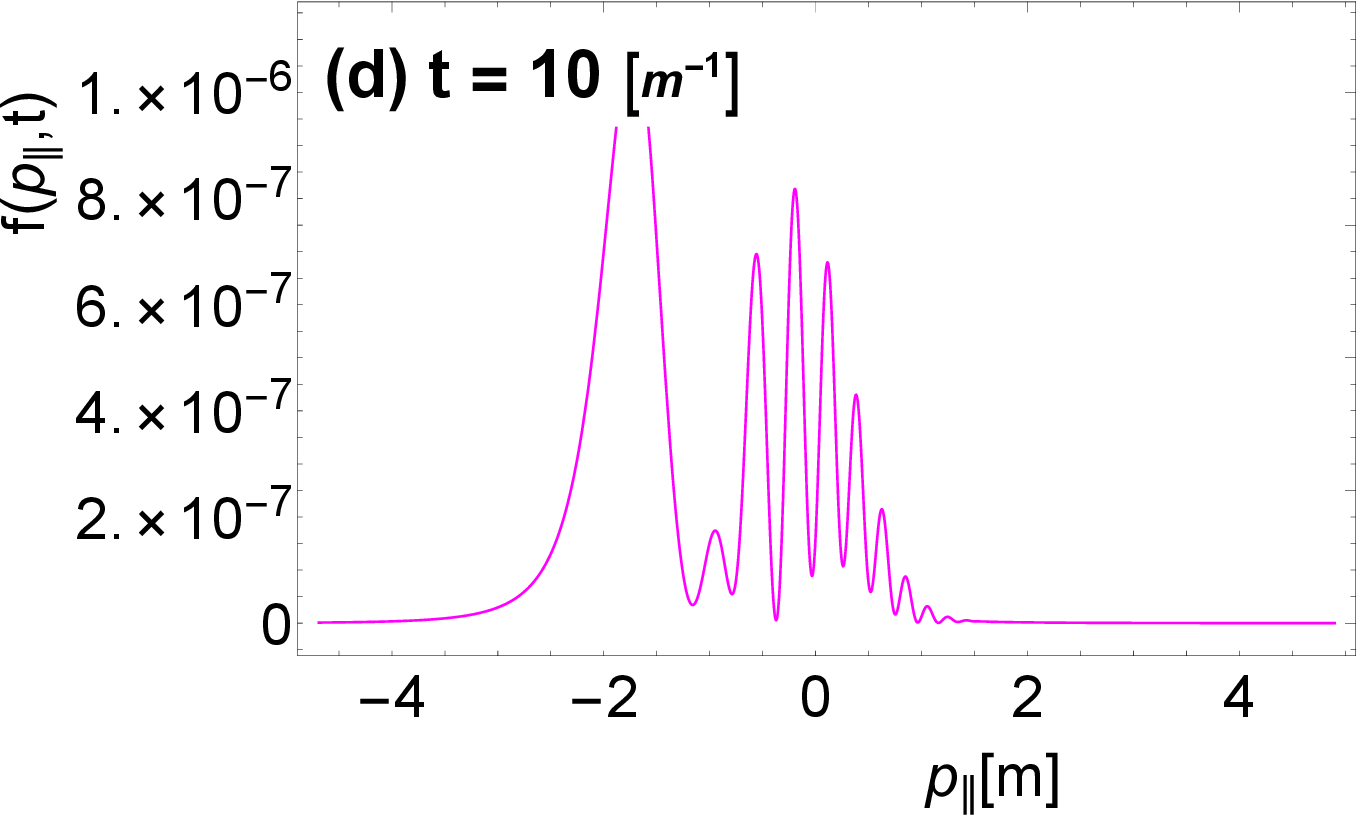}
\includegraphics[width =  1.9358802in]{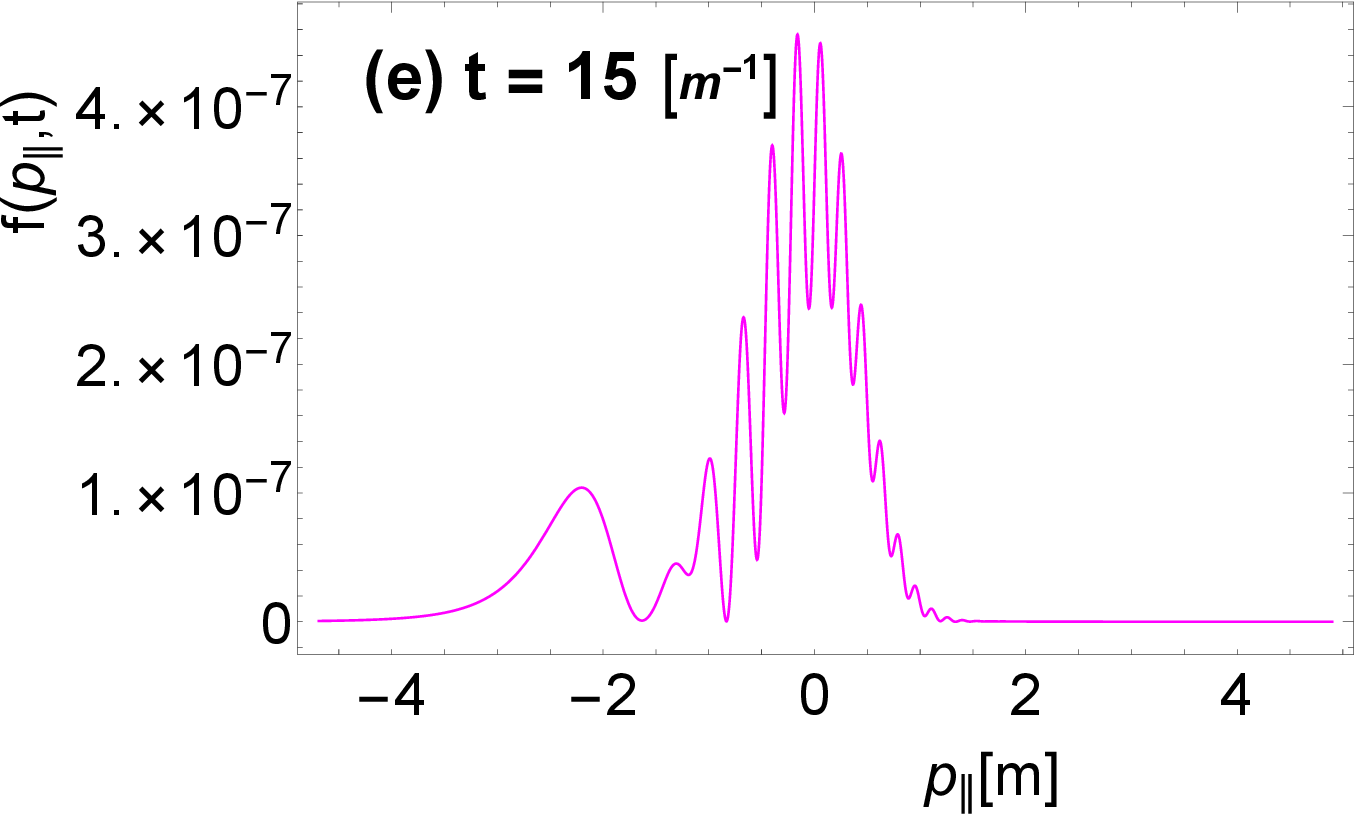}
\includegraphics[width =  1.9358802in]{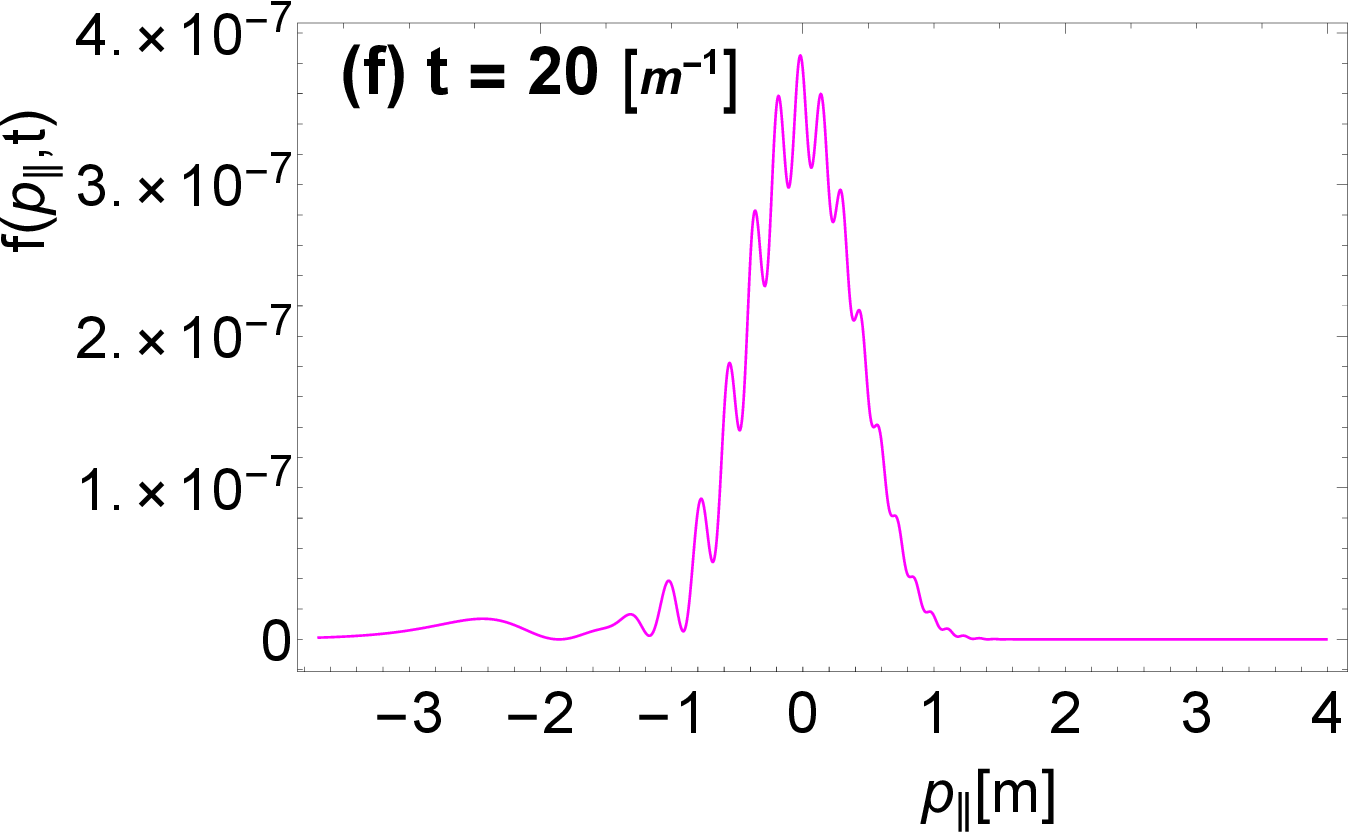}
\includegraphics[width =  1.9358802in]{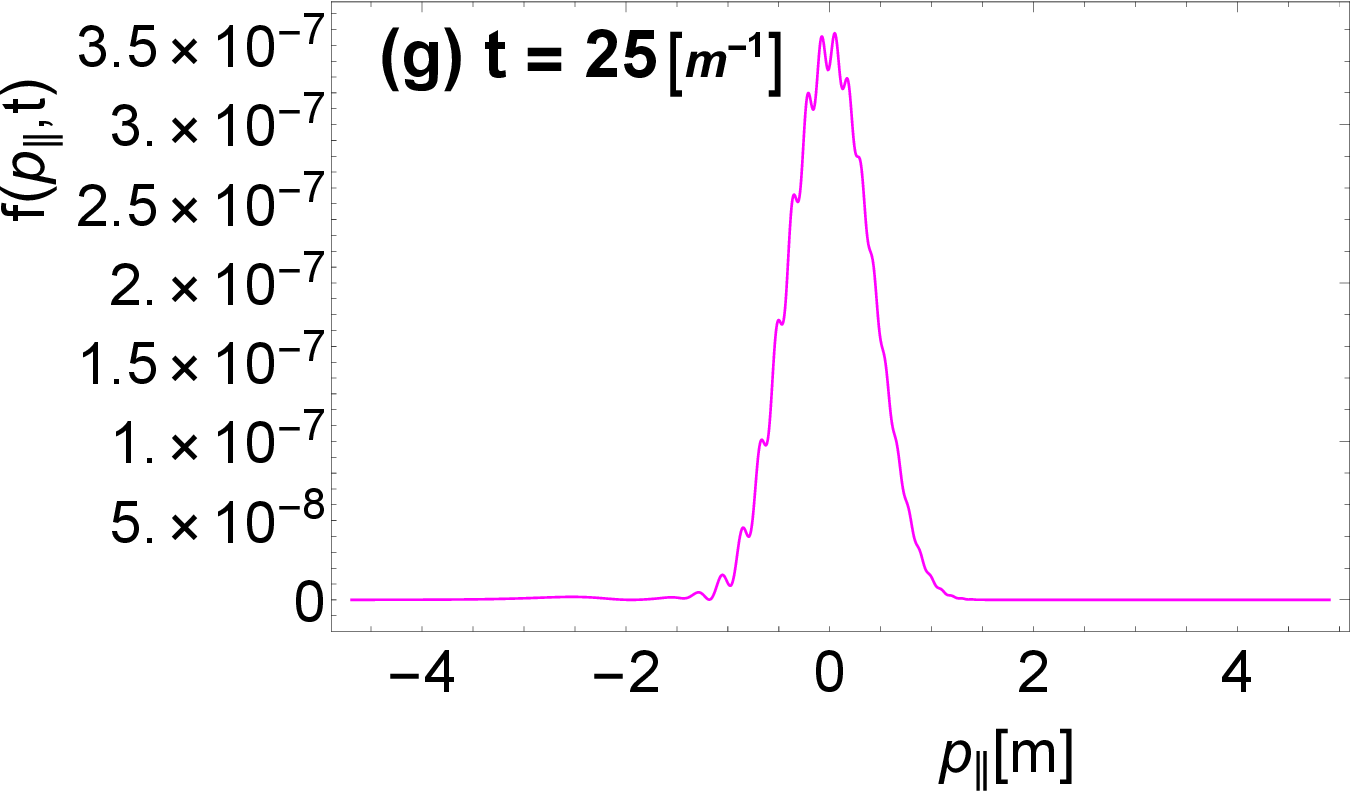}
\includegraphics[width =  1.9358802in]{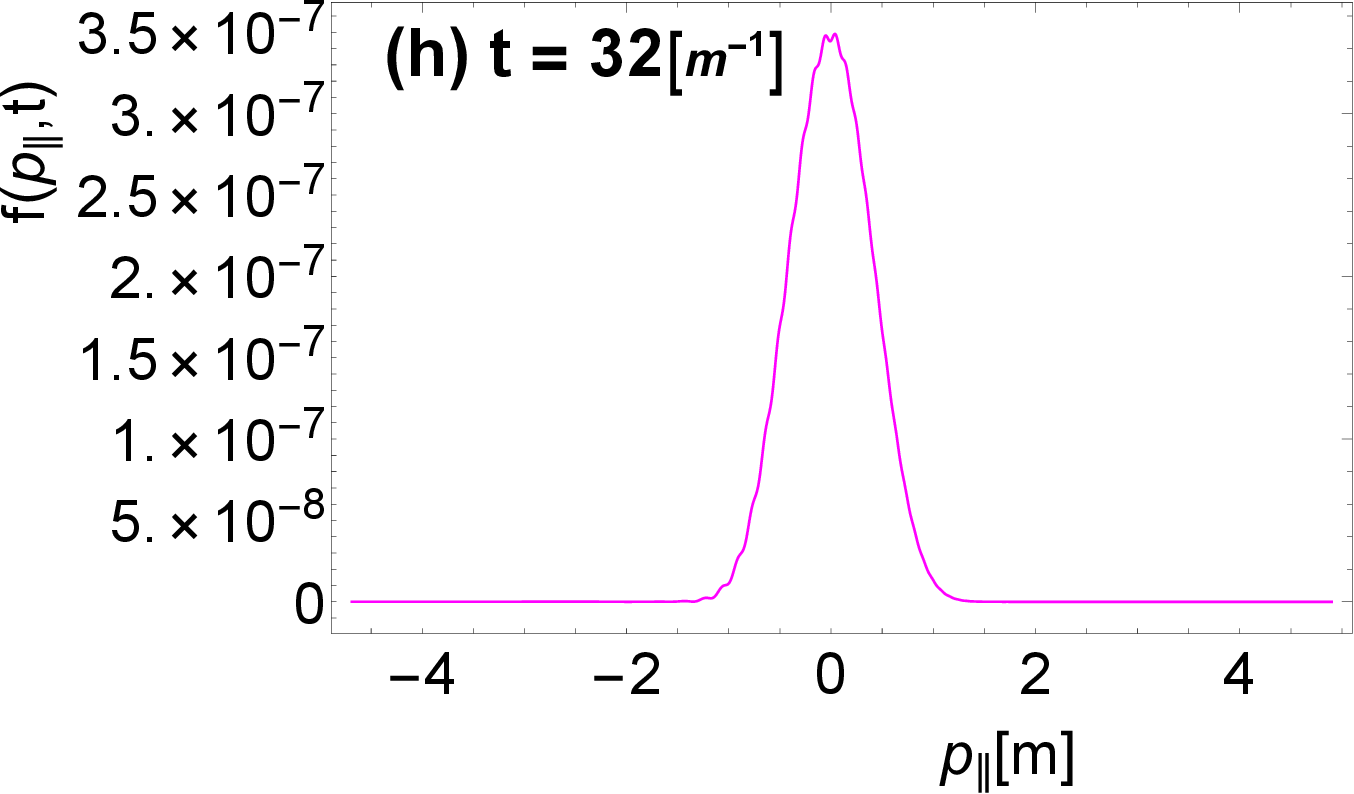}
\includegraphics[width =  1.9358802in]{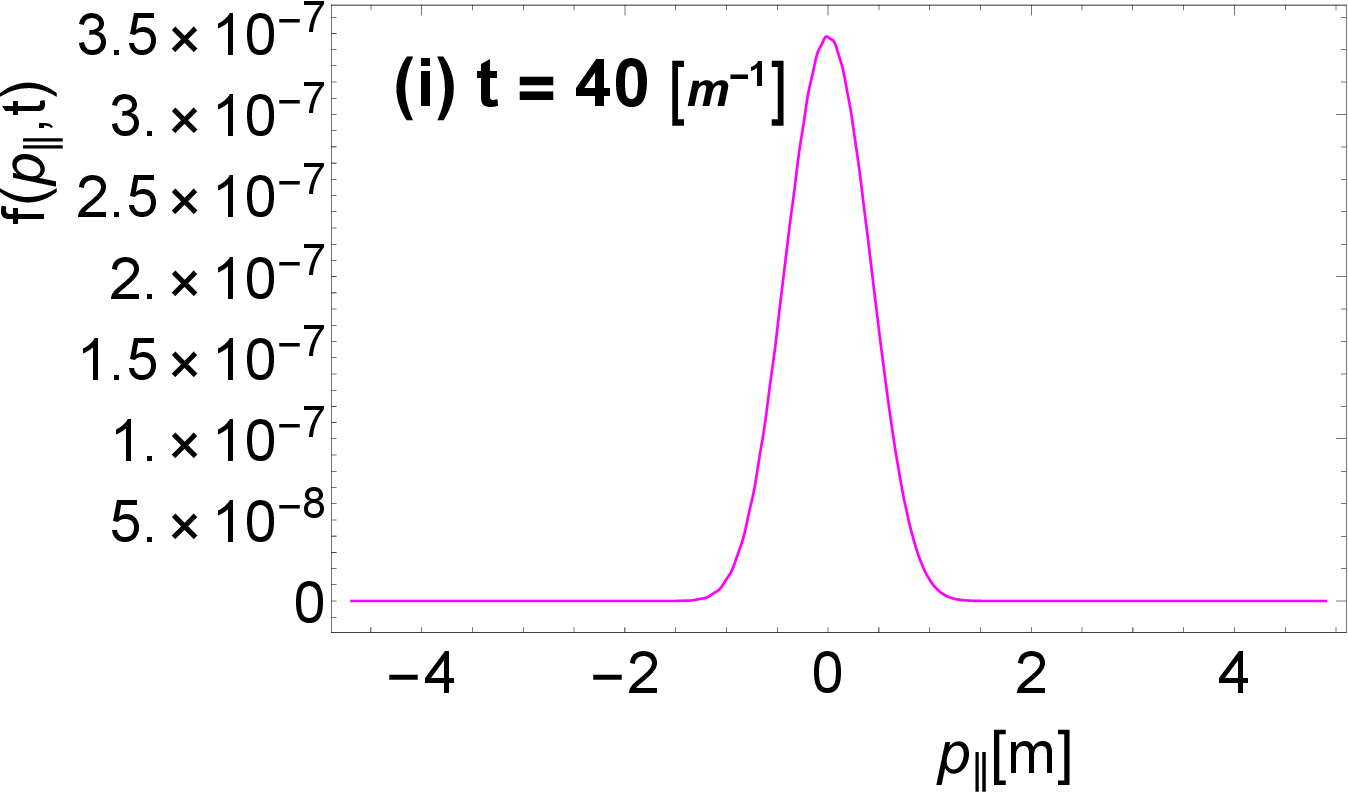}
}
\caption{Time evolution of momentum distribution function for second choice $\Omega(p_\parallel, t) = \omega(p_\parallel, t)$ and $V(p_\parallel, t) = - \frac{\dot{\omega}(p_\parallel,t)}{ 2 \omega(p_\parallel,t)}$. The transverse momentum is considered to be zero, and all the units are taken in the electron mass unit.The field parameters are  $E_0=0.2 E_c$ and $ \tau =10 [m^{-1}].$}
   	\label{firstorder}
\end{center}
\end{figure}

In Figure \ref{firstorder}, the time evolution of the momentum distribution function for the second choice, where $\Omega(p_\parallel, t) = \omega(p_\parallel, t)$ and $V(p_\parallel, t) = - \frac{\dot{\omega}(p_\parallel, t)}{2 \omega(p_\parallel, t)}$, is depicted.
At $t = -15 \, [m^{-1}]$, the spectra show two unequal peaks: one at $p_\parallel \approx 2 \, [m]$ with a large value, and another at $p_\parallel \approx 1 \, [m]$ (see fig.\ref{firstorder}(a)). As time proceeds and the electric field strength increases, the bi-modal structure of the spectra changes. We now observe a smooth, unimodal Gaussian-like profile with a peak located at $p_\parallel = 0 \, [m]$, as shown in Figure \ref{firstorder}(b) when the electric field is at its maximum at $t = 0$.
On the right side of the  Gaussian profile, there is deformation within the narrow range $-0.5< p_\parallel <1$ as depicted in Fig.\ref{firstorder}(c).
 After time  $t =\tau,$ when the electric field's magnitude approaches zero, a central peak emerges within the spectra. This peak manifests around a longitudinal momentum value of zero, accompanied by observable oscillations within a confined range  $-1 < p_\parallel <1,$ as demonstrated in Fig.\ref{firstorder}(d).
The central peak structure grows over time as the electric field vanishes. As a result, the Gaussian-like profile on the left side of the origin that was dominant earlier becomes diminished, and the central peak now dominates the peak and has oscillatory behavior at $t =15 [m^{-1}] $  as shown in figure \ref{firstorder}(e).
At $ t = 2 \tau,$we see that the left side peak $p_\parallel \approx -2 [m]$ is now hardly visible. The central peak shows oscillatory behavior, as shown in figure \ref{firstorder}(f), in a small window of longitudinal momentum where the electric field diminishes. Intriguingly, this oscillation exhibits an asymmetry, with its amplitude being more pronounced for negative longitudinal momentum than its positive longitudinal momentum.
At $t = 25 [m^{-1}],$ Only the dominant peak at $p_\parallel = 0$ persists, with a faint onset of oscillatory behavior superimposed on a Gaussian-like structure eventually, as depicted in figures \ref{firstorder}(g) to (h), the oscillation gradually fades away by $t = 32 [m^{-1}]$.
This type of similar trends discussed spectra for the first choice see figure \ref{lms_zero}(f) to (i).
At $ t = 40 [m^{-1}]$, oscillation becomes washed out, and spectra show a Gaussian-like structure  with$p_\parallel = 0 [m]. $
Comparing the trends of spectra for two different choices, as illustrated in Figures \ref{lms_zero} and \ref{firstorder}, we note distinct characteristics in the momentum spectra when the pairs are in the off-shell mass configuration, stemming from the choice of adiabatic basis functions.
Additionally, as particles reach the final on-shell configuration, the behavior of spectra becomes consistent in the absence of an electric field.
\subsection{ Temporal evolution of distribution function and dynamical scaling }
\begin{figure}[t]
\begin{center}
{\includegraphics[width =3.19358802in]{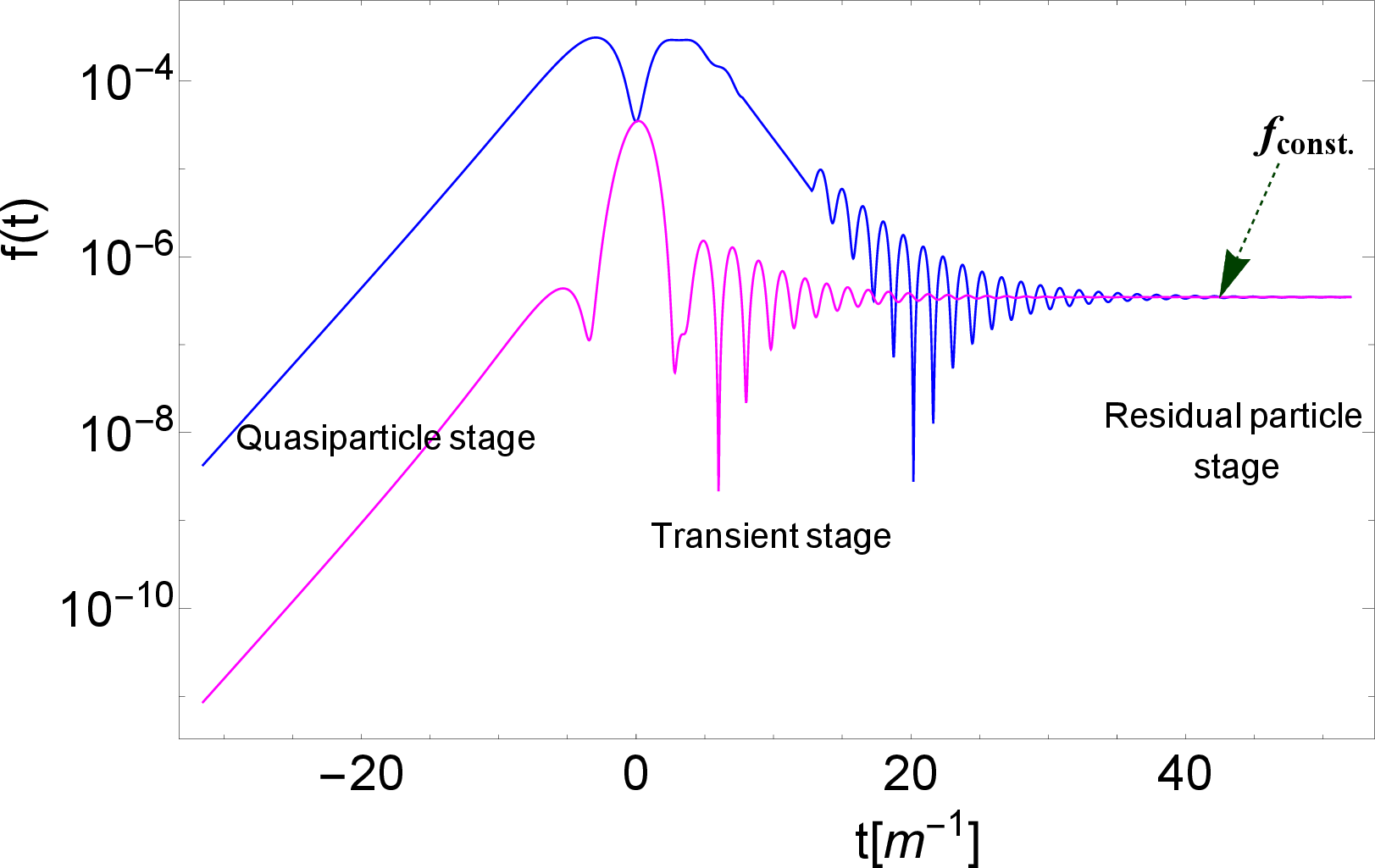}
\includegraphics[width =3.19358802in]{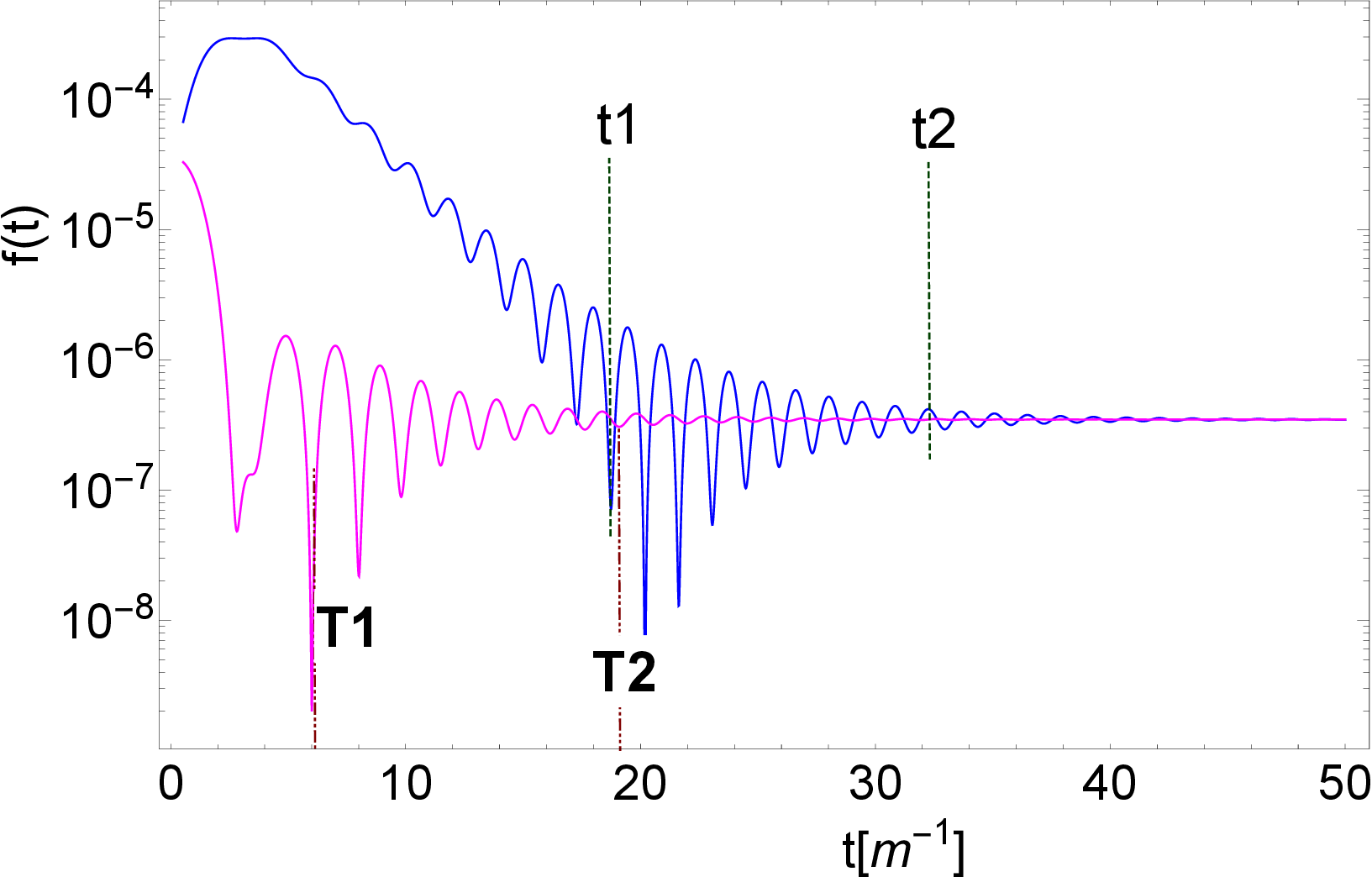}
}
\caption{Evolution of  distribution function  $f(\bm{p} =0, t)$  for two choices. The blue curve represent $f(t)$  for first  choice  of adiabatic freqency $\Omega(p_\parallel,t) = \omega(p_\parallel,t)$ and $V(p_\parallel,t) = 0$ and magenta curve for second choice  $\Omega(p_\parallel,t) = \omega(p_\parallel,t)$ and $V(p_\parallel,t) = -\frac{\dot{\omega}(\bm{p_\parallel},t)}{ \omega(p_\parallel,t)}$ .The momentum is considered to be zero, and all the units are taken in the electron mass unit.The field parameters are  $E_0=0.2 E_c$ and $ \tau =10 [m^{-1}].$}
\label{f_t}
\end{center}
\end{figure}

The influence of an external electric field renders the quantum vacuum unstable, leading to the generation of virtual particle-antiparticle pairs in an off-mass-shell state. These virtual charged particles undergo acceleration by the electric field, acquiring enough energy to transform into real particles in an on-shell mass state. Consequently, the distribution function $f(\bm{p},t)$ exhibits three temporal stages: $(i)$ the QPAP stage in the region of maximal external field values, $(ii)$ the transition region marked by fast oscillations, and $(iii)$ the final RPAP stage where $ f(t) $ approaches a constant residual value $f_{const.}.$, as depicted in Fig. \ref{f_t}. We plot the evolution of $f(t)$ concerning two different choices. The blue curves correspond to the basis 
$\Omega(\bm{p},t) = \omega(\bm{p},t)$ and $V(\bm{p},t)=0$ (first choice),  while the magenta  curves correspond to the basis $\Omega(\bm{p},t) = \omega(\bm{p},t)$ and $V(\bm{p},t)= - \frac{\dot{\omega}(p_\parallel,t)}{2\omega(p_\parallel,t)}$(second choice).
The magnitude of the distribution function in the QPAP and transient regions is suppressed in the second choice compared to the first one ( see left panel of Fig. \ref{f_t}).
Different also influences the behavior of fast oscillation in the transient region. To provide a qualitative context, one can define the time interval characterized by fast oscillations, bounded by the initial point $t_{in}$ where the oscillation of $f(t)$ first reaches the level of the residual particle stage. The time $t_{out}$ where the transient stage ends are when the average level of the oscillating $f(t)$ reaches the final state; the residual particle stage begins. We labeled this time $ t_{in} = t_1 (\textrm{first choice}), T_1  (\textrm{second choice})$  and $ t_{out} = t_2 (\textrm{first choice}), T_2(\textrm{second choice})$ as shown in right panel of Fig.\ref{f_t}. In the residual particle stage, quasiparticles become independent, and real particle-antiparticle pairs are observed with a lower value of  $f(t)$ than at the maximum electric field at $t = 0.$ We also noted that upon reaching the RPAP stage, different choices yield the same information about the pair production process when the electric field vanishes. 
\subsubsection{Momentum spectra and Dynamical Scaling  }
\begin{figure}[t]
\begin{center}
{\includegraphics[width =2.13559358802in]{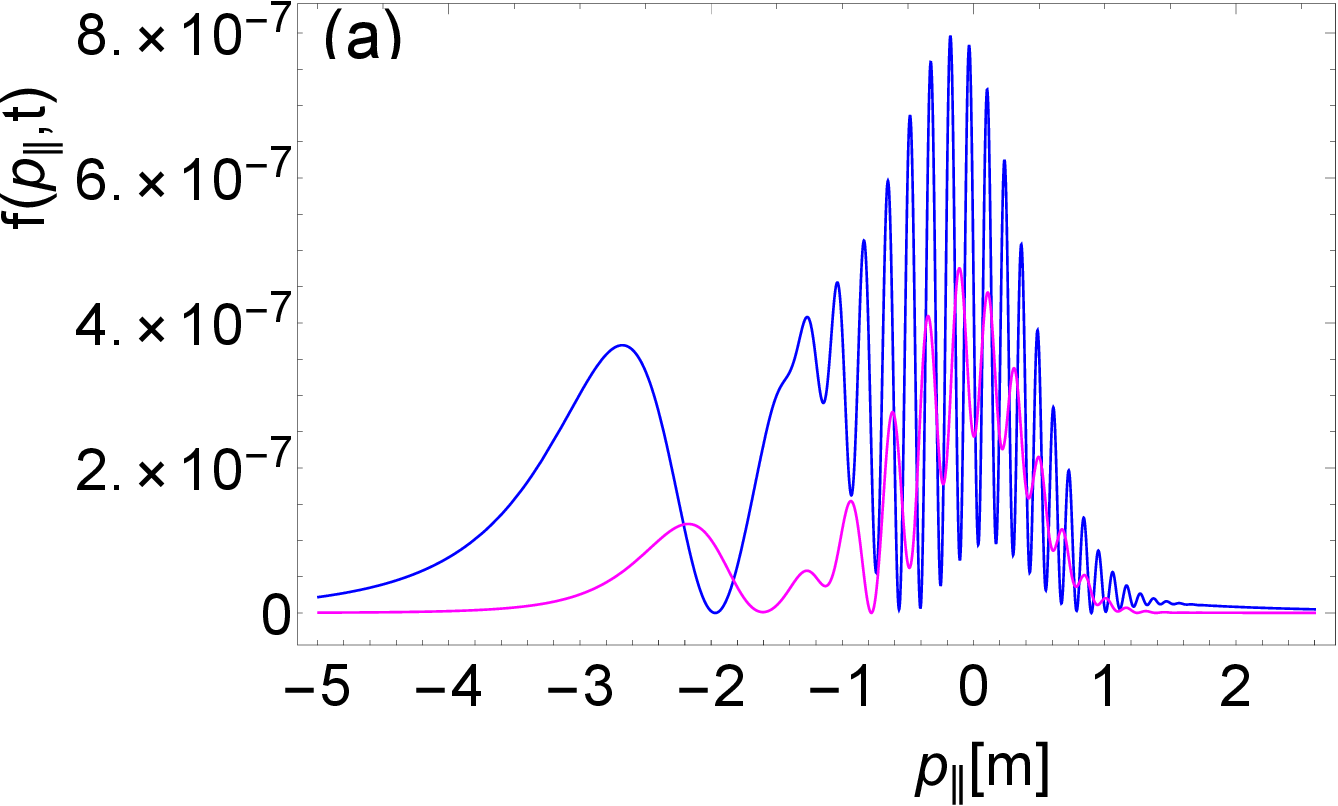}
\includegraphics[width =2.1359358802in]{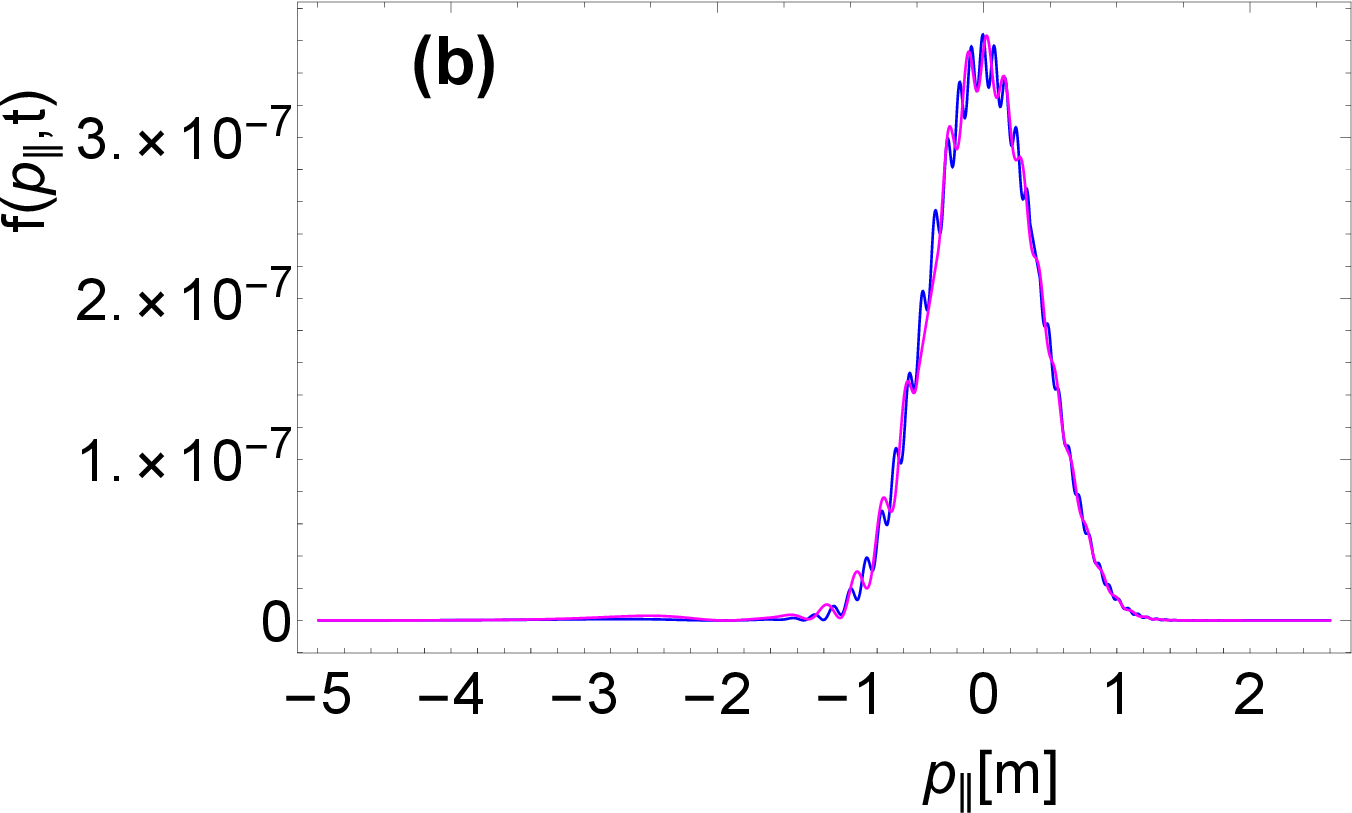}
\includegraphics[width =2.1359358802in]{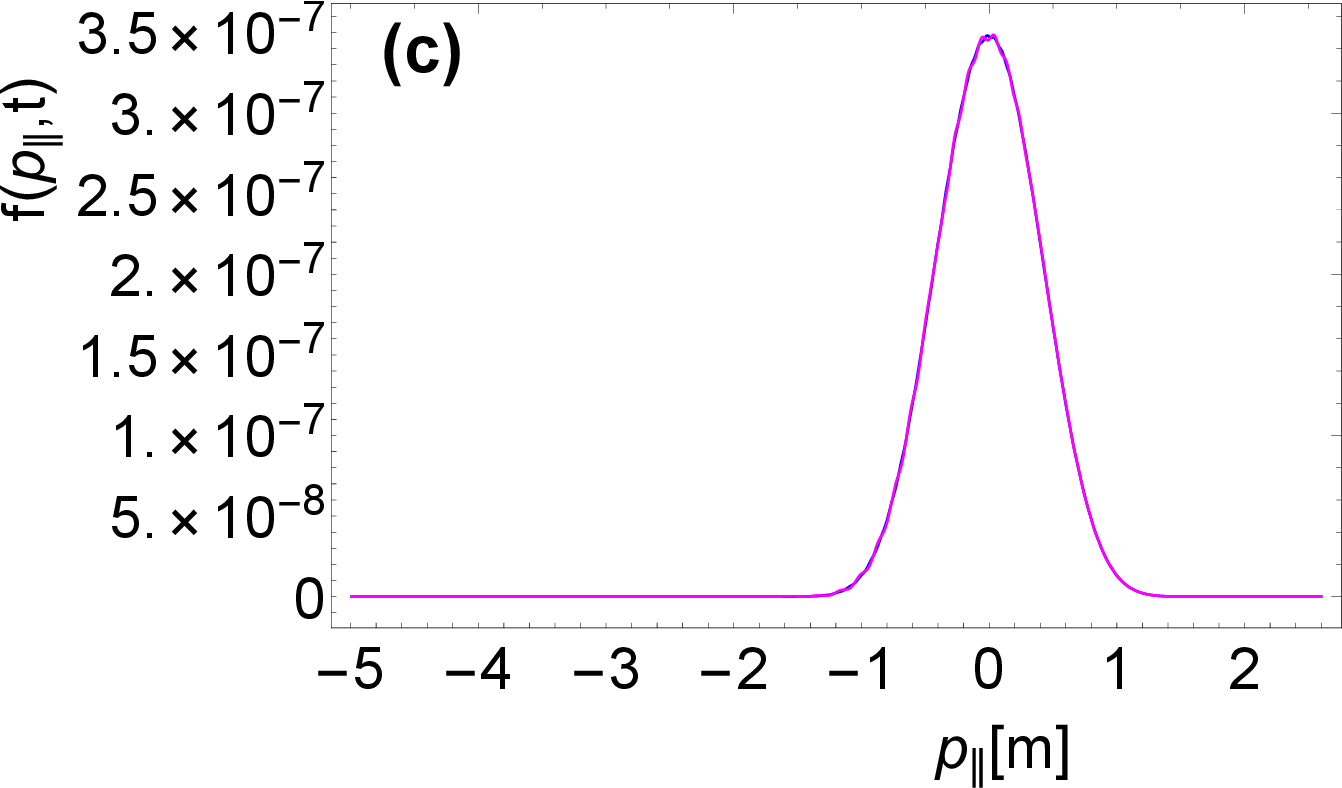}
}
\caption{Momentum spectra of created pairs at different times  $(a) t = \frac{3}{4} t_{2}, (b) t = \frac{5}{4} t_{2}$  and  $(c) t = \frac{7}{4} t_{2}$ for first choice ( blue curve) 
and spectra  at time  $(a) t = \frac{3}{4} T_{2}, (b) t = \frac{5}{4} T_{2}$  and  $(c) t = \frac{7}{4} T_{2}$  for second choice (magenta curve ).The field parameters are  $E_0=0.2 E_c$ and $ \tau =10 [m^{-1}].$ }
\label{btran_lms}
\end{center}
\end{figure}

In the previous section \ref{LMS_choice}, we discussed how the behavior of the momentum spectrum remains consistent across both choices of  adiabatic frequency functions as the electric field diminishes to zero at late times. However, the choice of adiabatic frequency functions, \((\Omega(p_\parallel, t), V(p_\parallel, t))\), significantly influences the system's behavior at earlier times, particularly in the QPAP and transient stages.
 A novel dynamical scaling is observed while analyzing the oscillatory momentum spectrum of the pairs created at intermediate times, calculated using the two adiabatic bases. In these bases, the same oscillatory momentum spectra are observed but at different times. However, when we scale the time by the point marking the end of the transient stage (or, onset of the residual stage) $t_{\text{out}}$ of dynamical evolution for each case of central momentum, the respective momentum spectra overlap. To illustrate this effect, we present the momentum spectra of created particles at specific times relative to the onset of the residual stage (or the conclusion of the transient stage). Specifically, we analyze three time points: \(t = \frac{3}{4} t_{\text{out}}\), near the beginning of the residual stage; \(t = \frac{5}{4} t_{\text{out}}\); and \(t = \frac{7}{4} t_{\text{out}}\), within the RPAP stage. It is important to emphasize that the time \(t_{\text{out}}\) varies depending on the chosen basis, which introduces a shift in the transition timing.
In Figure \ref{btran_lms}, at \(t = \frac{3}{4} t_{\text{out}}\), the momentum spectrum exhibits a two-peak structure. The first peak is small with a smooth profile, while the second peak has a Gaussian envelope accompanied by oscillatory effects. While the overall behavior of the two peaks is similar for both choices of bases, the amplitude of oscillations is noticeably larger for the first choice compared to the second, as shown in Figure \ref{btran_lms}(a). At \(t = \frac{5}{4} t_{\text{out}}\), well into the residual stage and beyond the transient stage, only the central peak at \(p_\parallel = 0[m]\) remains visible, now accompanied by onset oscillations. The second peak, however, becomes almost indistinguishable, as illustrated in Figure \ref{btran_lms}(b). It is important to note that the precise location of the oscillations depends on the chosen residual time and also varies with the momentum value. In this analysis, \(t_{\text{out}}\) is defined as the time corresponding to zero longitudinal momentum, marking the end of the transient region and the beginning of the residual stage.
As time progresses, the oscillation amplitude diminishes and vanishes entirely by \(t \approx 2 t_{\text{out}}\). Beyond this point, the spectra for both choices of adiabatic bases consistent, resulting in an identical final particle state, as shown in Figure \ref{btran_lms}(c). The qualitative features of the spectra remain consistent across both approaches, but differences in oscillation amplitude persist during earlier times. To investigate the origin of these variations in oscillations observed in the spectra, we analyze the approximate relation of the distribution function \(f(p_\parallel, t)\) for the two different adiabatic bases.

\begin{figure}[t]
\begin{center}
{
\includegraphics[width =  1.99in]{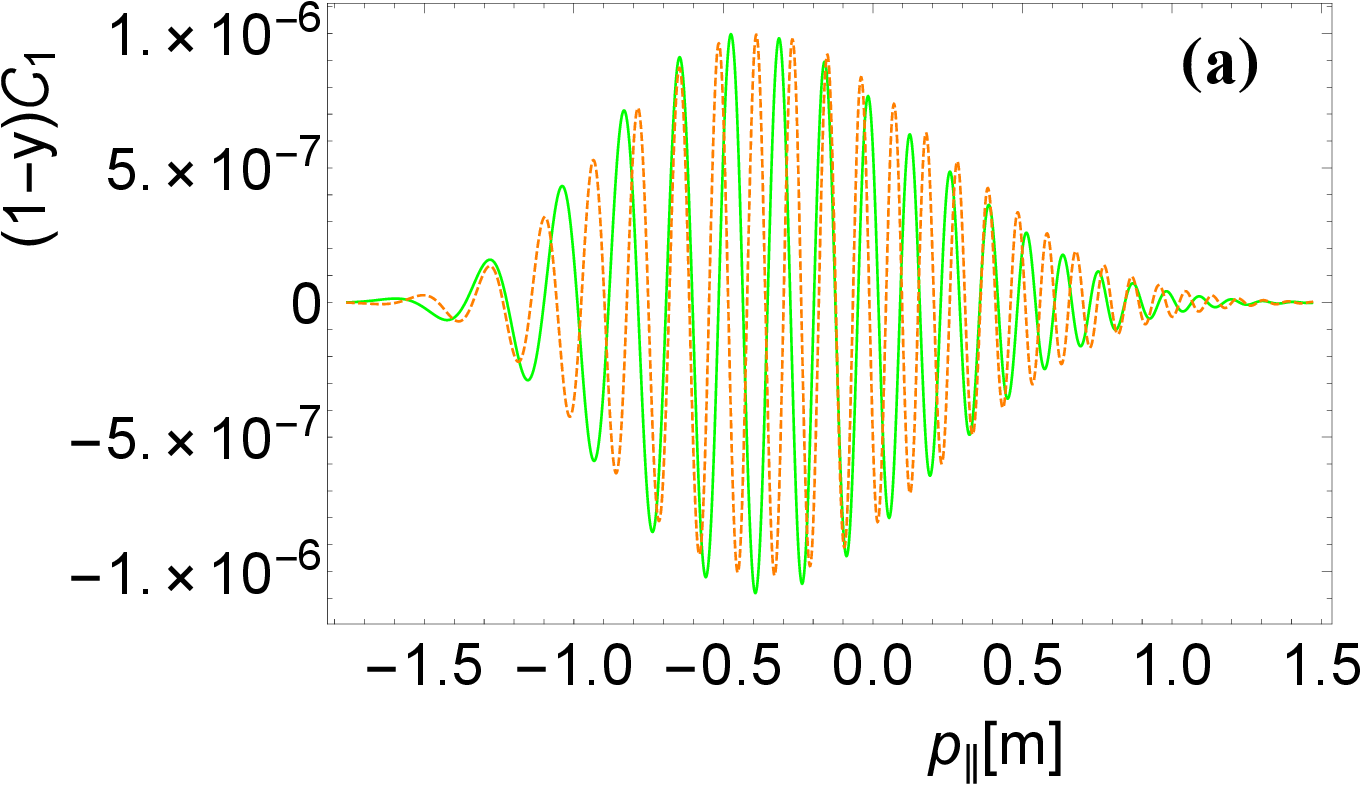}
\includegraphics[width =  1.99in]{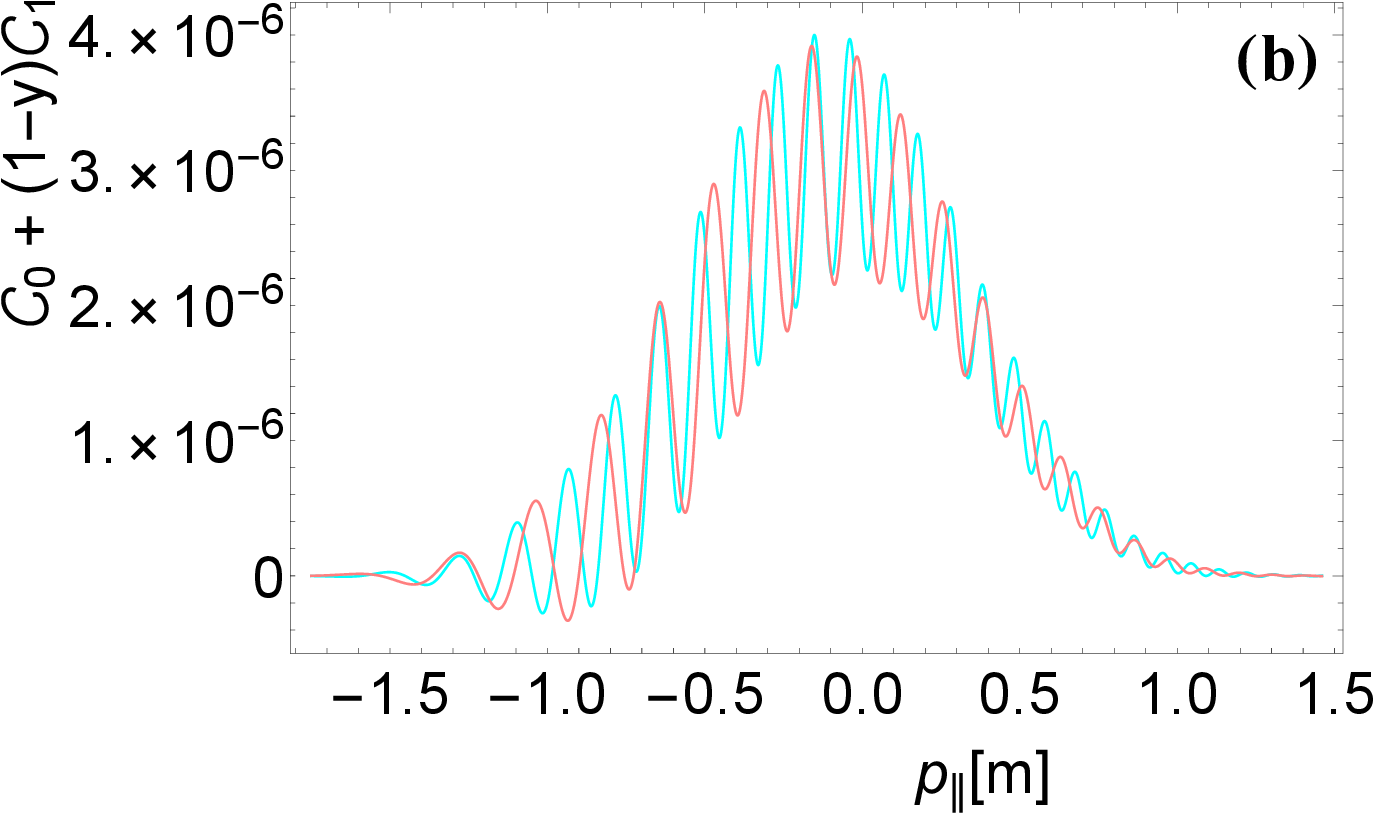}
\includegraphics[width =  1.99
in]{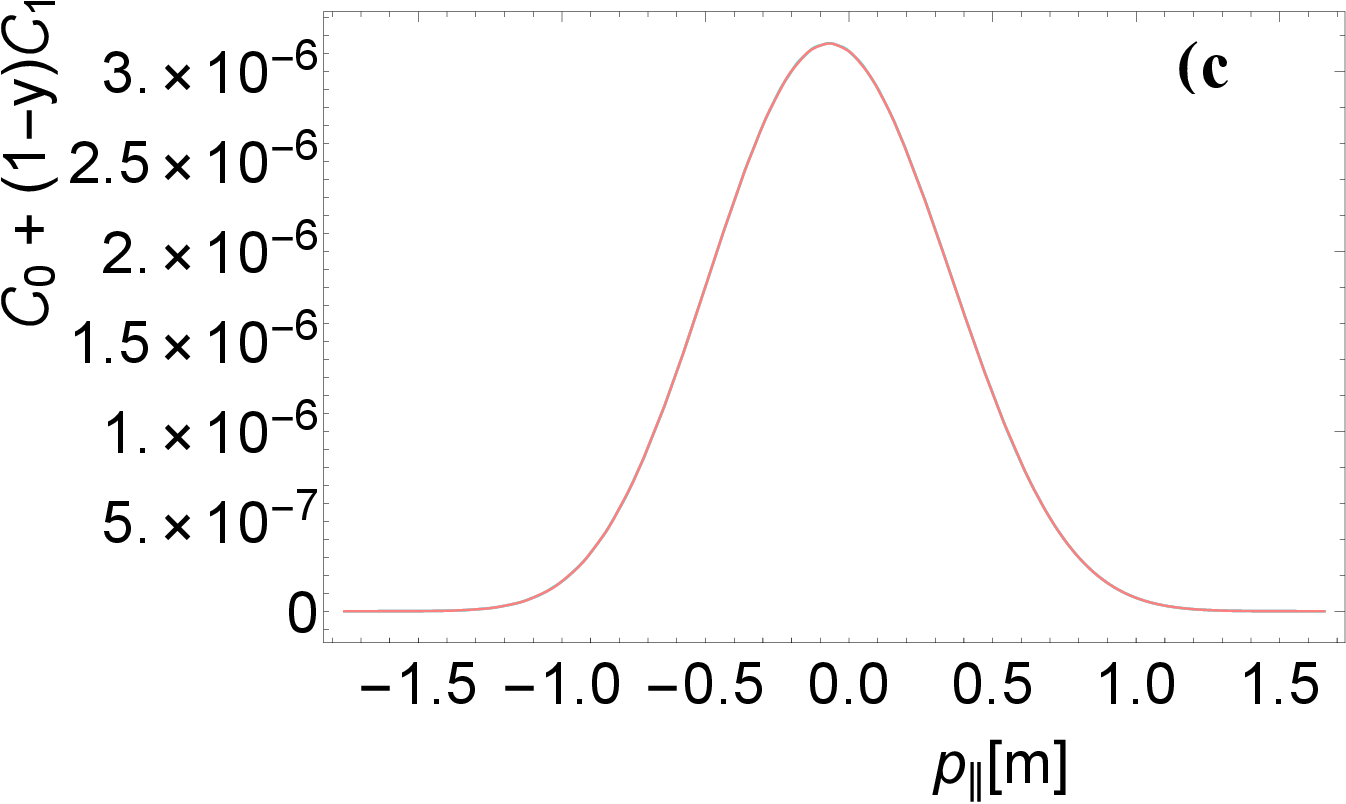}
}
\caption{ Time evolution of the components defined in Eq.\eqref{appdisfun} of the approximate analytical expression of  $f(\bm{p},t)$.
$\mathcal{C}_1(1-y)$ : first choice (orange) \& second choice (green) and $\mathcal{C}_0 + \mathcal{C}_1(1-y)$ : first choice(cyan) \& second choice(pink).The transverse momentum is considered to be zero, and all the units are taken in the electron mass unit.The field parameters are  $E_0=0.2E_c$ and $ \tau =10 [m^{-1}].$}
   	\label{comp_appr_terms}
\end{center}
\end{figure}

As discussed previously, the term $ (1-y) \mathcal{C}_1$ is mainly responsible for the quantum interference effect observed as an oscillation in the momentum spectra. In Figure \ref{comp_appr_terms}, we plot this term at finite time $t$, where  $ E( t) \rightarrow 0$ for both approaches. From Fig. \ref{comp_appr_terms}, we can observe that the first-order term of the approximate distribution function for both choices  qualitatively shows the same feature of a Gaussian envelope with sine or cosine oscillations,  at different time $t_{out}$ as explicitly confirmed by Eq. \eqref{appC1}. But, the nature is quantitatively different; the oscillation amplitude is more significant for the first choice of basis than the second choice for the same time  in figure not shown here.
In the late-time limit $(y \rightarrow 1)$, the coefficients of $(1 - y)$ in Eq. \eqref{pdfC1} can be further approximated by retaining only the dominant contribution and disregarding the others.
\begin{align}
    \mathcal{C}_1(p_\parallel) &\approx  2 |\Gamma_1 \overline{\Gamma_2}|   \nu_0  \omega_1  \Biggl[  \Bigl(1 + 4 \lambda^2 + \tau ( V_{1} + 2 \omega_1 \tau (\omega_1 - \omega_0  + w_1)
    - (\omega_0 - \omega_1 )^2 \tau  ) 
\Bigr) \cos{\Upsilon}   +  ( 2 w_1 - \tau \omega_1 V_1)   \sin{\Upsilon} \Biggr]
    \label{appC1}
\end{align}
The difference in oscillation amplitude observed at different times can be understood using the approximate expression, which incorporates the real function \(V(p_\parallel, t)\). This function influences the momentum spectra for different choices of bases, although the overall oscillatory pattern remains consistent as the system reaches the residual particle stage, where real particle formation occurs. As time progresses, the magnitude of \(V(p_\parallel, t)\) becomes significantly suppressed. Consequently, the zeroth-order term begins to dominate, causing the oscillatory behavior to diminish and eventually become less noticeable or vanish. However, this behavior remains finite and does not entirely disappear.

\subsubsection{Multi-photon regime}
\begin{figure}[t]
\begin{center}
{
\includegraphics[width =  2.029358802in]{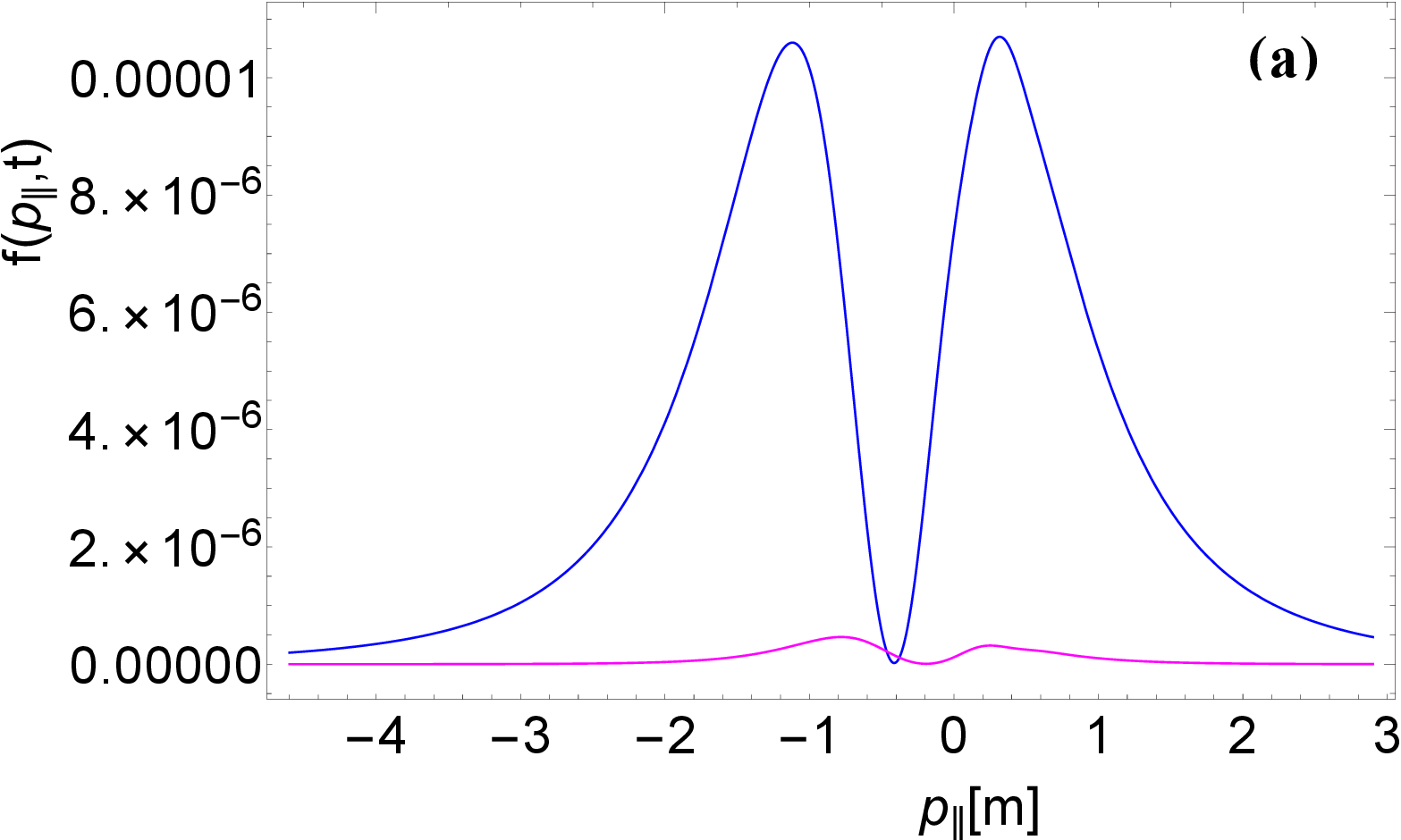}
\includegraphics[width =  2.029358802in]{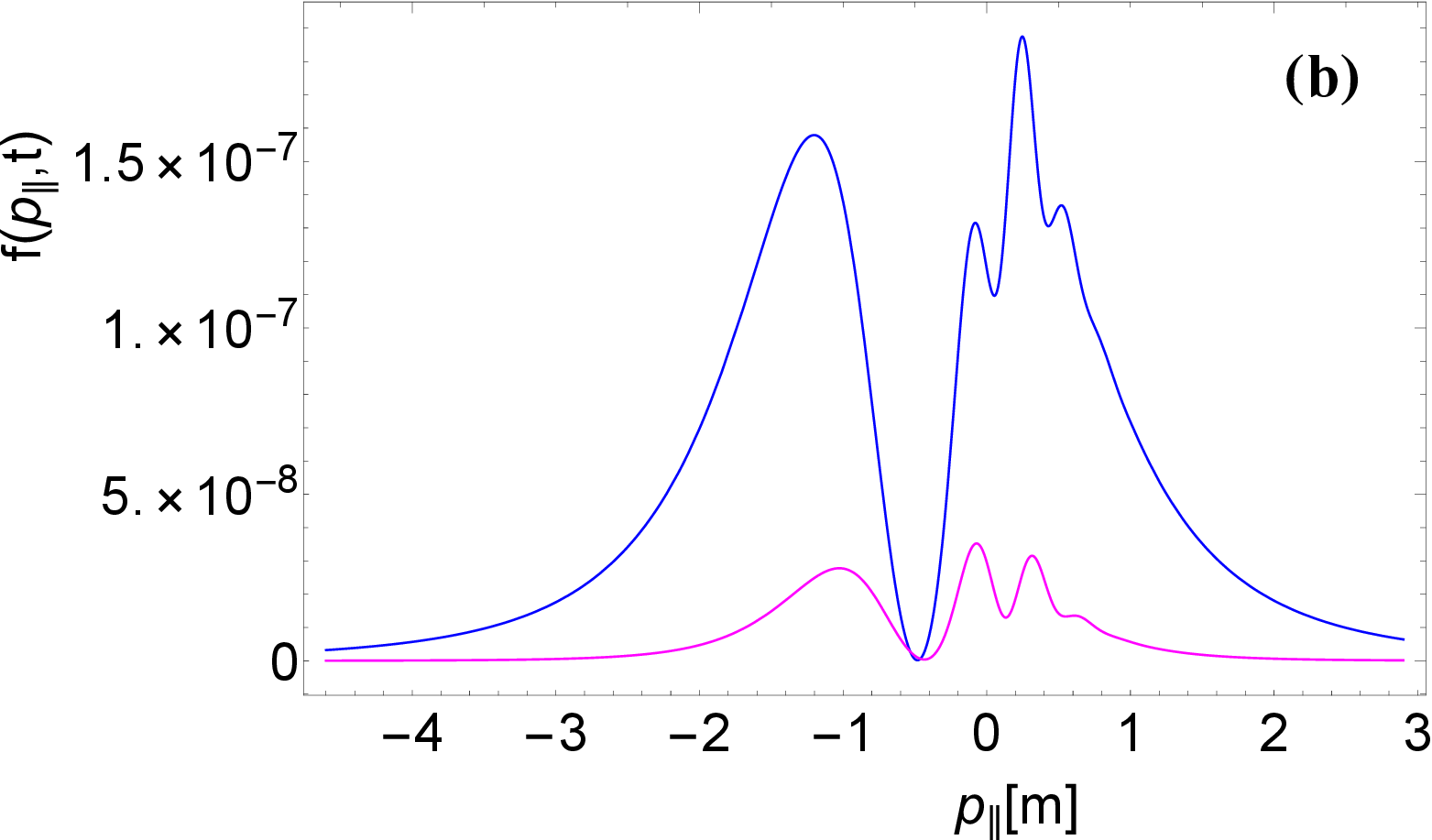}
\includegraphics[width =  2.029358802in]{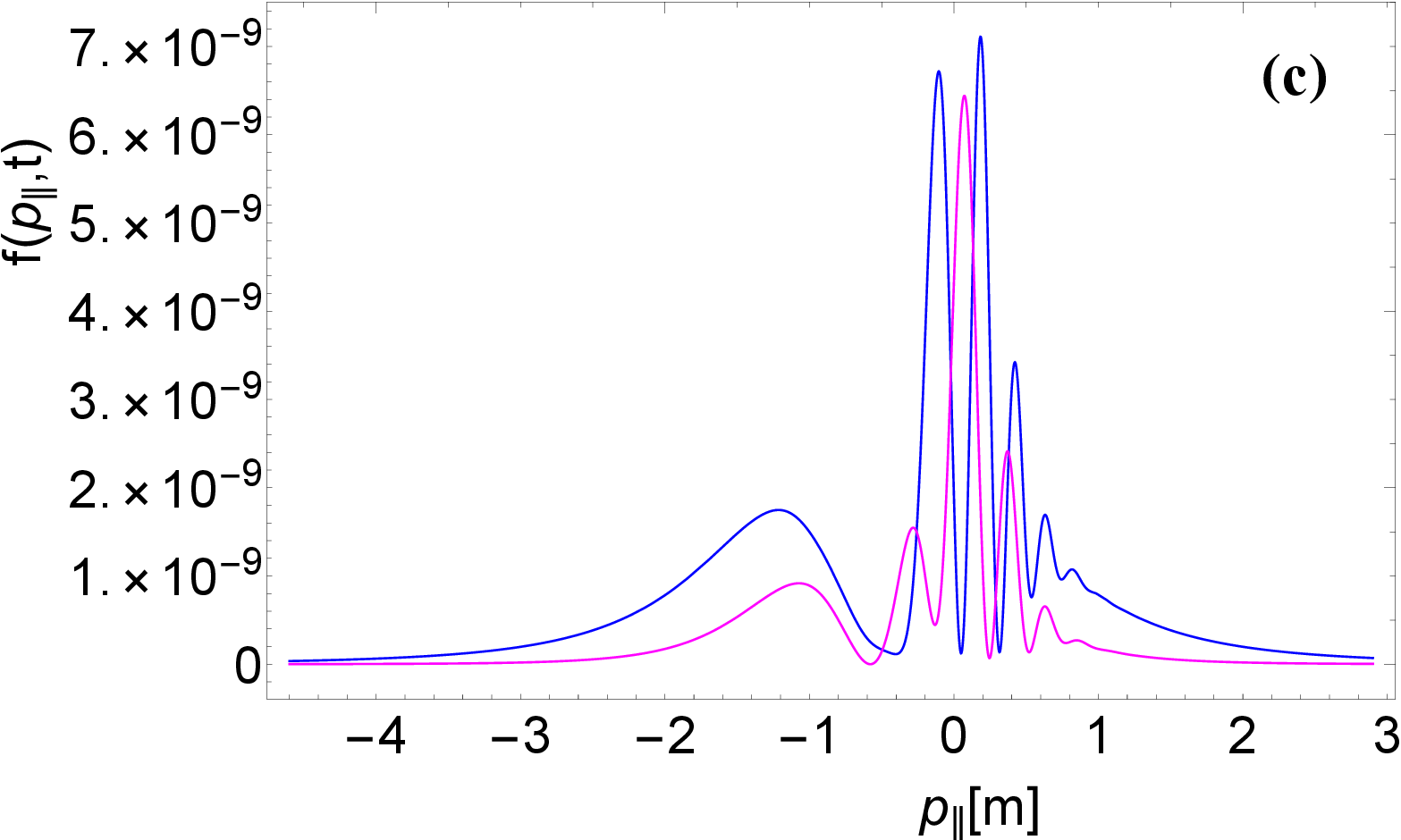} 
\includegraphics[width =  2.029358802in]{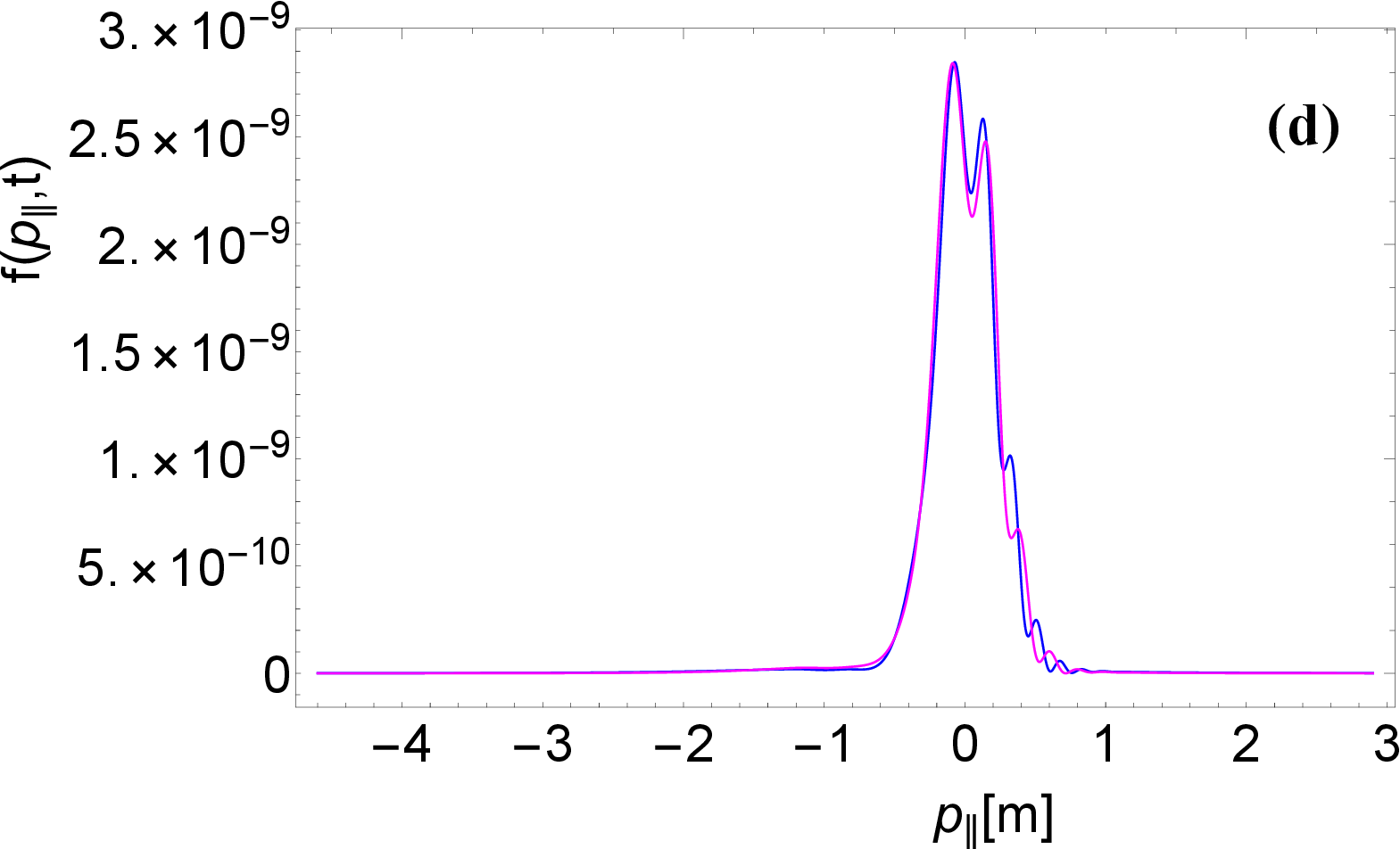} 
\includegraphics[width =  2.029358802in]{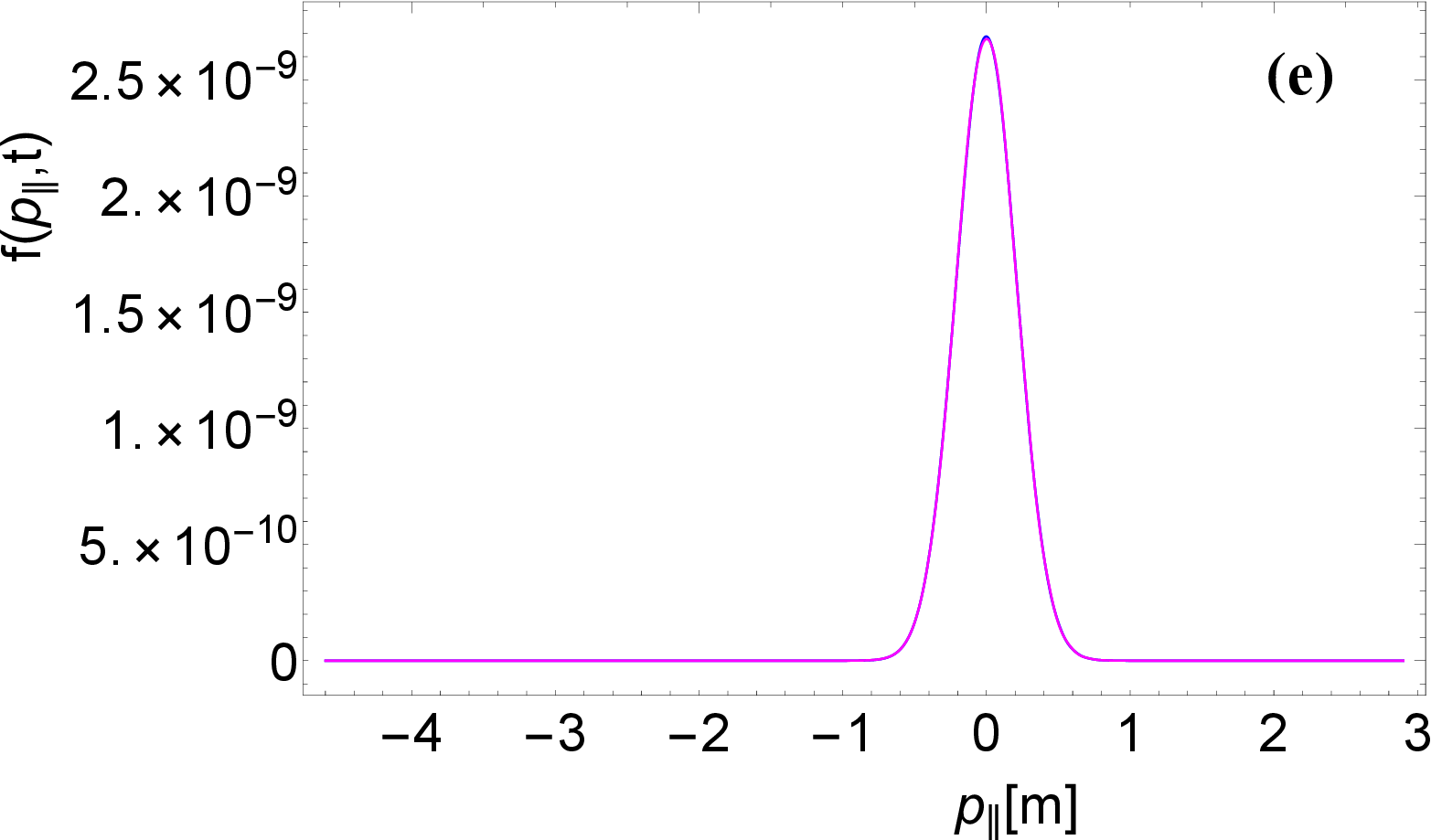} 
\includegraphics[width =  2.029358802in]{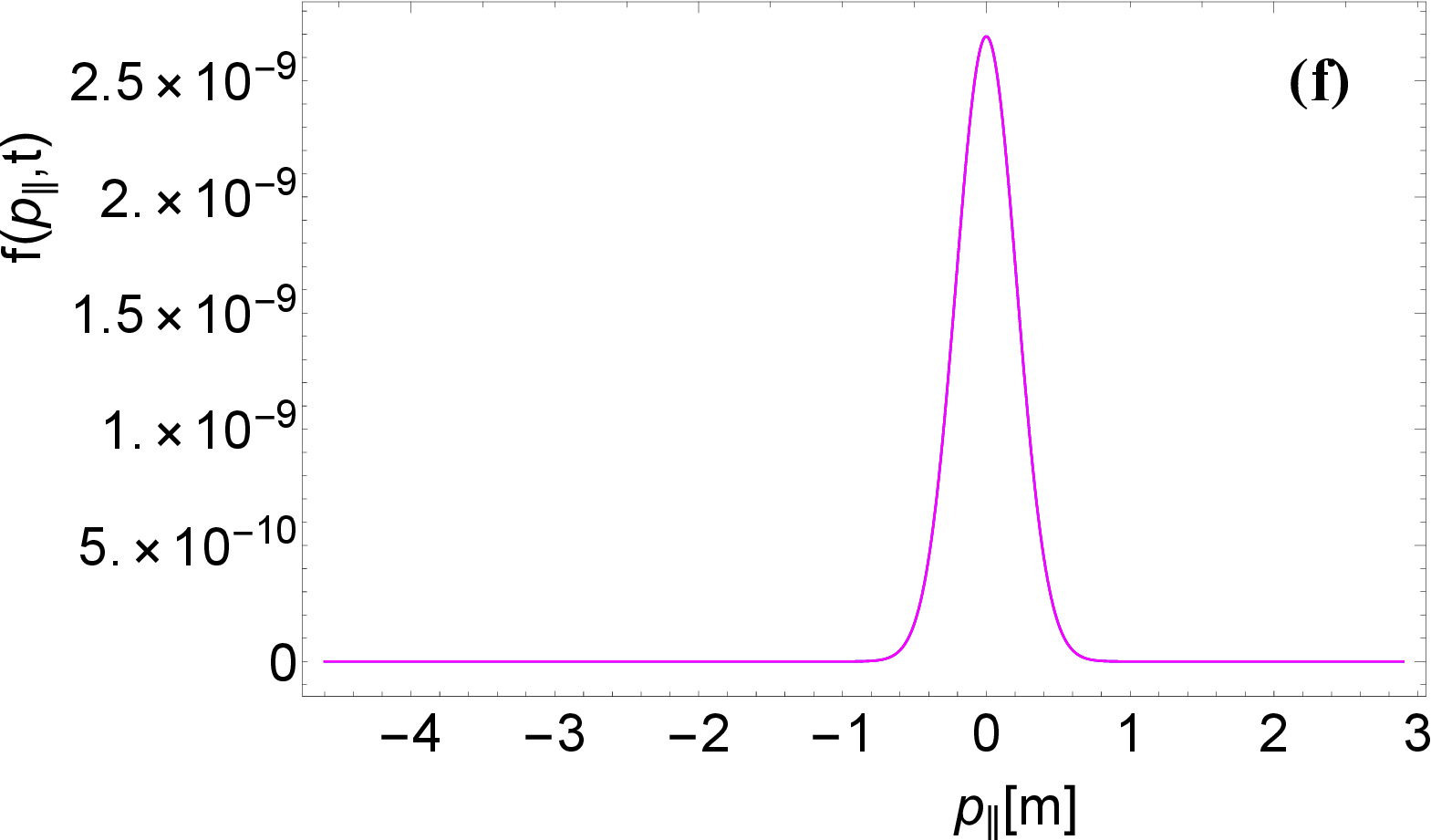} 
}
\caption{Time evolution of normalized momentum distribution function at different times.
(a) $t = \frac{1}{4} t_{out}.$ (b) $t = \frac{1}{2} t_{out}$ (c) $t =  \frac{3}{4} t_{out}$ (d) $t = t_{out}$ (e) $t =\frac{5}{4} t_{out}$ (f)$t=2 t_{out}.$ 
The blue curve represent $f(p_\parallel,t)$  for first  choice  $\Omega(p_\parallel,t) = \omega(p_\parallel,t)$ and $V(p_\parallel,t) = 0$ and magenta curve for second choice  $\Omega(p_\parallel,t) = \omega(p_\parallel,t)$ and $V(p_\parallel,t) = -\frac{\dot{\omega}(\bm{p_\parallel},t)}{ 2 \omega(p_\parallel,t)}$. The transverse momentum is considered to be zero, and all the units are taken in the electron mass unit.The field parameters are  $E_0=0.1E_c$ and $ \tau =5 [m^{-1}].$}
   	\label{multiphoton_a}
\end{center}
\end{figure}
In this section, we exmaine the momentum spectra of created pairs in the multiphoton regime. We choose laser pulse parameters to ensure that the Keldysh parameter, $\gamma \gg 1$. Figure \ref{multiphoton_a} displays the spectra for a short pulse duration $\tau = 5 \, [\text{m}^{-1}]$ and a pulse strength $E_0 = 0.1 E_c$. In this case, the Keldysh parameter is $\gamma = 2$, indicating that the system operates in the multiphoton regime, as expected from $n^{\text{th}}$-order perturbation theory, where $n$ represents the minimum number of photons required to surpass the pair creation threshold energy, i.e., $n\omega > 2m$ \cite{Mocken:2010uhp}.
To compare the momentum distribution functions for both cases, we aim to plot the spectra on a common time scale. To achieve this, we scale time by multiples of $t_{\text{out}}$, which corresponds to the RPAP  stage for each basis choice, as discussed in the previous section.
Figure \ref{multiphoton_a} presents the time evolution of the momentum distribution function, $f(p_\parallel, t)$, in momentum space for the two different cases.
For the first case, the spectra display two nearly equal peaks: one at $p_\parallel = -1.1 \, [\text{m}]$ and another at $p_\parallel = 0.32 \, [\text{m}]$, with the distribution being symmetric about $p_\parallel \approx -0.4$. In contrast, for the second case, the spectra also exhibit a bimodal profile, with unequal peaks: one at $p_\parallel = -0.77 \, [\text{m}]$ and another at $p_\parallel = 0.25 \, [\text{m}]$, but the magnitude of the function is less pronounced compared to the first case and is hardly visible, as shown in Fig.~\ref{multiphoton_a}(a). As time progresses, the spectra, which initially show a smooth bimodal structure, begin to change rapidly. At $t \approx t_{\text{out}}/2$, both choices exhibit nearly identical spectral trends. The right-side peak splits, forming three smaller peaks, and the left-side peak at $p_\parallel \approx -1.1 \, [\text{m}]$ is suppressed in the first case. For the second case, there is still one peak on the left side of the origin at $p_\parallel = -1 \, [\text{m}]$, though it is less dominant, while the right side of the spectrum shows a multi-modal structure, as seen in Fig.~\ref{multiphoton_a}(b).
In Fig.~\ref{multiphoton_a}(c), the momentum spectrum has developed a multi-modal profile, with the peak at $p_\parallel \approx 0$ being much more prominent than the smaller, unequal peaks observed in both cases. While there is a slight difference in magnitude, the overall behavior is qualitatively similar. This occurs at $t = 0.75 t_{\text{out}}$, a time when the quasiparticles have reached the stage where they transition to real particles. Finally, at $t = t_{\text{out}}$, the momentum spectra show a dip at $p_\parallel \approx 0$ and the merging of the multi-modal peaks. This merging occurs as the distinct peaks from both approaches fade, marking the transition to the residual particle stage. At this point, the dependence on the different choices of mode functions becomes irrelevant, and we observe an ambiguity in the distribution function's definition due to the real time-dependent function $V(p_\parallel, t)$ no longer playing a significant role. As a result, the spectra for both approaches exhibit a smooth unimodal profile with a peak occurring at $p_\parallel = 0 $, as depicted in Fig.~\ref{multiphoton_a}(f).

An intriguing qualitative contrast emerges when comparing the scenario with $\gamma = 0.5$ to the current situation. The figure highlights this distinct behavior, showing a multi-modal pattern at a specific moment. This pattern, marked by more than two peaks, occurs at $t \approx \tau$.

%
\section{Conclusion}
We investigate pair creation from the vacuum under a spatially homogeneous, time-dependent Sauter pulsed field within the framework of scalar quantum electrodynamics (QED). Employing the standard Bogoliubov transformation approach, we derive the single-particle distribution function, which is inherently linked to the adiabatic mode functions defining the choice of basis. Since the exact mode functions \(\Phi_{\bm{p}}^{(\pm)}(t)\) for the Sauter pulsed electric field are well-known in the literature, we use these to find the expression for the particle distribution function. This expression depends on the choice of basis, a well-established aspect that we further explore. Specifically, the distribution function’s evolution depends on the selected time-dependent functions \(\Omega(\bm{p}, t)\) and \(V(\bm{p}, t)\) in the adiabatic mode functions.
Our analysis reveals that the temporal behavior of the particle distribution function and the momentum spectrum of created pairs in the sub-critical field limit exhibit distinct features, depending on the choice of adiabatic basis. Despite these differences, the transition from virtual particles to real particles occurs in three distinct temporal stages, common across all bases. However, the timing of the transient stage varies: for one basis, the transient stage appears later, leading to a slower approach to the constant value of the Real Particle-Antiparticle Pair (RPAP) state. This suggests that the choice of functions $\Omega(\bm{p}, t)$ and $V(\bm{p}, t)$ influences the timing of transitions, even though the overall qualitative physics remains unaffected.

We define the initiation time of the transient stage as $t_{\text{in}}$ and the end of the transient stage as $t_{\text{out}}$, which marks the beginning of the RPAP stage. For different basis choices, the timescales $( t_{\text{in}}, t_{\text{out}} )$ generally differ. However, we find that after $t = t_{\text{out}}$, the distribution function behaves similarly across all bases. This is because, in the RPAP state, the distribution function becomes constant, and the particles satisfy the on-mass-shell condition, with real particle-antiparticle pairs emerging from the vacuum.

We also study the time evolution of the momentum spectra of created pairs, using the standard choice of $\Omega(\bm{p}, t) = \omega(\bm{p}, t)$ and $V(\bm{p}, t) = 0$. The momentum spectrum displays a multi-structured evolution, consistent with previous observations for fermions in spatially homogeneous, time-dependent electric fields (e.g., \cite{deepak2022, Sah:2023udt, Diez:2022ywi}). Near the RPAP stage, the spectra exhibit oscillatory features due to quantum interference when the electric field is nearly zero. These oscillations form a central Gaussian peak with superimposed oscillations, creating a quantum interference pattern that diminishes over time. We derive approximate analytical expressions for $f(\bm{p}, t)$ at finite times, showing that the momentum spectra consist of three distinct functional behaviors. The oscillatory behavior results from the interplay between first- and zeroth-order terms in the distribution function, with the former being suppressed over time, causing the oscillations to fade.

When comparing the momentum spectra for different adiabatic bases, we observe varying behaviors during the QPAP  and transient regions. However, after the transient stage ends, the spectra converge to a similar form. Notably, quantum interference in the RPAP stage appears earlier for the second basis choice, as the transition to real particles occurs more rapidly. To illustrate this, we observe a novel dynamical scaling when analyzing the oscillatory momentum spectrum of pairs created at finite times. For each adiabatic basis, we observe the same oscillatory momentum spectra, but at different times. When we scale time by the point marking the end of the transient stage (or the initiation of the RPAP stage), $t_{\text{out}}$, for each case of central momentum, the respective momentum spectra overlap.

Determining $t_{\text{out}}$ precisely is challenging because it depends on momentum, so we approximate it for both bases. At $t = 2 t_{\text{out}}$, the momentum spectra exhibit a smooth, Gaussian-like profile for both bases, highlighting the consistency of the adiabatic basis approach at sufficiently large times.

We further show that the first-order term $C_1$ and the zeroth-order term $C_0$ in the approximate distribution function contribute to the observed oscillatory behavior at $t = t_{\text{out}}$ for both bases. However, the timing of these oscillations differs between bases, reflecting variations in the temporal progression of pair production stages.

 We find that the choice of $\Omega(\bm{p}, t)$ and $V(p_\parallel, t)$ induces a temporal shift in the creation of real particle-antiparticle pairs. This shift is analogous to the changes in momentum scales observed under different gauges, where the canonical momentum reflects the physical field description, and the kinetic momentum incorporates the effects of the external field.

In the multiphoton regime ($\gamma > 1$), the momentum spectra at finite times exhibit multimodal structures in the RPAP state for both adiabatic basis choices. This contrasts with the unimodal profiles typical of the tunneling regime ($\gamma < 1$). By comparing time relative to the residual stage, we observe that both bases produce a multiphoton signature in the momentum spectra.

As discussed in the literature, the dynamics of pair production are complicated by the choice of basis, and no existing theoretical framework fully describes the evolution of particles and antiparticles over all times. For instance, the quantum kinetic equation formalism for time-dependent, strong-field QED still faces these challenges, despite being widely accepted. However, it is crucial to understand the information embedded within the chosen basis, particularly with respect to the pair creation process in a vacuum under strong fields at finite times. The theoretical understanding of pair production in a  time-dependent electric field is generally limited to asymptotic time ($t \to \infty$). However, infinite time is not physically meaningful, and researchers typically use the late-time limit ($t \gg \tau$), where the electric field vanishes. In this context, our study identifies the critical time scale at which real particle-antiparticle pairs can be created from the vacuum under a time-dependent pulsed field. This work emphasizes the importance of studying the dynamics of pair production in different basis. It opens the door for further studies on realistic laser pulse configurations, simulations, and experiments. Understanding the dynamical scaling and extracting key insights from such analyses will guide future simulations and experimental efforts on vacuum pair production in strong-field QED.


\section{Acknowledgments}

Deepak acknowledges the financial assistance provided by the Raja Ramanna Center for Advanced Technology (RRCAT) and the Homi Bhabha National Institute (HBNI) for carrying out this research work.
\section{Appendix}
\label{appendix1}
We introduce it here in anticipation of encountering the Gamma function $\Gamma(z)$ in the subsequent content. The Gamma function typically obeys the following relationship:
 \begin{align}
        \Gamma(1+z) = z \Gamma(z), &&
         \Gamma(1-z) \Gamma(z) = \frac{\pi}{\sin(\pi z)},
          \label{GmI1}
 \end{align}
 from which we can derive the  following useful relations,
  \begin{align}
      | \Gamma( \ii z)|^2 = \frac{\pi }{z \sinh{(\pi z)}}, &&
      | \Gamma(1+ \ii z)|^2 = \frac{\pi z}{\sinh{(\pi z)}}, &&
        |\Gamma(\frac{1}{2} + \ii z)|^2 &= \frac{\pi}{\cosh{(\pi z)}}
        \label{GmI2}
  \end{align}
Using the mathematical identities \eqref{GmI2}, we can compute $|\Gamma_1|^2$ and $|\Gamma_2|^2$ as 
 \begin{align}
       | \Gamma_1|^2 &=  \bet{\frac{ \Gamma ( c) \Gamma (c-a-b-1) }{\Gamma (c-a) \Gamma (c-b)}}^2  \nonumber \\ 
       &= \frac{\omega_0}{\omega_1(1+ \tau^2 \omega_1^2)}
       \Biggl(  \frac{\cosh{(\frac{\pi }{2} (  2 \lambda - \tau (\omega_0 + \omega_1 ))} \cosh{(\frac{\pi }{2} (  2 \lambda + \tau (\omega_0 + \omega_1 )))} }{\sinh{(\pi \tau \omega_0)} \sinh{(\pi \tau \omega_1)}}\Biggr) \nonumber \\
         | \Gamma_2|^2 &= \bet{\frac{ \Gamma ( c) \Gamma (a+b-c) }{\Gamma (a) \Gamma (b)}}^2
         \nonumber \\ 
       &= \frac{\omega_0}{\omega_1}
       \Biggl(  \frac{\cosh{(\frac{\pi}{2} ( \tau (\omega_1- \omega_0 ) -2 \lambda)} \cosh{(\frac{\pi \tau}{2} (  \tau (\omega_1 -  \omega_0 ) +2 \lambda))} }{\sinh{(\pi \tau \omega_0)} \sinh{(\pi \tau \omega_1)}}\Biggr)
       \label{gm1gm2}
 \end{align}
When computing expressions like $\Gamma_1 \bar{\Gamma_2}$, approximate methods prove advantageous. A frequently utilized technique entails utilizing Stirling's formula for the Gamma function \cite{article}, offering a simpler yet effective approach to assess the desired expression.
\begin{align}
     \Gamma(z) \approx  z^{z - 1/2} e^{-z}\sqrt{2 \pi }
\end{align}
Then, we derive the set of equations employing Stirling's formula for the Gamma function, which are used to determine the Gamma function in the computation of the particle distribution function.
\begin{align}
     \Gamma(1+ \ii x) &\sim \sqrt{2 \pi}  e^{(\frac{1}{2} ln(x) - \frac{\pi}{2} x )+ \ii ( x (ln(x) -1) + \frac{\pi}{4})}
     \nonumber \\
     \Gamma(-\ii x) &\sim \sqrt{2 \pi}  e^{( \frac{\pi}{2} x -\frac{1}{2} ln(x)  )+ \ii ( x(1- ln(x) ) - \frac{\pi}{4})}
     \nonumber \\
     \Gamma(\ii x) &\sim \sqrt{2 \pi}  e^{( -\frac{\pi}{2} x -\frac{1}{2} ln(x)  )+ \ii ( x(ln(x) -1) - \frac{\pi}{4})}
     \end{align}
 So,
      \begin{align}
            \Gamma_1 \Bar{\Gamma_2} &= \left(  \frac{\Gamma(c) \Gamma(c-a-b-1)}{\Gamma(c-a) \Gamma(c-b)}\right) \overline{ \Bigl(  \frac{\Gamma(c) \Gamma(a+b-c)}{\Gamma(a) \Gamma(b)}} \Bigr) 
     \end{align}
Subsequently, following certain algebraic manipulations, we obtain :          
 \begin{align}
    \Gamma_1 \overline{\Gamma_2} &= |\Gamma_1 \overline{\Gamma_2}| e^{\ii \mathrm{\varrho}}
      \end{align}
      where,
\begin{align}
      |\Gamma_1 \overline{\Gamma_2}|  &= \frac{\omega_0}{\omega_1} 
    \frac{1}{ \sinh{(\pi \tau \omega_0)} \sinh{(\pi \tau \omega_1)}\sqrt{1 + \omega_1^2 \tau^2}}
      \sqrt{\frac{\cosh{(\frac{\pi }{2} (  2 \lambda - \tau (\omega_0 + \omega_1 ))} \cosh{(\frac{\pi }{2} (  2 \lambda + \tau (\omega_0 + \omega_1 )))}}{\cosh{(\frac{\pi}{2} ( \tau (\omega_1- \omega_0 ) -2 \lambda)} \cosh{(\frac{\pi \tau}{2} (  \tau (\omega_1 -  \omega_0 ) +2 \lambda))} }}
  \end{align}
\begin{align}
    \mathrm{\rho} &=    \frac{1}{4} \biggl[  2 \pi  + 4 \tau \omega_1 - 4 \tan^{-1}(\tau \omega_1 )   - 4 \tau \omega_1 ( -1  + \ln\tau \omega_1) - 4 \tau \omega_1 \ln(\tau \omega_1)  \nonumber\\
&   - ( 2 \lambda  +  (
\omega_0 - \omega_1 ) \tau ) ( -2  +   \ln{(\frac{1}{4}  + (\lambda + \frac{\tau}{2} (\omega_0 - \omega_1))^2)} ) + ( 2 \lambda  + ( -\omega_0 + \omega_1) \tau) ( -2  \nonumber\\
& +   \ln{(\frac{1}{4}  + (\lambda + \frac{\tau}{2} (-\omega_0 + \omega_1))^2)} ) - ( 2 \lambda - (\omega_0 + \omega_1) \tau)  ( -2  +   \ln{(\frac{1}{4}(1  + (\lambda + \frac{\tau}{2} (\omega_0 - \omega_1))^2))} ) \nonumber\\
& + ( 2 \lambda + (\omega_0 + \omega_1) \tau)( -2  +   \ln{(\frac{1}{4}(1  + (\lambda + \frac{\tau}{2} (\omega_0 + \omega_1))^2))} ) 
    \biggr]
    \end{align}
\bibliography{main}

\begin{thebibliography}{61}
\expandafter\ifx\csname natexlab\endcsname\relax\def\natexlab#1{#1}\fi
\expandafter\ifx\csname bibnamefont\endcsname\relax
  \def\bibnamefont#1{#1}\fi
\expandafter\ifx\csname bibfnamefont\endcsname\relax
  \def\bibfnamefont#1{#1}\fi
\expandafter\ifx\csname citenamefont\endcsname\relax
  \def\citenamefont#1{#1}\fi
\expandafter\ifx\csname url\endcsname\relax
  \def\url#1{\texttt{#1}}\fi
\expandafter\ifx\csname urlprefix\endcsname\relax\def\urlprefix{URL }\fi
\providecommand{\bibinfo}[2]{#2}
\providecommand{\eprint}[2][]{\url{#2}}

\bibitem[{\citenamefont{Sauter}(1931)}]{Sauter:1931zz}
\bibinfo{author}{\bibfnamefont{F.}~\bibnamefont{Sauter}}, \bibinfo{journal}{Z. Phys.} \textbf{\bibinfo{volume}{69}}, \bibinfo{pages}{742} (\bibinfo{year}{1931}).

\bibitem[{\citenamefont{Sauter}(1932)}]{Sauter:1932gsa}
\bibinfo{author}{\bibfnamefont{F.}~\bibnamefont{Sauter}}, \bibinfo{journal}{Z. Phys.} \textbf{\bibinfo{volume}{73}}, \bibinfo{pages}{547} (\bibinfo{year}{1932}).

\bibitem[{\citenamefont{Heisenberg and Euler}(1936)}]{Heisenberg:1936nmg}
\bibinfo{author}{\bibfnamefont{W.}~\bibnamefont{Heisenberg}} \bibnamefont{and} \bibinfo{author}{\bibfnamefont{H.}~\bibnamefont{Euler}}, \bibinfo{journal}{Z. Phys.} \textbf{\bibinfo{volume}{98}}, \bibinfo{pages}{714} (\bibinfo{year}{1936}), \eprint{physics/0605038}.

\bibitem[{\citenamefont{Weisskopf}(1936)}]{Weisskopf:1936hya}
\bibinfo{author}{\bibfnamefont{V.}~\bibnamefont{Weisskopf}}, \bibinfo{journal}{Kong. Dan. Vid. Sel. Mat. Fys. Med.} \textbf{\bibinfo{volume}{14N6}}, \bibinfo{pages}{1} (\bibinfo{year}{1936}).

\bibitem[{\citenamefont{Schwinger}(1951)}]{Schwinger:1951nm}
\bibinfo{author}{\bibfnamefont{J.~S.} \bibnamefont{Schwinger}}, \bibinfo{journal}{Phys. Rev.} \textbf{\bibinfo{volume}{82}}, \bibinfo{pages}{664} (\bibinfo{year}{1951}).

\bibitem[{\citenamefont{Hawking}(1974)}]{Hawking:1974rv}
\bibinfo{author}{\bibfnamefont{S.~W.} \bibnamefont{Hawking}}, \bibinfo{journal}{Nature} \textbf{\bibinfo{volume}{248}}, \bibinfo{pages}{30} (\bibinfo{year}{1974}).

\bibitem[{\citenamefont{Leonhardt}(2018)}]{Leonhardt:2016qdi}
\bibinfo{author}{\bibfnamefont{U.}~\bibnamefont{Leonhardt}}, \bibinfo{journal}{Annalen Phys.} \textbf{\bibinfo{volume}{530}}, \bibinfo{pages}{1700114} (\bibinfo{year}{2018}), \eprint{1609.03803}.

\bibitem[{\citenamefont{Yakimenko et~al.}(2019)}]{Yakimenko:2018kih}
\bibinfo{author}{\bibfnamefont{V.}~\bibnamefont{Yakimenko}} \bibnamefont{et~al.}, \bibinfo{journal}{Phys. Rev. Lett.} \textbf{\bibinfo{volume}{122}}, \bibinfo{pages}{190404} (\bibinfo{year}{2019}), \eprint{1807.09271}.

\bibitem[{\citenamefont{Marklund et~al.}(2022)\citenamefont{Marklund, Blackburn, Gonoskov, Magnusson, Bulanov, and Ilderton}}]{Marklund:2022gki}
\bibinfo{author}{\bibfnamefont{M.}~\bibnamefont{Marklund}}, \bibinfo{author}{\bibfnamefont{T.~G.} \bibnamefont{Blackburn}}, \bibinfo{author}{\bibfnamefont{A.}~\bibnamefont{Gonoskov}}, \bibinfo{author}{\bibfnamefont{J.}~\bibnamefont{Magnusson}}, \bibinfo{author}{\bibfnamefont{S.~S.} \bibnamefont{Bulanov}}, \bibnamefont{and} \bibinfo{author}{\bibfnamefont{A.}~\bibnamefont{Ilderton}} (\bibinfo{year}{2022}), \eprint{2209.11720}.

\bibitem[{\citenamefont{Vincenti}(2019)}]{Vincenti:2018nzj}
\bibinfo{author}{\bibfnamefont{H.}~\bibnamefont{Vincenti}}, \bibinfo{journal}{Phys. Rev. Lett.} \textbf{\bibinfo{volume}{123}}, \bibinfo{pages}{105001} (\bibinfo{year}{2019}), \eprint{1812.05357}.

\bibitem[{\citenamefont{Fedotov et~al.}(2023)\citenamefont{Fedotov, Ilderton, Karbstein, King, Seipt, Taya, and Torgrimsson}}]{Fedotov:2022ely}
\bibinfo{author}{\bibfnamefont{A.}~\bibnamefont{Fedotov}}, \bibinfo{author}{\bibfnamefont{A.}~\bibnamefont{Ilderton}}, \bibinfo{author}{\bibfnamefont{F.}~\bibnamefont{Karbstein}}, \bibinfo{author}{\bibfnamefont{B.}~\bibnamefont{King}}, \bibinfo{author}{\bibfnamefont{D.}~\bibnamefont{Seipt}}, \bibinfo{author}{\bibfnamefont{H.}~\bibnamefont{Taya}}, \bibnamefont{and} \bibinfo{author}{\bibfnamefont{G.}~\bibnamefont{Torgrimsson}}, \bibinfo{journal}{Phys. Rept.} \textbf{\bibinfo{volume}{1010}}, \bibinfo{pages}{1} (\bibinfo{year}{2023}), \eprint{2203.00019}.

\bibitem[{\citenamefont{Berdyugin et~al.}(2022)}]{Berdyugin:2021njg}
\bibinfo{author}{\bibfnamefont{A.~I.} \bibnamefont{Berdyugin}} \bibnamefont{et~al.}, \bibinfo{journal}{Science} \textbf{\bibinfo{volume}{375}}, \bibinfo{pages}{430} (\bibinfo{year}{2022}), \eprint{2106.12609}.

\bibitem[{\citenamefont{Schmitt et~al.}(2023)}]{Schmitt:2022pkd}
\bibinfo{author}{\bibfnamefont{A.}~\bibnamefont{Schmitt}} \bibnamefont{et~al.}, \bibinfo{journal}{Nature Phys.} \textbf{\bibinfo{volume}{19}}, \bibinfo{pages}{830} (\bibinfo{year}{2023}), \eprint{2207.13400}.

\bibitem[{\citenamefont{Parker}(1969)}]{Parker:1969au}
\bibinfo{author}{\bibfnamefont{L.}~\bibnamefont{Parker}}, \bibinfo{journal}{Phys. Rev.} \textbf{\bibinfo{volume}{183}}, \bibinfo{pages}{1057} (\bibinfo{year}{1969}).

\bibitem[{\citenamefont{Lueders and Roberts}(1990)}]{Lueders:1990np}
\bibinfo{author}{\bibfnamefont{C.}~\bibnamefont{Lueders}} \bibnamefont{and} \bibinfo{author}{\bibfnamefont{J.~E.} \bibnamefont{Roberts}}, \bibinfo{journal}{Commun. Math. Phys.} \textbf{\bibinfo{volume}{134}}, \bibinfo{pages}{29} (\bibinfo{year}{1990}).

\bibitem[{\citenamefont{Fulling}(1979)}]{Fulling:1979ac}
\bibinfo{author}{\bibfnamefont{S.~A.} \bibnamefont{Fulling}}, \bibinfo{journal}{Gen. Rel. Grav.} \textbf{\bibinfo{volume}{10}}, \bibinfo{pages}{807} (\bibinfo{year}{1979}).

\bibitem[{\citenamefont{Fahn et~al.}(2019)\citenamefont{Fahn, Giesel, and Kobler}}]{Fahn:2018ahm}
\bibinfo{author}{\bibfnamefont{M.~J.} \bibnamefont{Fahn}}, \bibinfo{author}{\bibfnamefont{K.}~\bibnamefont{Giesel}}, \bibnamefont{and} \bibinfo{author}{\bibfnamefont{M.}~\bibnamefont{Kobler}}, \bibinfo{journal}{Universe} \textbf{\bibinfo{volume}{5}}, \bibinfo{pages}{170} (\bibinfo{year}{2019}), \eprint{1812.11122}.

\bibitem[{\citenamefont{Agullo et~al.}(2015)\citenamefont{Agullo, Nelson, and Ashtekar}}]{Agullo:2014ica}
\bibinfo{author}{\bibfnamefont{I.}~\bibnamefont{Agullo}}, \bibinfo{author}{\bibfnamefont{W.}~\bibnamefont{Nelson}}, \bibnamefont{and} \bibinfo{author}{\bibfnamefont{A.}~\bibnamefont{Ashtekar}}, \bibinfo{journal}{Phys. Rev. D} \textbf{\bibinfo{volume}{91}}, \bibinfo{pages}{064051} (\bibinfo{year}{2015}), \eprint{1412.3524}.

\bibitem[{\citenamefont{Dabrowski and Dunne}(2016)}]{Dabrowski:2016tsx}
\bibinfo{author}{\bibfnamefont{R.}~\bibnamefont{Dabrowski}} \bibnamefont{and} \bibinfo{author}{\bibfnamefont{G.~V.} \bibnamefont{Dunne}}, \bibinfo{journal}{Phys. Rev. D} \textbf{\bibinfo{volume}{94}}, \bibinfo{pages}{065005} (\bibinfo{year}{2016}), \eprint{1606.00902}.

\bibitem[{\citenamefont{Schmidt et~al.}(1998)\citenamefont{Schmidt, Blaschke, Ropke, Smolyansky, Prozorkevich, and Toneev}}]{Schmidt:1998vi}
\bibinfo{author}{\bibfnamefont{S.~M.} \bibnamefont{Schmidt}}, \bibinfo{author}{\bibfnamefont{D.}~\bibnamefont{Blaschke}}, \bibinfo{author}{\bibfnamefont{G.}~\bibnamefont{Ropke}}, \bibinfo{author}{\bibfnamefont{S.~A.} \bibnamefont{Smolyansky}}, \bibinfo{author}{\bibfnamefont{A.~V.} \bibnamefont{Prozorkevich}}, \bibnamefont{and} \bibinfo{author}{\bibfnamefont{V.~D.} \bibnamefont{Toneev}}, \bibinfo{journal}{Int. J. Mod. Phys. E} \textbf{\bibinfo{volume}{7}}, \bibinfo{pages}{709} (\bibinfo{year}{1998}), \eprint{hep-ph/9809227}.

\bibitem[{\citenamefont{Schmidt et~al.}(1999)\citenamefont{Schmidt, Blaschke, Ropke, Prozorkevich, Smolyansky, and Toneev}}]{Schmidt:1998zh}
\bibinfo{author}{\bibfnamefont{S.~M.} \bibnamefont{Schmidt}}, \bibinfo{author}{\bibfnamefont{D.}~\bibnamefont{Blaschke}}, \bibinfo{author}{\bibfnamefont{G.}~\bibnamefont{Ropke}}, \bibinfo{author}{\bibfnamefont{A.~V.} \bibnamefont{Prozorkevich}}, \bibinfo{author}{\bibfnamefont{S.~A.} \bibnamefont{Smolyansky}}, \bibnamefont{and} \bibinfo{author}{\bibfnamefont{V.~D.} \bibnamefont{Toneev}}, \bibinfo{journal}{Phys. Rev. D} \textbf{\bibinfo{volume}{59}}, \bibinfo{pages}{094005} (\bibinfo{year}{1999}), \eprint{hep-ph/9810452}.

\bibitem[{\citenamefont{Aleksandrov et~al.}(2020)\citenamefont{Aleksandrov, Dmitriev, Sevostyanov, and Smolyansky}}]{Aleksandrov:2020mez}
\bibinfo{author}{\bibfnamefont{I.~A.} \bibnamefont{Aleksandrov}}, \bibinfo{author}{\bibfnamefont{V.~V.} \bibnamefont{Dmitriev}}, \bibinfo{author}{\bibfnamefont{D.~G.} \bibnamefont{Sevostyanov}}, \bibnamefont{and} \bibinfo{author}{\bibfnamefont{S.~A.} \bibnamefont{Smolyansky}}, \bibinfo{journal}{Eur. Phys. J. ST} \textbf{\bibinfo{volume}{229}}, \bibinfo{pages}{3469} (\bibinfo{year}{2020}), \eprint{2004.02179}.

\bibitem[{\citenamefont{Smolyansky et~al.}(2019)\citenamefont{Smolyansky, Panferov, Blaschke, and Gevorgyan}}]{Smolyansky:2019dqd}
\bibinfo{author}{\bibfnamefont{S.~A.} \bibnamefont{Smolyansky}}, \bibinfo{author}{\bibfnamefont{A.~D.} \bibnamefont{Panferov}}, \bibinfo{author}{\bibfnamefont{D.~B.} \bibnamefont{Blaschke}}, \bibnamefont{and} \bibinfo{author}{\bibfnamefont{N.~T.} \bibnamefont{Gevorgyan}}, \bibinfo{journal}{Particles} \textbf{\bibinfo{volume}{2}}, \bibinfo{pages}{208} (\bibinfo{year}{2019}).

\bibitem[{\citenamefont{Smolyansky et~al.}(2010)\citenamefont{Smolyansky, Blaschke, Chertilin, Roepke, and Tarakanov}}]{Smolyansky:2010as}
\bibinfo{author}{\bibfnamefont{S.~A.} \bibnamefont{Smolyansky}}, \bibinfo{author}{\bibfnamefont{D.~B.} \bibnamefont{Blaschke}}, \bibinfo{author}{\bibfnamefont{A.~V.} \bibnamefont{Chertilin}}, \bibinfo{author}{\bibfnamefont{G.}~\bibnamefont{Roepke}}, \bibnamefont{and} \bibinfo{author}{\bibfnamefont{A.~V.} \bibnamefont{Tarakanov}} (\bibinfo{year}{2010}), \eprint{1012.0559}.

\bibitem[{\citenamefont{Blaschke et~al.}(2011{\natexlab{a}})\citenamefont{Blaschke, Ropke, Schmidt, Smolyansky, and Tarakanov}}]{Blaschke:2010vs}
\bibinfo{author}{\bibfnamefont{D.~B.} \bibnamefont{Blaschke}}, \bibinfo{author}{\bibfnamefont{G.}~\bibnamefont{Ropke}}, \bibinfo{author}{\bibfnamefont{S.~M.} \bibnamefont{Schmidt}}, \bibinfo{author}{\bibfnamefont{S.~A.} \bibnamefont{Smolyansky}}, \bibnamefont{and} \bibinfo{author}{\bibfnamefont{A.~V.} \bibnamefont{Tarakanov}}, \bibinfo{journal}{Contrib. Plasma Phys.} \textbf{\bibinfo{volume}{51}}, \bibinfo{pages}{451} (\bibinfo{year}{2011}{\natexlab{a}}), \eprint{1006.1098}.

\bibitem[{\citenamefont{Blaschke et~al.}(2011{\natexlab{b}})\citenamefont{Blaschke, Ropke, Dmitriev, Smolyansky, and Tarakanov}}]{Blaschke:2011af}
\bibinfo{author}{\bibfnamefont{D.~B.} \bibnamefont{Blaschke}}, \bibinfo{author}{\bibfnamefont{G.}~\bibnamefont{Ropke}}, \bibinfo{author}{\bibfnamefont{V.~V.} \bibnamefont{Dmitriev}}, \bibinfo{author}{\bibfnamefont{S.~A.} \bibnamefont{Smolyansky}}, \bibnamefont{and} \bibinfo{author}{\bibfnamefont{A.~V.} \bibnamefont{Tarakanov}} (\bibinfo{year}{2011}{\natexlab{b}}), \eprint{1101.6021}.

\bibitem[{\citenamefont{Di~Piazza et~al.}(2006)\citenamefont{Di~Piazza, Hatsagortsyan, and Keitel}}]{DiPiazza:2006pr}
\bibinfo{author}{\bibfnamefont{A.}~\bibnamefont{Di~Piazza}}, \bibinfo{author}{\bibfnamefont{K.~Z.} \bibnamefont{Hatsagortsyan}}, \bibnamefont{and} \bibinfo{author}{\bibfnamefont{C.~H.} \bibnamefont{Keitel}}, \bibinfo{journal}{Phys. Rev. Lett.} \textbf{\bibinfo{volume}{97}}, \bibinfo{pages}{083603} (\bibinfo{year}{2006}), \eprint{hep-ph/0602039}.

\bibitem[{\citenamefont{Anderson et~al.}(2005)\citenamefont{Anderson, Molina-Paris, and Mottola}}]{Anderson:2005hi}
\bibinfo{author}{\bibfnamefont{P.~R.} \bibnamefont{Anderson}}, \bibinfo{author}{\bibfnamefont{C.}~\bibnamefont{Molina-Paris}}, \bibnamefont{and} \bibinfo{author}{\bibfnamefont{E.}~\bibnamefont{Mottola}}, \bibinfo{journal}{Phys. Rev. D} \textbf{\bibinfo{volume}{72}}, \bibinfo{pages}{043515} (\bibinfo{year}{2005}), \eprint{hep-th/0504134}.

\bibitem[{\citenamefont{Habib et~al.}(2000)\citenamefont{Habib, Molina-Paris, and Mottola}}]{Habib:1999cs}
\bibinfo{author}{\bibfnamefont{S.}~\bibnamefont{Habib}}, \bibinfo{author}{\bibfnamefont{C.}~\bibnamefont{Molina-Paris}}, \bibnamefont{and} \bibinfo{author}{\bibfnamefont{E.}~\bibnamefont{Mottola}}, \bibinfo{journal}{Phys. Rev. D} \textbf{\bibinfo{volume}{61}}, \bibinfo{pages}{024010} (\bibinfo{year}{2000}), \eprint{gr-qc/9906120}.

\bibitem[{\citenamefont{Kluger et~al.}(1998)\citenamefont{Kluger, Mottola, and Eisenberg}}]{Kluger:1998bm}
\bibinfo{author}{\bibfnamefont{Y.}~\bibnamefont{Kluger}}, \bibinfo{author}{\bibfnamefont{E.}~\bibnamefont{Mottola}}, \bibnamefont{and} \bibinfo{author}{\bibfnamefont{J.~M.} \bibnamefont{Eisenberg}}, \bibinfo{journal}{Phys. Rev. D} \textbf{\bibinfo{volume}{58}}, \bibinfo{pages}{125015} (\bibinfo{year}{1998}), \eprint{hep-ph/9803372}.

\bibitem[{\citenamefont{Kluger et~al.}(1993)\citenamefont{Kluger, Eisenberg, and Svetitsky}}]{Kluger:1992md}
\bibinfo{author}{\bibfnamefont{Y.}~\bibnamefont{Kluger}}, \bibinfo{author}{\bibfnamefont{J.~M.} \bibnamefont{Eisenberg}}, \bibnamefont{and} \bibinfo{author}{\bibfnamefont{B.}~\bibnamefont{Svetitsky}}, \bibinfo{journal}{Int. J. Mod. Phys. E} \textbf{\bibinfo{volume}{2}}, \bibinfo{pages}{333} (\bibinfo{year}{1993}), \eprint{hep-ph/0311293}.

\bibitem[{\citenamefont{Blaschke et~al.}(2022)\citenamefont{Blaschke, Dmitriev, Gevorgyan, Mahato, Panferov, Smolyansky, and Tseryupa}}]{Blaschke:2022ppg}
\bibinfo{author}{\bibfnamefont{D.~B.} \bibnamefont{Blaschke}}, \bibinfo{author}{\bibfnamefont{V.~V.} \bibnamefont{Dmitriev}}, \bibinfo{author}{\bibfnamefont{N.~T.} \bibnamefont{Gevorgyan}}, \bibinfo{author}{\bibfnamefont{B.}~\bibnamefont{Mahato}}, \bibinfo{author}{\bibfnamefont{A.~D.} \bibnamefont{Panferov}}, \bibinfo{author}{\bibfnamefont{S.~A.} \bibnamefont{Smolyansky}}, \bibnamefont{and} \bibinfo{author}{\bibfnamefont{V.~A.} \bibnamefont{Tseryupa}}, \bibinfo{journal}{Springer Proc. Phys.} \textbf{\bibinfo{volume}{281}}, \bibinfo{pages}{187} (\bibinfo{year}{2022}), \eprint{2201.10594}.

\bibitem[{\citenamefont{Panferov et~al.}(2019)\citenamefont{Panferov, Smolyansky, Blaschke, and Gevorgyan}}]{Panferov:2019stq}
\bibinfo{author}{\bibfnamefont{A.}~\bibnamefont{Panferov}}, \bibinfo{author}{\bibfnamefont{S.}~\bibnamefont{Smolyansky}}, \bibinfo{author}{\bibfnamefont{D.}~\bibnamefont{Blaschke}}, \bibnamefont{and} \bibinfo{author}{\bibfnamefont{N.}~\bibnamefont{Gevorgyan}}, \bibinfo{journal}{EPJ Web Conf.} \textbf{\bibinfo{volume}{204}}, \bibinfo{pages}{06008} (\bibinfo{year}{2019}), \eprint{1901.01395}.

\bibitem[{\citenamefont{Jiang et~al.}(2024)\citenamefont{Jiang, Li, and Li}}]{Jiang:2024ilt}
\bibinfo{author}{\bibfnamefont{R.~Z.} \bibnamefont{Jiang}}, \bibinfo{author}{\bibfnamefont{Z.~L.} \bibnamefont{Li}}, \bibnamefont{and} \bibinfo{author}{\bibfnamefont{Y.~J.} \bibnamefont{Li}} (\bibinfo{year}{2024}), \eprint{2410.15313}.

\bibitem[{\citenamefont{Gavrilov and Gitman}(1996)}]{Gavrilov:1996pz}
\bibinfo{author}{\bibfnamefont{S.~P.} \bibnamefont{Gavrilov}} \bibnamefont{and} \bibinfo{author}{\bibfnamefont{D.~M.} \bibnamefont{Gitman}}, \bibinfo{journal}{Phys. Rev. D} \textbf{\bibinfo{volume}{53}}, \bibinfo{pages}{7162} (\bibinfo{year}{1996}), \eprint{hep-th/9603152}.

\bibitem[{\citenamefont{Gelis and Tanji}(2016)}]{Gelis:2015kya}
\bibinfo{author}{\bibfnamefont{F.}~\bibnamefont{Gelis}} \bibnamefont{and} \bibinfo{author}{\bibfnamefont{N.}~\bibnamefont{Tanji}}, \bibinfo{journal}{Prog. Part. Nucl. Phys.} \textbf{\bibinfo{volume}{87}}, \bibinfo{pages}{1} (\bibinfo{year}{2016}), \eprint{1510.05451}.

\bibitem[{\citenamefont{Popov}(1971)}]{Popov:1971iga}
\bibinfo{author}{\bibfnamefont{V.~S.} \bibnamefont{Popov}}, \bibinfo{journal}{Zh. Eksp. Teor. Fiz.} \textbf{\bibinfo{volume}{61}}, \bibinfo{pages}{1334} (\bibinfo{year}{1971}).

\bibitem[{\citenamefont{Blaschke et~al.}(2015)\citenamefont{Blaschke, Juchnowski, Panferov, and Smolyansky}}]{Blaschke:2014fca}
\bibinfo{author}{\bibfnamefont{D.}~\bibnamefont{Blaschke}}, \bibinfo{author}{\bibfnamefont{L.}~\bibnamefont{Juchnowski}}, \bibinfo{author}{\bibfnamefont{A.}~\bibnamefont{Panferov}}, \bibnamefont{and} \bibinfo{author}{\bibfnamefont{S.}~\bibnamefont{Smolyansky}}, \bibinfo{journal}{Phys. Part. Nucl.} \textbf{\bibinfo{volume}{46}}, \bibinfo{pages}{797} (\bibinfo{year}{2015}), \eprint{1412.6372}.

\bibitem[{\citenamefont{Cooper and Mottola}(1989)}]{Cooper:1989kf}
\bibinfo{author}{\bibfnamefont{F.}~\bibnamefont{Cooper}} \bibnamefont{and} \bibinfo{author}{\bibfnamefont{E.}~\bibnamefont{Mottola}}, \bibinfo{journal}{Phys. Rev. D} \textbf{\bibinfo{volume}{40}}, \bibinfo{pages}{456} (\bibinfo{year}{1989}).

\bibitem[{\citenamefont{Kluger et~al.}(1991)\citenamefont{Kluger, Eisenberg, Svetitsky, Cooper, and Mottola}}]{Kluger:1991ib}
\bibinfo{author}{\bibfnamefont{Y.}~\bibnamefont{Kluger}}, \bibinfo{author}{\bibfnamefont{J.~M.} \bibnamefont{Eisenberg}}, \bibinfo{author}{\bibfnamefont{B.}~\bibnamefont{Svetitsky}}, \bibinfo{author}{\bibfnamefont{F.}~\bibnamefont{Cooper}}, \bibnamefont{and} \bibinfo{author}{\bibfnamefont{E.}~\bibnamefont{Mottola}}, \bibinfo{journal}{Phys. Rev. Lett.} \textbf{\bibinfo{volume}{67}}, \bibinfo{pages}{2427} (\bibinfo{year}{1991}).

\bibitem[{\citenamefont{Greiner and Reinhardt}(1992)}]{Greiner:1992bv}
\bibinfo{author}{\bibfnamefont{W.}~\bibnamefont{Greiner}} \bibnamefont{and} \bibinfo{author}{\bibfnamefont{J.}~\bibnamefont{Reinhardt}}, \emph{\bibinfo{title}{{Quantum electrodynamics}}} (\bibinfo{year}{1992}), ISBN \bibinfo{isbn}{978-3-540-87560-4}.

\bibitem[{\citenamefont{Reinhardt and Greiner}(1994)}]{Reinhardt:1994aha}
\bibinfo{author}{\bibfnamefont{J.}~\bibnamefont{Reinhardt}} \bibnamefont{and} \bibinfo{author}{\bibfnamefont{W.}~\bibnamefont{Greiner}}, \bibinfo{journal}{Lect. Notes Phys.} \textbf{\bibinfo{volume}{440}}, \bibinfo{pages}{153} (\bibinfo{year}{1994}).

\bibitem[{\citenamefont{Banerjee and Singh}(2019)}]{Banerjee:2018fbw}
\bibinfo{author}{\bibfnamefont{C.}~\bibnamefont{Banerjee}} \bibnamefont{and} \bibinfo{author}{\bibfnamefont{M.~P.} \bibnamefont{Singh}}, \bibinfo{journal}{Phys. Rev. D} \textbf{\bibinfo{volume}{100}}, \bibinfo{pages}{056016} (\bibinfo{year}{2019}), \eprint{1809.06901}.

\bibitem[{\citenamefont{Winitzki}(2005)}]{Winitzki:2005rw}
\bibinfo{author}{\bibfnamefont{S.}~\bibnamefont{Winitzki}}, \bibinfo{journal}{Phys. Rev. D} \textbf{\bibinfo{volume}{72}}, \bibinfo{pages}{104011} (\bibinfo{year}{2005}), \eprint{gr-qc/0510001}.

\bibitem[{\citenamefont{Dabrowski and Dunne}(2014)}]{Dabrowski:2014ica}
\bibinfo{author}{\bibfnamefont{R.}~\bibnamefont{Dabrowski}} \bibnamefont{and} \bibinfo{author}{\bibfnamefont{G.~V.} \bibnamefont{Dunne}}, \bibinfo{journal}{Phys. Rev. D} \textbf{\bibinfo{volume}{90}}, \bibinfo{pages}{025021} (\bibinfo{year}{2014}), \eprint{1405.0302}.

\bibitem[{\citenamefont{Barry}(1989)}]{Barry:1989zz}
\bibinfo{author}{\bibfnamefont{M.~V.} \bibnamefont{Barry}}, \bibinfo{journal}{Proc. Roy. Soc. Lond. A} \textbf{\bibinfo{volume}{422}}, \bibinfo{pages}{7} (\bibinfo{year}{1989}).

\bibitem[{\citenamefont{Dumlu and Dunne}(2010)}]{Dumlu:2010ua}
\bibinfo{author}{\bibfnamefont{C.~K.} \bibnamefont{Dumlu}} \bibnamefont{and} \bibinfo{author}{\bibfnamefont{G.~V.} \bibnamefont{Dunne}}, \bibinfo{journal}{Phys. Rev. Lett.} \textbf{\bibinfo{volume}{104}}, \bibinfo{pages}{250402} (\bibinfo{year}{2010}), \eprint{1004.2509}.

\bibitem[{\citenamefont{Brezin and Itzykson}(1970)}]{Brezin:1970xf}
\bibinfo{author}{\bibfnamefont{E.}~\bibnamefont{Brezin}} \bibnamefont{and} \bibinfo{author}{\bibfnamefont{C.}~\bibnamefont{Itzykson}}, \bibinfo{journal}{Phys. Rev. D} \textbf{\bibinfo{volume}{2}}, \bibinfo{pages}{1191} (\bibinfo{year}{1970}).

\bibitem[{\citenamefont{Anderson et~al.}(2018)\citenamefont{Anderson, Mottola, and Sanders}}]{Anderson:2017hts}
\bibinfo{author}{\bibfnamefont{P.~R.} \bibnamefont{Anderson}}, \bibinfo{author}{\bibfnamefont{E.}~\bibnamefont{Mottola}}, \bibnamefont{and} \bibinfo{author}{\bibfnamefont{D.~H.} \bibnamefont{Sanders}}, \bibinfo{journal}{Phys. Rev. D} \textbf{\bibinfo{volume}{97}}, \bibinfo{pages}{065016} (\bibinfo{year}{2018}), \eprint{1712.04522}.

\bibitem[{\citenamefont{Anderson and Mottola}(2014)}]{Anderson:2013ila}
\bibinfo{author}{\bibfnamefont{P.~R.} \bibnamefont{Anderson}} \bibnamefont{and} \bibinfo{author}{\bibfnamefont{E.}~\bibnamefont{Mottola}}, \bibinfo{journal}{Phys. Rev. D} \textbf{\bibinfo{volume}{89}}, \bibinfo{pages}{104038} (\bibinfo{year}{2014}), \eprint{1310.0030}.

\bibitem[{\citenamefont{Abramowitz and Stegun}(1964)}]{abramowitz}
\bibinfo{author}{\bibfnamefont{M.}~\bibnamefont{Abramowitz}} \bibnamefont{and} \bibinfo{author}{\bibfnamefont{I.~A.} \bibnamefont{Stegun}}, \emph{\bibinfo{title}{Handbook of Mathematical Functions with Formulas, Graphs, and Mathematical Tables}} (\bibinfo{publisher}{Dover}, \bibinfo{address}{New York}, \bibinfo{year}{1964}), \bibinfo{edition}{ninth dover printing} ed.

\bibitem[{\citenamefont{Blaschke et~al.}(2011{\natexlab{c}})\citenamefont{Blaschke, Dmitriev, Ropke, and Smolyansky}}]{Blaschke:2011is}
\bibinfo{author}{\bibfnamefont{D.~B.} \bibnamefont{Blaschke}}, \bibinfo{author}{\bibfnamefont{V.~V.} \bibnamefont{Dmitriev}}, \bibinfo{author}{\bibfnamefont{G.}~\bibnamefont{Ropke}}, \bibnamefont{and} \bibinfo{author}{\bibfnamefont{S.~A.} \bibnamefont{Smolyansky}}, \bibinfo{journal}{Phys. Rev. D} \textbf{\bibinfo{volume}{84}}, \bibinfo{pages}{085028} (\bibinfo{year}{2011}{\natexlab{c}}), \eprint{1105.5397}.

\bibitem[{\citenamefont{Keldysh}(1965)}]{Keldysh:1965ojf}
\bibinfo{author}{\bibfnamefont{L.~V.} \bibnamefont{Keldysh}}, \bibinfo{journal}{J. Exp. Theor. Phys.} \textbf{\bibinfo{volume}{20}}, \bibinfo{pages}{1307} (\bibinfo{year}{1965}).

\bibitem[{\citenamefont{Sah and Singh}(2022)}]{deepak2022}
\bibinfo{author}{\bibfnamefont{D.}~\bibnamefont{Sah}} \bibnamefont{and} \bibinfo{author}{\bibfnamefont{M.~P.} \bibnamefont{Singh}}, \bibinfo{journal}{Workshop QED Laser Plasmas (QLASP22),Max Planck Institute for the Physics of Complex Systems, Dresden, Germany}  (\bibinfo{year}{2022}), \urlprefix\url{https://www.pks.mpg.de/qlasp22}.

\bibitem[{\citenamefont{Deepak and Singh}(2024)}]{10.1007/978-981-97-0289-3_275}
\bibinfo{author}{\bibnamefont{Deepak}} \bibnamefont{and} \bibinfo{author}{\bibfnamefont{M.~P.} \bibnamefont{Singh}}, in \emph{\bibinfo{booktitle}{Proceedings of the XXV DAE-BRNS High Energy Physics (HEP) Symposium 2022, 12--16 December, Mohali, India}}, edited by \bibinfo{editor}{\bibfnamefont{S.}~\bibnamefont{Jena}}, \bibinfo{editor}{\bibfnamefont{A.}~\bibnamefont{Shivaji}}, \bibinfo{editor}{\bibfnamefont{V.}~\bibnamefont{Bhardwaj}}, \bibinfo{editor}{\bibfnamefont{K.}~\bibnamefont{Lochan}}, \bibinfo{editor}{\bibfnamefont{H.~K.} \bibnamefont{Jassal}}, \bibinfo{editor}{\bibfnamefont{A.}~\bibnamefont{Joseph}}, \bibnamefont{and} \bibinfo{editor}{\bibfnamefont{P.}~\bibnamefont{Khuswaha}} (\bibinfo{publisher}{Springer Nature Singapore}, \bibinfo{address}{Singapore}, \bibinfo{year}{2024}), pp. \bibinfo{pages}{1023--1025}, ISBN \bibinfo{isbn}{978-981-97-0289-3}.

\bibitem[{\citenamefont{Diez et~al.}(2023)\citenamefont{Diez, Alkofer, and Kohlf\"urst}}]{Diez:2022ywi}
\bibinfo{author}{\bibfnamefont{M.}~\bibnamefont{Diez}}, \bibinfo{author}{\bibfnamefont{R.}~\bibnamefont{Alkofer}}, \bibnamefont{and} \bibinfo{author}{\bibfnamefont{C.}~\bibnamefont{Kohlf\"urst}}, \bibinfo{journal}{Phys. Lett. B} \textbf{\bibinfo{volume}{844}}, \bibinfo{pages}{138063} (\bibinfo{year}{2023}), \eprint{2211.07510}.

\bibitem[{\citenamefont{Sah and Singh}(2023{\natexlab{a}})}]{Sah:2023jlz}
\bibinfo{author}{\bibfnamefont{D.}~\bibnamefont{Sah}} \bibnamefont{and} \bibinfo{author}{\bibfnamefont{M.~P.} \bibnamefont{Singh}} (\bibinfo{year}{2023}{\natexlab{a}}), \eprint{2309.12079}.

\bibitem[{\citenamefont{Dumlu and Dunne}(2011)}]{Dumlu:2011rr}
\bibinfo{author}{\bibfnamefont{C.~K.} \bibnamefont{Dumlu}} \bibnamefont{and} \bibinfo{author}{\bibfnamefont{G.~V.} \bibnamefont{Dunne}}, \bibinfo{journal}{Phys. Rev. D} \textbf{\bibinfo{volume}{83}}, \bibinfo{pages}{065028} (\bibinfo{year}{2011}), \eprint{1102.2899}.

\bibitem[{\citenamefont{Mocken et~al.}(2010)\citenamefont{Mocken, Ruf, M\"uller, and Keitel}}]{Mocken:2010uhp}
\bibinfo{author}{\bibfnamefont{G.~R.} \bibnamefont{Mocken}}, \bibinfo{author}{\bibfnamefont{M.}~\bibnamefont{Ruf}}, \bibinfo{author}{\bibfnamefont{C.}~\bibnamefont{M\"uller}}, \bibnamefont{and} \bibinfo{author}{\bibfnamefont{C.~H.} \bibnamefont{Keitel}}, \bibinfo{journal}{Phys. Rev. A} \textbf{\bibinfo{volume}{81}}, \bibinfo{pages}{022122} (\bibinfo{year}{2010}).

\bibitem[{\citenamefont{Sah and Singh}(2023{\natexlab{b}})}]{Sah:2023udt}
\bibinfo{author}{\bibfnamefont{D.}~\bibnamefont{Sah}} \bibnamefont{and} \bibinfo{author}{\bibfnamefont{M.~P.} \bibnamefont{Singh}} (\bibinfo{year}{2023}{\natexlab{b}}), \eprint{2301.06545}.

\bibitem[{\citenamefont{Olver et~al.}(2010)\citenamefont{Olver, Lozier, Boisvert, and Clark}}]{article}
\bibinfo{author}{\bibfnamefont{F.}~\bibnamefont{Olver}}, \bibinfo{author}{\bibfnamefont{D.}~\bibnamefont{Lozier}}, \bibinfo{author}{\bibfnamefont{R.}~\bibnamefont{Boisvert}}, \bibnamefont{and} \bibinfo{author}{\bibfnamefont{C.}~\bibnamefont{Clark}} (\bibinfo{year}{2010}).

\end{thebibliography}

\end{document}